%% file: main.tex
\documentclass[]{aastex631}

\usepackage{amssymb}
\usepackage{amsmath}
\usepackage{gensymb}
\usepackage{wasysym}
\usepackage{chemformula}

\begin{document}

\title{Effect of Equation of State and Cutoff Density in Smoothed Particle Hydrodynamics Simulations of the Moon-Forming Giant Impact}

\correspondingauthor{Scott D. Hull}
\email{shull4@ur.rochester.edu}

\author[0000-0003-4216-7733]{Scott D. Hull}
\affiliation{Department of Earth and Environmental Sciences, University of Rochester, P.O. Box 270221, Rochester, NY 14627, USA}

\author[0000-0001-5014-0448]{Miki Nakajima}
\affiliation{Department of Earth and Environmental Sciences, University of Rochester, P.O. Box 270221, Rochester, NY 14627, USA}

\author[0000-0002-6638-7223]{Natsuki Hosono}
\affiliation{Center for Planetary Science, Integrated Research Center of Kobe University, 7-1-48 Minatojima-Minamimachi, Chuo-ku, Kobe 650-0047, Japan}
\affiliation{Center for Mathematical Science and Advanced Technology, Japan Agency for Marine-Earth Science and Technology, 3173-25, Showa-machi, Kanazawa-ku Kanagawa, Yokohama 236-0001, Japan}

\author[0000-0002-2342-3458]{Robin M. Canup}
\affiliation{Planetary Science Directorate, Southwest Research Institute, Boulder, CO, USA}

\author[0000-0001-7098-8198]{Rene Gassm\"{o}ller}
\affiliation{Department of Geological Sciences, University of Florida, Gainesville, FL 32611, USA}




\begin{abstract}
\input{abstract}
\end{abstract}

\section{Introduction}
\label{intro}
\input{introduction}
\section{Model}
\label{model}
\input{model}
\section{Results}
\label{results}
\input{results}
\section{Discussion}
\label{discussion}
\input{discussion}
\section{Conclusions}
\label{conclusions}
\input{conclusions}
\section{Acknowledgements}
\label{acknowledgement}
\input{acknowledgement}
%
\clearpage
\appendix
\section{Effect of Density on Smoothing Length and the Kernel}
\label{appendix_A}
\input{appendix_A}
\clearpage
\section{Results of \texorpdfstring{$b=0.75$}{b=0.75} Simulations}
\label{appendix_B}
\input{appendix_B}
\clearpage
\section{Additional Results}
\label{appendix_C}
\input{appendix_C}
\clearpage
\bibliography{main}{}
\bibliographystyle{aasjournal}



\end{document}

%% file: abstract.tex
The amount of vapor in the impact-generated protolunar disk carries implications for the dynamics, devolatilization, and moderately volatile element (MVE) isotope fractionation during lunar formation.  The equation of state (EoS) used in simulations of the giant impact is required to calculate the vapor mass fraction (VMF) of the modeled protolunar disk.  Recently, a new version of M-ANEOS (``Stewart M-ANEOS'') was released with an improved treatment of heat capacity and expanded experimental Hugoniot.  Here, we compare this new M-ANEOS version with a previous version (``N-SPH M-ANEOS'') and assess the resulting differences in smoothed particle hydrodynamics (SPH) simulations.  We find that Stewart M-ANEOS results in cooler disks with smaller values of VMF and results in differences in disk mass that are dependent on the initial impact angle.  We also assess the implications of the minimum ``cutoff'' density ($\rho_{c}$), similar to a maximum smoothing length, that is set as a fast-computing alternative to an iteratively calculated smoothing length.  We find that the low particle resolution of the disk typically results in $>40\%$ of disk particles falling to $\rho_c$, influencing the dynamical evolution and VMF of the disk.  Our results show that choice of EoS, $\rho_{c}$, and particle resolution can cause the VMF and disk mass to vary by tens of percent.  Moreover, small values of $\rho_{c}$ produce disks that are prone to numerical instability and artificial shocks.  We recommend that future giant impact SPH studies review smoothing methods and ensure the thermodynamic stability of the disk over simulated time.

%% file: introduction.tex
\subsection{Origin of the Moon}
\label{intro:OriginoftheMoon}
The origin of the Moon remains an active but unsolved area of research.  Any model of the lunar origin must satisfy the fundamental observables of the contemporary Earth-Moon system, including:

\begin{enumerate}
    \item Sufficient material to reproduce the masses of the Earth ($M_\oplus = 5.97 \times 10^{24}$ kg) and the Moon ($M_{\rm L} = 7.35 \times 10^{22}$ kg).
    \item Combined angular momentum of the protoearth and the protolunar source material that has a lower bound of the observed, uniquely high Earth-Moon angular momentum ($L_{\rm EM} = 3.5 \times 10^{34}$ $\rm kg \cdot m^2 / s$), where a reasonable excess could be lost over time through additional processes like evection resonance with the Sun \citep{touma1998resonances, cuk2012making, zahnle2015tethered, tian2017coupled, rufu2020tidal}.
    \item Near-identical ratios of key isotope groups between the Earth and Moon that suggest a common material origin.
    \item A significant bulk \ch{Fe} depletion in the Moon as exhibited by its small core.
    \item A broad depletion in volatile elements and an enrichment of heavy moderately volatile isotopes.
\end{enumerate}

The last three criteria have been the particular focus of significant research since the Apollo sample returns revealed the geochemistry of extensive portions of the Moon and are covered in more detail in Section \ref{intro:composition}.

The oldest models of lunar formation include fission from the Earth due to an initially rapid rotation \citep{darwin1880xx, darwin1898tides}, co-accretion with the Earth \citep{weidenschilling1986origin}, and gravitational capture of the Moon as a rogue asteroid \citep{buck1982lunar, nakazawa1983origin}.  However, none of these models are able to sufficiently satisfy all of the observables listed above.

In the 1970s, a giant impact origin of the Moon was proposed.  In what has become known as the canonical framework, a Mars-sized body, ``Theia'', impacted the Earth at an oblique angle and produced a protolunar debris disk in a stable Earth orbit \citep[e.g.,][]{hartmann1975satellite, cameron1976origin, canup2001origin, nakajima2021scaling}.  Such an impact would have been low-energy (occurring at roughly the mutual escape velocity of the protoearth and Theia) and of high accretion efficiency, meaning that very little mass escaped the Earth-Moon system following the impact.  More recent developments in the giant impact hypothesis have arisen largely due to increasing computational power, which allows for the direct simulation of the impact and the following tens of hours of disk evolution.  Smoothed particle hydrodynamics (SPH) \citep{lucy1977numerical, gingold1977smoothed, monaghan1982particle, monaghan1983shock} is a particle simulation method that is commonly used to model impacts.  The governing SPH equations act as convolution functions which map the conservation equations onto a proximity-weighted kernel function, thereby allowing for fluid-like behavior to be simulated and the giant impact dynamics to be modelled in detail.  While alternative grid-based and meshless methods exist \citep[e.g.,][]{crawford2010giant, canup2013lunar, hopkins2015new, deng2019primordial}, SPH simulations are the numerical basis for the majority of giant impact investigations \citep[e.g.,][]{canup2001origin, canup2004dynamics}.

SPH simulations of the canonical giant impact successfully reproduce the masses and dynamics of the observed Earth-Moon system.  The low impact velocity ($\sim 9$ km/s) and oblique impact angles $\sim 45 \degree$ generally produce disks with masses between $1-2 M_{\rm L}$ and disk angular momenta $> 0.18 L_{\rm EM}$, which is a lower bound on disk angular momentum if the Moon formed from debris exterior to the Roche limit \citep[]{canup2004simulations, salmon2012lunar}.  However, a well-known weakness of the canonical model is that the resulting protolunar disk is usually composed of $> 70 \%$ Theia silicate, which contradicts the isotopic similarity between Earth and the Moon unless Theia and the Earth had identical abundances of several key isotope systems \citep{mastrobuono2015primordial, dauphas2017isotopic, mastrobuono2017composition}.  Attempts to reconcile Theia's isotopic composition with the canonical setup \citep{meier2014origin, nielsen2021isotopic} and alternative setups to the canonical giant impact model have been implemented in order to satisfy the isotopic constraints, although often with worse dynamical matches to the Earth-Moon system than the canonical model.  Such alternative models include the fast-spinning protoearth \citep{cuk2012making}, a hit and run impact \citep{reufer2012hit}, a half Earths impact \citep{canup2012forming}, a synestia \citep{lock2018origin}, multiple impacts \citep{rufu2017multiple}, the magma ocean protoearth \citep{hosono2019terrestrial}, hit, run, and return impacts \citep{asphaug2021collision}, and immediate protolunar seed formation \citep{kegerreis2022immediate}.  To date, no single model of the giant impact has completely satisfied both the dynamical and compositional constraints of the observable Earth-Moon system.

\subsection{Volatile Composition of the Moon and MVE Isotope Fractionation} \label{intro:composition}

The Earth and Moon share near-identical abundances of key isotope ratios, including the well-studied $\rm ^{16}O,^{17}O,^{18}O$ ternary isotope system \citep{wiechert2001oxygen, spicuzza2007oxygen, hallis2010oxygen, young2016oxygen} and the $\rm ^{182}W$ isotope system, making for a strong argument that the Moon's precursor material was sourced primarily from the BSE after the Earth had differentiated much of its present-day core.  The Earth and Moon also share bulk compositional similarities, albeit with some important differences.  The BSE and bulk silicate Moon (BSM) share approximately chondritic proportions of the major refractory elements with the exception of \ch{FeO}, which is thought to be enriched in the BSM by a factor of 1.5-1.6 \citep{ringwood1977composition}.  However, the bulk Moon itself is depleted in \ch{Fe} relative to the bulk Earth by at least a factor of 2 owing to its small core \citep{jones2000geochemical}.  Additionally, while the BSE is depleted in volatiles relative to chondrites, the BSM is depleted in volatiles relative to the BSE, implying significant vaporization and more devolatilization than the BSE during lunar formation if the protolunar material was originally chondritic in volatile abundances \citep{ringwood1977basaltic, wolf1980moon, jones2000geochemical, day2014evaporative}.

Further insight into the formation of the Moon is given by the moderately volatile elements (MVEs).  MVEs are a class of elements with condensation temperatures between \ch{Mg}, \ch{Fe}, \ch{Si} and \ch{FeS} \citep{palme1988moderately}, making them particularly sensitive to evaporation and condensation during planet formation.  Along with the BSM's volatile depletion relative to the BSE, lunar sample analyses have shown that the Moon is also enriched in heavy MVE isotopes relative to the BSE, notably $\rm \delta$\ch{^{41}K} \citep{wang2016potassium, nie2019vapor, tian2020potassium, dauphas2022extent}, $\rm \delta$\ch{^{87}Rb} \citep{pringle2017rubidium, nie2019vapor, dauphas2022extent}, $\rm \delta$\ch{^{66}Zn} \citep{paniello2012zinc, kato2015extensive}, $\rm \delta$\ch{^{65}Cu} \citep{herzog2009isotopic}, $\rm \delta$\ch{^{71}Ga} \citep{kato2017gallium, wimpenny2022gallium}.  Tin ($\rm \delta$\ch{^{124}Sn}) is the only MVE that has a light isotope enrichment in the BSM \citep{wang2019tin}.  The heavy MVE isotope enrichment in the Moon, along with its broad volatile depletion, likely occurred by evaporative mass loss and kinetic isotope fractionation from the protolunar material \citep[e.g.,][]{canup2015protolunar, day2017late, nie2019vapor, tang2020evaporation, charnoz2021tidal}.

\subsection{Combining Dynamics and Geochemistry in the Disk} \label{intro:combine}

Vaporization during lunar formation is also important to the dynamical evolution of the disk.  Dynamical considerations of the disk's vapor fraction include magnetorotational instability and propensity for mixing in the protolunar disk \citep{carballido2016magneto, nie2019vapor}, hydrodynamic escape of volatiles \citep{desch2013isotopic, nakajima2015melting}, pressure support \citep{lock2018origin}, and changes in the distribution of angular momentum owing to condensation \citep{lock2019giant}.  Moreover, the disk evolution model proposed by \cite{nakajima2022large} suggests that forming a large Moon in a disk with a high vapor mass fraction (VMF), including a synestia, is difficult because lunar seeds would experience strong gas drag from the disk and fall onto the Earth, thereby failing to grow into a Moon.  This model supports less energetic impacts that produce disks with smaller VMF, such as the canonical model (VMF $\sim 10-30\%$, \citealt{nakajima2014investigation}) compared to more energetic models with high disk VMF, such as half-Earths model and fast-spinning Earth model (VMF $\sim 80-100\%$, \cite{nakajima2014investigation}).  Moreover, VMF is a function of the disk's pressure and temperature and controls the volatile speciation in the disk, some of which escaped and resulted in the Moon's volatile depletion and MVE isotope fractionation \citep[e.g.,][]{canup2015protolunar, lock2018origin, charnoz2021tidal}.  Therefore, accurate models of vapor production are crucial for resolving the origin of the Moon.

It is not straightforward to identify the VMF of the Moon-forming disk, as liquid-vapor mixtures are not numerically resolvable in SPH giant impact simulations owing to their extreme density differences \citep{ritchie2001multiphase, ruiz2022dealing}.  However, VMF can be obtained from SPH simulations by post-processing the thermodynamic state of the disk, which is given by the chosen equation of state (EoS).  This makes the calculated VMF highly dependent on the EoS as well. VMF can also depend on the particle resolution of the SPH simulation, as discussed in Section \ref{outcomes-b075}.

Here we investigate the thermodynamic state and VMF of the resulting disks from SPH simulations of the canonical giant impact using a newly released version of the Modified Analytical Equation of State (M-ANEOS) \citep{stewart2020shock}.  We compare these results against simulations using a previous version of M-ANEOS.  We also investigate dynamical and thermodynamic differences in the disks resulting from the minimum ``cutoff'' density ($\rho_{c}$), which is a computationally expedient method for resolving the maximum smoothing length in low particle resolution regions of the simulation.  In total, we present and compare 18 SPH simulations of the canonical giant impact across two different impact angles, two different EoS, four values of $\rho_{c}$, and using resolutions between $10^5$-$10^7$ particles.

%% file: model.tex
\subsection{FDPS SPH} \label{intro-fdps-sph}

We conduct simulations of the canonical Moon-forming giant impact using SPH.  SPH is a Lagrangian numerical method which is composed of discrete, point-like particles that interact with one another through a proximity-weighted, smoothing kernel function $W (\mathbf{r}_{i} - \mathbf{r}_{j}, h_{i})$, where $\mathbf{r}$ are the particle position coordinates and $h_{i}$ is the particle-dependent smoothing length.  At each timestep and for each particle, the system evolves in response to forces by convolution of parameters $A$ (e.g., density, energy) onto $W$ or its gradient \citep{monaghan1992smoothed},
\begin{equation}
    A_{i}(\mathbf{r}_{i}) = \sum_{j}^{N} \frac{m_{j} A_{j} W(\mathbf{r}_{i} - \mathbf{r}_{j}, h_{i})}{\rho_{j}} \label{sph_equation},
\end{equation}
where $i$ is the given particle, $j$ are neighboring particles within the kernel, $N$ is the total number of particles in the simulation, $m_{j}$ is the particle mass, and $\rho_{j}$ is the particle density.  Equation \eqref{sph_equation} is repeated for every particle in the system.  For example, a particle's density ($\rho_{i}$) can be calculated using equation \eqref{sph_equation} as,
\begin{equation}
    \rho_{i} = \sum_{{j}}^{N} m_{j} W (\mathbf{r}_{i} - \mathbf{r}_{j}, h_{i}), \label{sph-density-dis}
\end{equation}
where $W$ acts as a volume element for the particle.  We use a Wendland C6 kernel with a variable $h$ and a support radius, such that modified smoothing length is $H_{i} = 2.5 h_{i}$ (see Appendix \ref{appendix_A}).  The calculation of $h_{i}$ for each particle is inversely proportional to $\rho_{i}$ via the relationship $h_{i} = 1.2 \left (m_{i} / \rho_{i} \right )^{1/3}$, requiring equation \eqref{sph-density-dis} and the equation for $h_{i}$ to be solved together in a short iterative loop.  We note that values of $h_{i} = 1.866 \left (m_{i} / \rho_{i} \right )^{1/3}$ and $H_{i} = 2.449 h_{i}$ have been previously proposed for the Wendland C6 kernel \citep{dehnen2012improving}, although we do not expect these differences to undermine the results presented here.  We do not force $h_i$ to overlap a set number of neighboring particles, as smoothing lengths in the disk can become very large and cause unrealistic spreading and mass loss in the disk \citep{canup2013lunar} as well as a slowdown in computational efficiency \citep{genda2011merging}.

SPH directly solves for the conservation of mass, momentum, and energy using equation \eqref{sph_equation} and its differential forms \citep{monaghan1992smoothed}.  To close the equations and obtain pressure and thermodynamic variables, an analytical or tabular EoS needs to be implemented with $\rho$ and internal energy, $U$, as inputs.  Generally, giant impact simulations can only run for a couple of simulated days before the build-up of numerical error causes disk particles to lose angular momentum and fall inwards towards Earth \citep{canup2013lunar}.  We therefore run our SPH models for 50 simulated hours with an insignificant numerical viscosity expected.

Here, we introduce our SPH code, FDPS SPH\footnote{The FDPS SPH package can be cloned from \url{https://github.com/NatsukiHosono/FDPS_SPH}.}, based on the Framework for Developing a Particle Simulator (FDPS) \citep{hosono2016giant, hosono2017unconvergence, iwasawa2016implementation, namekata2018fortran}.  FPDS SPH is a native C++ package and is capable of scaling to supercomputers with large multithreading or multiprocessing environments.  In our simulations, we have found that FDPS SPH can currently run giant impact simulations composed of $10^6$ particles at a rate of 1.1 simulated hours per wall clock hour with 200 cores on the BlueHive Cluster at the University of Rochester.  To benchmark its performance and prove its numerical stability, FDPS SPH has been validated against the analytical solution for a shock tube.  FDPS SPH advances time via leapfrog integration over a Courant–Friedrichs–Lewy (CFL) timestep stability criterion and employs the SPH viscosity implementation assessed by \cite{monaghan1997sph}.  Timesteps are typically on the order of seconds.  We also utilize the Balsara switch for modulating artificial viscosity and avoiding interparticle penetration \citep{balsara1995neumann}.

\subsection{Cutoff Density, $\rho_{c}$} \label{model-intro}
FDPS SPH requires a ``cutoff density'', $\rho_{c}$, which is the minimum density allowed for all particles in the simulation.  This method has also been implemented in SPH codes used by \cite{genda2011merging} and \cite{nakajima2014investigation, nakajima2015melting}.  $\rho_c$ is enforced at the beginning of the loop per timestep following the calculation, such that the calculation $h_i$ and all other parameters follow. The density calculation given by equation \ref{sph-density-dis} is dependent on the availability and proximity of neighboring particles.  When a particle becomes isolated and there are not a sufficient number of neighbors within a maximum smoothing length, such as in low resolution structures, $\rho_{i}$ can fall to just the contribution of just a few nearby particles or the single, given particle.  The specified value of $\rho_{c}$ is meant to mitigate this phenomenon and prevent numerical errors.  As smoothing length is a function of density, setting a value of $\rho_c$ also defines a maximum smoothing length in the simulation for isolated particles.  In contrast to SPH codes that require expensive smoothing length convergence routines to capture a specific number of neighboring particles, $\rho_c$ is a comparatively simple and inexpensive way to control the maximum extent of the smoothing length.  Values of $\rho_c = 5$ $\rm kg/m^3$ and $\rho_c = 2000$ $\rm kg/m^3$ correspond to $H_i = 3203$ km and $H_i = 434$ km respectively for a $10^6$ particle simulation.  The implementation of $\rho_c$ helps to prevent particles from interacting with one another at unrealistically large distances, avoid artificial spreading of the disk, keep particles from reaching unphysically small densities and temperatures in the low-resolution disk, and to optimize parallel computing without the expensive computational overhead of searching for neighboring particles.

A numerical weakness of simulated canonical giant impact disks is that they represent only $\sim 1$-$2\%$ of the total mass as most particles compose the protoearth, thereby making the disks very low resolution features in SPH.  The lack of particles causes the disk density to fall to the specified value of $\rho_{c}$, therefore making the specified value of $\rho_{c}$ highly relevant to the outcome of the disk.  To our knowledge, the effects of $\rho_c$ or a maximum smoothing length on the thermodynamic state of the disk have not been assessed in previous studies of SPH giant impacts.  We therefore assess simulations across a range of different values of $\rho_{c}$ (5, 500, 1000, and 2000 $\rm kg/m^3$).  We review the effects of low particle resolution and $\rho_{c}$ as they relate to SPH in more detail in Section \ref{treatment-rho-c}.

\subsection{Choice of EoS} 
\label{choice-of-eos}

Giant impact hydrocodes require an EoS that can span a wide range of temperatures, pressures, and densities suited to complex geological materials.  ANEOS (ANalytical Equation of State) was developed in the 1970s as an analytical method for treating molecular species without the restraints of tabular data and is able to capture phase state transitions \citep{thompson1972improvements}.  ANEOS is derived from the Helmholtz free energy, $F$, as the sum of temperature-independent contributions, $F_{\rm cold}(\rho)$, a thermal contribution, $F_{\rm thermal} (\rho, T)$, and an electronic contribution, $F_{\rm electronic} (\rho, T)$ \citep{thompson1972improvements, melosh2007hydrocode},
\begin{equation}
    F (\rho, T) = F_{\rm cold} (\rho) + F_{\rm thermal} (\rho, T) + F_{\rm electronic} (\rho, T), \label{helmholtz_aneos}
\end{equation}

A limitation of ANEOS is that the vapor species is assumed to be ionized as an ideal monatomic mixture, which incurs a significant energy cost to produce relative to molecular vapors.  However, the vapor in the post-impact disk is expected to be a complex mixture of monatomic and molecular gasses \citep{visscher2013chemistry}.  This deficiency in ANEOS motivated the development of M-ANEOS \citep[Modified ANalytical Equation of State,][]{melosh2007hydrocode}, which treats the vapor as molecular clusters by including their molecular Helmholtz free energy as a function of each species' partition function.  M-ANEOS has since become the ubiquitous version of ANEOS for modelling the Moon-forming giant impact \citep[e.g.,][]{canup2004simulations, reufer2012hit, nakajima2014investigation, nakajima2015melting}.

Recently, the treatment of heat capacity in M-ANEOS has come under scrutiny because M-ANEOS relies on a hard-coded Dulong-Petit limit for resolving heat capacity in the thermal term, $F_{\rm thermal} (\rho, T)$.  This limit underestimates the heat capacity of liquid forsterite, which can produce erroneous temperatures \citep{stewart2020shock}.  To compensate for these shortcomings, \cite{stewart2020shock} modified M-ANEOS to include a user-defined parameter, $f_{cv}$, in the heat capacity term in $F_{\rm thermal} (\rho, T)$,
\begin{equation}
    F_{\rm thermal} (\rho, T) = N_0 kT \left [f_{cv} \left (3 \ln{\left ( 1 - e^{-\theta / T} \right )} - \mathcal{D} (\theta / T) \right ) + \frac{3}{2} \frac{1}{{b'}} \ln{\left ( 1 + \Psi^{b'} \right )} \right ], \label{helmholtz_aneos_thermal}
\end{equation}
where $N_0$ is the number of atoms per unit mass, $k$ is the Boltzmann constant, $T$ is temperature, $\theta$ is the Debye temperature, $\mathcal{D}$ is a Debye function, ${b'}$ is a parameter to adjust the critical point, and $\Psi$ is dimensionless function that relates phase states.  Combined with newly-expanded Hugoniot data up to 950 GPa \citep{root2018principal} and the inclusion of a melt curve, \cite{stewart2020shock} found that a fitted value of $f_{cv} = 1.35$ for forsterite and $f_{cv} = 1.33$ for iron improved convergence to the experimental Hugoniot.

Here, we compare this new version of M-ANEOS, which we refer to as ``Stewart M-ANEOS'', to a previous version, ``N-SPH M-ANEOS'', which is implemented in \cite{nakajima2014investigation, nakajima2015melting} as well as in the GADGET2 code \citep{marcus2011role, cuk2012making}.  Given the computational expense of the M-ANEOS subroutines, both EoS are implemented in FDPS SPH as 120x120 grid-based tables whose grid spacings become finer with decreasing density.  These thermodynamic results are interpolated on the tables as a function of $\rho$ and $U$ as provided by SPH.  N-SPH M-ANEOS does not include the heat capacity correction found in Stewart M-ANEOS and has erroneous relationships between particle velocity and temperature extrapolated above its sub-200 GPa experimental Hugoniot \citep{root2018principal}, both contributing to systematically higher disk temperatures and VMF than Stewart M-ANEOS. Recently, \cite{hosono2022influence} provided the first SPH implementation of Stewart M-ANEOS for a preexisting magma ocean and underlying solid rock during the giant impact while assessing its ability to the reproduce the refractory isotopic similarities between the Moon and the Earth.  While their study found that Stewart M-ANEOS placed too much impactor material into the disk to explain the Moon's composition, they did not assess specific dynamical differences or vapor production.

\subsection{Initial Conditions for SPH Canonical Giant Impact Simulations} \label{model-initial-conditions}

\begin{deluxetable}{c c c c c c c c c}
    \tabletypesize{\scriptsize}
    \tablewidth{0pt}
    \tablecaption{Canonical Giant Impact Model Run Parameters \label{table:GI_initial_conditions}}
    \tablehead{
    \colhead{Run Name} & \colhead{$b$} & \colhead{EoS} & \colhead{$\rho_{c}$ ($\rm kg/m^3$)} & \colhead{$N_{\rm tot}$} & \colhead{$N_{\rm tar}$} & \colhead{$R_{\rm tar}$ (km)} & \colhead{$R_{\rm imp}$ (km)} & \colhead{$v_{\rm imp}$ (m/s)}
    }
    \startdata
    {5b073S} & {0.73} & {Stewart M-ANEOS} & {5} & {1149900} & {1000400} & {6288.54} & {3523.86} & {9098.06} \\
    {5b073N} & {0.73} & {N-SPH M-ANEOS} & {5} & {1149900} & {1000400} & {6209.29} & {3479.23} & {9156.44} \\ 
    {5b075S} & {0.75} & {Stewart M-ANEOS} & {5} & {1149900} & {1000400} & {6288.54} & {3523.86} & {9098.06} \\ 
    {5b075N} & {0.75} & {N-SPH M-ANEOS} & {5} & {1149900} & {1000400} & {6209.29} & {3479.23} & {9156.44} \\ 
    {500b073S} & {0.73} & {Stewart M-ANEOS} & {500} & {1149900} & {1000400} & {6288.89} & {3524.14} & {9097.77} \\
    {500b073N} & {0.73} & {N-SPH M-ANEOS} & {500} & {1149900} & {1000400} & {6208.94} & {3479.44} & {9156.10} \\
    {500b075S} & {0.75} & {Stewart M-ANEOS} & {500} & {1149900} & {1000400} & {6288.89} & {3524.14} & {9097.77} \\
    {500b075N} & {0.75} & {N-SPH M-ANEOS} & {500} & {1149900} & {1000400} & {6208.94} & {3479.44} & {9156.10} \\
    {1000b073S} & {0.73} & {Stewart M-ANEOS} & {1000} & {1149900} & {1000400} & {6288.72} & {3524.01} & {9097.91} \\
    {1000b073N} & {0.73} & {N-SPH M-ANEOS} & {1000} & {1149900} & {1000400} & {6208.60} & {3479.25} & {9156.35} \\
    {1000b075S} & {0.75} & {Stewart M-ANEOS} & {1000} & {1149900} & {1000400} & {6288.72} & {3524.01} & {9097.91} \\
    {1000b075N} & {0.75} & {N-SPH M-ANEOS} & {1000} & {1149900} & {1000400} & {6208.60} & {3479.25} & {9156.35} \\
    {2000b073S} & {0.73} & {Stewart M-ANEOS} & {2000} & {1149900} & {1000400} & {6288.57} & {3523.82} & {9098.06} \\
    {2000b073N} & {0.73} & {N-SPH M-ANEOS} & {2000} & {1149900} & {1000400} & {6208.44} & {3479.60} & {9156.26} \\
    {2000b075S} & {0.75} & {Stewart M-ANEOS} & {2000} & {1149900} & {1000400} & {6288.57} & {3523.82} & {9098.06} \\
    {2000b075N} & {0.75} & {N-SPH M-ANEOS} & {2000} & {1149900} & {1000400} & {6208.44} & {3479.60} & {9156.26} \\
    {5b073S-high} & {0.73} & {Stewart M-ANEOS} & {5} & {10355722} & {9009522} & {6440.09} & {3565.52} & {9010.12} \\
    {2000b075N-low} & {0.75} & {N-SPH M-ANEOS} & {2000} & {114943} & {100000} & {6208.44} & {3479.60} & {9156.26} \\
    \enddata
\tablecomments{The setup parameters for each canonical giant impact SPH simulation.  The naming convention of each run gives the value of $\rho_c$, the impact parameter, and the EoS (``S'' for Stewart M-ANEOS, ``N'' for N-SPH M-ANEOS).  The impact velocity $v_{\rm imp}$ is always set as the mutual escape velocity between Earth and Theia, $v_{\rm esc}$.  For canonical impacts, the mass of the target protoearth is $M_{\rm tar} \sim M_\oplus = 5.29 \times 10^{24}$ kg and the mass of the impactor, Theia, is $M_{\rm imp} \sim M_{\rm Mars} = 7.91 \times 10^{23}$ kg, where $M_{\rm Mars}$ is the Martian mass.  $b$ is the scaled impact parameter, $\rho_{c}$ is the minimum "cutoff" density for every particle, $N_{\rm tot}$ is the total number of particles in the simulation, $R_{\rm tar}$ is the radius of the target protoearth, and $R_{\rm imp}$ is the radius of the impactor.}
\end{deluxetable}

We run 18 total SPH simulations of the canonical giant impact with varying EoS, $\rho_{c}$, particle resolution, and impact parameter, $b$, which is explained below (Table \ref{model-initial-conditions}).  The values of $\rho_{c}$ include $5$, $500$, $1000$, and $2000$ $\rm kg/m^3$, while we subset simulations between two different impact parameters: $b=0.73$ and $b=0.75$.  These simulations were performed on dedicated nodes on the University of Rochester's BlueHive supercomputer using the standard density-differential SPH (SSPH) method.  For each value of $\rho_{c}$ and both Stewart and N-SPH M-ANEOS, a target and impactor with a $30$ ${\rm wt.}\%$ iron core and $70$ ${\rm wt.}\%$ forsterite mantle were initially created with a spherical close-pack algorithm and allowed to settle for $\sim 2.5$ simulated days under the influence of self-gravity and the enforcement of material-specific isentropic conditions ($S_{\rm forsterite} = 3165$ J/kg/K, $S_{\rm iron} = 1500$ J/kg/K).  This process is similar to the ``warm start'' described by \cite{canup2004simulations}.  We assume a target of $M_{\rm tar} \sim 1 M_\oplus$ and $R_{\rm tar} \sim 1 R_{\oplus}$, with the combined mass of the target and the impactor of $M_{\rm tot} = 1.019 M_\oplus$ and an impactor with radius $R_{\rm imp} \sim R_{\rm Mars}$. All particles in the simulation are assigned equal masses regardless of their material type to avoid disproportionate gravitation among particles.

Once a target and impactor are generated, their relative positions and velocities are treated as a two-body problem and reverse-time integrated to a distance of $3 R_{\rm tar}$ such that the impact velocity is equal to the mutual escape velocity of Earth and Theia at the specified value of $b$.  The impact geometry is given by the scaled impact parameter $b = \sin{\theta}$, where $\theta$ is the impact angle on the target's equatorial plane.  A value of $b = 0$ gives a head-on impact, whereas a value of $b=1$ gives a perfectly grazing impact.  For the canonical setup, the impact velocity is typically $v_{\rm imp} \approx 9$ km/s and $\theta \approx 45 \degree$.  For our simulations, we choose $b=0.73$ and $b=0.75$ as two values of $b$ which have previously produced successful disks to give precedent for our analysis.  We neglect the effects of tidal distortion on the pre-impact trajectory of Theia, which we find to be minimal at our initial separation of $3 R_{\rm tar}$.  From these initial particle coordinates and velocities, the giant impact simulation begins.  Each giant impact is run to 50 simulated hours to track and compare the evolution of the disk before numerical viscosity begins to influence the simulation.  The iterative process for identifying planet, disk, and escaping particles relies on calculating the Keplerian orbital elements of each particle and is similar to that described by \cite{canup2001scaling}.  We run the iterative calculation twice, adjusting the bulk density of the protoearth accordingly on the second iteration to obtain our final solution.  We have produced results similar to those found in giant impact SPH simulations presented by \cite{canup2004simulations} and \cite{nakajima2014investigation}.

\subsection{Analysis of the Disk} \label{model-analysis}

Following the identification of the disk particles, we assess the VMF of the disk.  SPH does not internally differentiate between liquid and vapor and this is therefore achieved by post-processing SPH results.  VMF is given by a lever rule on the EoS-dependent liquid-vapor phase curves,
\begin{equation}
    {\rm VMF} = \frac{1}{N_{\rm disk}} \sum_{i}^{N_{\rm disk}} \frac{S_{i}(T_{i}) - S_{{l}}(T_{i})}{S_{v}(T_{i}) - S_{l}(T_{i})}, \label{vmf-eq}
\end{equation}
where $N_{\rm disk}$ is the total number of disk particles, $S_{i}(T_{i})$ is the specific entropy of the disk particle at its corresponding temperature $T_{i}$, $S_{l}(T_{i})$ is the corresponding entropy of the liquid phase curve at $T_{i}$, and $S_{v}(T_{i})$ is the corresponding specific entropy of the vapor phase curve at $T_{i}$.  We refer to specific entropy as simply entropy hereon.  To simplify visualization and the calculation of VMF, we define supercritical particles as those with temperatures that exceed the critical point and do not directly consider the pressure component as it does not meaningfully change the results of the VMF calculation.  The range of disk pressures ensures that the overwhelming majority of disk particles with temperatures above the critical point are either truly supercritical or otherwise purely vapor (see Appendix \ref{appendix_C} and \cite{caracas2023no}, their Figure 3).  All supercritical particles are treated as fully vaporized and are summed into equation \eqref{vmf-eq} accordingly.

Some previous SPH giant impact studies have computed circular disk particle orbits with equivalent angular momentum to the SPH-calculated orbit when resolving which particles will ultimately end up in the disk \citep[e.g.,][]{canup2004simulations, canup2013lunar, nakajima2014investigation}.  The physical justification for this process is that mutual collisions of mass in the disk will dampen orbital eccentricities faster than angular momentum can be transported \citep{canup2004dynamics, canup2013lunar}.  As a circular orbit is the lowest energy configuration for mutually gravitating objects, the decrease in orbital energy associated with the process is associated with an increase in the internal energy and entropy of disk material.  We therefore calculate the decrease in orbital energy associated with the equivalent circular orbits of disk particles \citep[e.g.,][]{canup2004simulations, canup2013lunar, nakajima2014investigation}.  Following the method of \cite{nakajima2014investigation}, we assume that this decrease in orbital energy is entirely converted to heat as the difference in orbital energy between the initially inclined, eccentric orbit and the final, equivalent circular orbit, $\Delta U_{i}$,
\begin{equation}
    \Delta U_{i} = -\frac{G M_\oplus}{2 a_{i}} \left (1 - \frac{1}{\left ( 1 - e_{i}^2 \right ) \cos{I_{i}}^2} \right),
\end{equation}
where $a_{i}$ is the particle's orbital semi-major axis, $e_{i}$ is its eccentricity, and $I_{i}$ is its inclination.  We apply this process to all disk particles regardless of phase state.  The total disk entropy gain due to orbital circularization, $\Delta S_{\rm circ}$, is given by the second law of thermodynamics,
\begin{equation}
    \Delta S_{\rm circ} = \sum_{i}^N \frac{\Delta U_{i}}{T_{i}},
\end{equation}
In agreement with \cite{nakajima2014investigation}, we find that orbital circularization of disk particles results in a $\sim 10\%$ mean entropy gain to the disk, although our results show that this effect can produce most of the disk's vapor as disk particles are shifted further into pure vapor, supercritical, or mixed-phase regions of the phase diagram.  For purposes of visualization and assessment we calculate the effects of the disks' circularization on entropy and VMF at each timestep given that this behavior is not captured directly in SPH.  However, the true effect of this feature is transient over the first 10-20 hours and approaches the steady-state solution given by the end of our simulations.

%% file: results.tex
\begin{deluxetable*}{r c c c c c c c c c}
    \tabletypesize{\scriptsize}
    \tablewidth{0pt}
    \tablecaption{Canonical Giant Impact Model Results ($b=0.73$) \label{table:gi_b073_outputs}}
    \tablehead{
        \colhead{} & \colhead{5b073S} & \colhead{5b073S-high} & \colhead{500b073S} & \colhead{1000b073S} & \colhead{2000b073S} & \colhead{5b073N} & \colhead{500b073N} & \colhead{1000b073N} & \colhead{2000b073N}
    }
    \startdata
    {Protoearth Mass ($M_\oplus$)} & {1.0} & {1.0} & {1.01} & {1.0} & {0.99} & {1.0} & {1.01} & {1.01} & {1.0} \\ 
    {Disk Mass ($M_{\rm L}$)} & {1.3} & {1.52} & {1.02} & {1.45} & {1.85} & {1.07} & {0.83} & {0.87} & {1.08} \\ 
    {Escaping Mass ($M_{\rm L}$)} & {0.25} & {0.12} & {0.11} & {0.13} & {0.09} & {0.11} & {0.09} & {0.15} & {0.08} \\ 
    {$N_{\rm Protoearth}$} & {1128298} & {10149772} & {1134333} & {1127908} & {1122888} & {1133538} & {1137158} & {1135655} & {1133783} \\ 
    {$N_{\rm disk}$} & {18085} & {190477} & {14104} & {20147} & {25713} & {14867} & {11554} & {12144} & {14981} \\ 
    {$N_{\rm escape}$} & {3517} & {15473} & {1463} & {1845} & {1299} & {1595} & {1188} & {2101} & {1136} \\ 
    {$N_{\rm total}$} & {1149900} & {10355722} & {1149900} & {1149900} & {1149900} & {1150000} & {1149900} & {1149900} & {1149900} \\ 
    {Disk Mass $\geq$ $R_{\rm Roche}$ ($M_{\rm L}$)} & {1.18} & {1.52} & {0.95} & {1.24} & {1.84} & {1.06} & {0.78} & {0.81} & {1.01} \\ 
    {$N_{\rm disk}$ $\geq$ $R_{\rm Roche}$} & {16450} & {190443} & {13162} & {17269} & {25565} & {14707} & {10872} & {11222} & {14040} \\ 
    {Disk Iron Mass Fraction ($\%$)} & {2.0} & {1.21} & {0.6} & {0.52} & {1.14} & {1.49} & {2.83} & {3.21} & {0.35} \\ 
    {Disk Iron Mass Fraction $\geq$ $R_{\rm Roche}$ ($\%$)} & {1.66} & {1.21} & {0.52} & {0.34} & {1.14} & {1.42} & {2.34} & {2.5} & {0.31} \\ 
    {Protoearth Avg. $\rho$ ($\rm kg/m^3$)} & {1035.02} & {1032.7} & {1026.25} & {1034.62} & {1037.84} & {1026.87} & {1019.68} & {1020.5} & {1025.86} \\ 
    {Protoearth $a$ (km)} & {11262.56} & {11275.09} & {11346.69} & {11264.41} & {11224.34} & {11338.63} & {11403.31} & {11392.54} & {11346.97} \\ 
    {Protoearth $b$ (km)} & {10856.91} & {10845.8} & {10845.69} & {10853.8} & {10849.07} & {10846.9} & {10834.39} & {10831.79} & {10844.01} \\ 
    {$L_{\rm disk}$ ($L_{\rm EM}$)} & {0.34} & {0.37} & {0.28} & {0.37} & {0.42} & {0.29} & {0.21} & {0.21} & {0.29} \\ 
    {$L_{\rm disk}$ $\geq R_{\rm Roche}$ ($L_{\rm EM}$)} & {0.32} & {0.37} & {0.27} & {0.34} & {0.42} & {0.28} & {0.2} & {0.2} & {0.28} \\ 
    {$L_{\rm total}$ ($L_{\rm EM}$)} & {1.41} & {1.43} & {1.43} & {1.41} & {1.4} & {1.42} & {1.43} & {1.43} & {1.43} \\ 
    {Disk VMF (w/ circ.)  ($\%$)} & {35.73} & {21.52} & {19.30} & {11.26} & {10.75} & {25.70} & {27.34} & {29.53} & {10.88} \\ 
    {Disk VMF (w/o circ.)  ($\%$)} & {29.40} & {10.75} & {2.87} & {1.49} & {2.54} & {23.59} & {18.23} & {16.68} & {2.73} \\ 
    {Avg. $S_{\rm disk}$ (w/ circ.) (J/kg/K)} & {6155.63} & {5295.08} & {5002.56} & {4638.46} & {4539.32} & {5293.26} & {4833.46} & {4796.38} & {4323.49} \\ 
    {Avg. $S_{\rm disk}$ (w/o circ.) (J/kg/K)} & {5815.31} & {4765.32} & {4255.76} & {4183.49} & {4121.33} & {5185.21} & {4499.65} & {4366.93} & {4046.38} \\ 
    {Avg. $\Delta S_{\rm circ}$ (J/kg/K)} & {340.2} & {529.95} & {746.75} & {454.74} & {417.76} & {108.05} & {333.71} & {429.45} & {277.08} \\ 
    {$M_{\rm M}$ ($M_{\rm L}$)} & {2.06} & {2.13} & {1.77} & {2.18} & {2.23} & {1.75} & {1.25} & {1.2} & {1.77} \\ 
    {Disk Theia Mass Fraction ($\%$)} & {63.13} & {66.67} & {66.78} & {69.79} & {69.4} & {69.97} & {71.7} & {74.1} & {72.95} \\ 
    {Avg. $T_{\rm disk}$ (K)} & {6345.79} & {4748.05} & {3664.25} & {3465.15} & {3444.84} & {6004.06} & {6637.25} & {6280.91} & {4342.08}
    \enddata
\tablecomments{End-state simulation outputs (50 simulated hours) corresponding to an impact parameter $b=0.73$ with setups given by Table \ref{table:GI_initial_conditions}.  Here, $M_\oplus$ is the mass of the Earth, $M_{\rm L}$ is the mass of the Moon, $N_{\rm i}$ is the number of particles, $R_{\rm Roche}$ is the Earth's Roche radius, $\rho$ is density, $a$ is the post-impact protoearth's equatorial radius at 50 hours, $b$ is the post-impact protoearth's polar radius at 50 hours, $L_{\rm i}$ is the angular momentum, VMF is the vapor mass fraction, the ``w/ circ'' indicates the inclusion of the orbital circularization of disk particles described in Section \ref{model-analysis} whereas ``w/o circ'' does not, $\Delta S_{\rm circ}$ is the average entropy gain per particle due to orbital circularization, $S_{\rm i}$ is the entropy, $M_{\rm M}$ is the predicted mass of the Moon given by equation 2 from \cite{canup2001origin}, and $T_{\rm disk}$ is the temperature of the disk.}
\end{deluxetable*}
\begin{deluxetable*}{r c c c c c c c c c}
    \tabletypesize{\scriptsize}
    \tablewidth{0pt}
    \tablecaption{Canonical Giant Impact Model Results ($b=0.75$) \label{table:gi_b075_outputs}}
    \tablehead{
        \colhead{} & \colhead{5b075S} & \colhead{500b075S} & \colhead{1000b075S} & \colhead{2000b075S} & \colhead{5b075N} & \colhead{500b075N} & \colhead{1000b075N} & \colhead{2000b075N} & \colhead{2000b075N-low}
    }
    \startdata
    {Protoearth Mass ($M_\oplus$)} & {1.0} & {1.0} & {1.01} & {1.0} & {1.0} & {1.0} & {1.01} & {1.01} & {1.0} \\ 
    {Disk Mass ($M_{\rm L}$)} & {0.47} & {0.78} & {0.26} & {0.83} & {0.48} & {0.95} & {0.73} & {0.74} & {1.19} \\ 
    {Escaping Mass ($M_{\rm L}$)} & {0.7} & {0.57} & {0.86} & {0.55} & {0.9} & {0.27} & {0.39} & {0.38} & {0.49} \\ 
    {$N_{\rm Protoearth}$} & {1133651} & {1131165} & {1134448} & {1130695} & {1130900} & {1132941} & {1134369} & {1134430} & {112563} \\ 
    {$N_{\rm disk}$} & {6522} & {10807} & {3580} & {11593} & {6600} & {13210} & {10120} & {10212} & {1656} \\ 
    {$N_{\rm escape}$} & {9727} & {7928} & {11872} & {7612} & {12500} & {3749} & {5411} & {5258} & {681} \\ 
    {$N_{\rm total}$} & {1149900} & {1149900} & {1149900} & {1149900} & {1150000} & {1149900} & {1149900} & {1149900} & {114900} \\ 
    {Disk Mass $\geq$ $R_{\rm Roche}$ ($M_{\rm L}$)} & {0.44} & {0.67} & {0.22} & {0.45} & {0.47} & {0.91} & {0.67} & {0.68} & {1.07} \\ 
    {$N_{\rm disk}$ $\geq$ $R_{\rm Roche}$} & {6134} & {9250} & {3119} & {6236} & {6488} & {12583} & {9345} & {9383} & {1487} \\ 
    {Disk Iron Mass Fraction ($\%$)} & {3.65} & {3.23} & {0.92} & {2.13} & {7.09} & {1.68} & {1.82} & {1.29} & {6.58} \\ 
    {Disk Iron Mass Fraction $\geq$ $R_{\rm Roche}$ ($\%$)} & {3.37} & {2.94} & {0.7} & {0.61} & {6.92} & {1.54} & {1.58} & {1.14} & {6.4} \\ 
    {Protoearth Avg. $\rho$ ($\rm kg/m^3$)} & {1020.62} & {1023.54} & {1018.55} & {1023.44} & {1023.59} & {1019.96} & {1018.77} & {1020.58} & {1025.79} \\ 
    {Protoearth $a$ (km)} & {11384.81} & {11355.39} & {11402.78} & {11355.13} & {11353.99} & {11387.98} & {11401.47} & {11388.05} & {11322.14} \\ 
    {Protoearth $b$ (km)} & {10826.06} & {10827.45} & {10821.57} & {10824.42} & {10826.97} & {10820.32} & {10820.97} & {10827.73} & {10819.38} \\ 
    {$L_{\rm disk}$ ($L_{\rm EM}$)} & {0.12} & {0.17} & {0.06} & {0.18} & {0.1} & {0.21} & {0.17} & {0.18} & {0.27} \\ 
    {$L_{\rm disk}$ $\geq R_{\rm Roche}$ ($L_{\rm EM}$)} & {0.11} & {0.15} & {0.05} & {0.11} & {0.1} & {0.2} & {0.16} & {0.17} & {0.24} \\ 
    {$L_{\rm total}$ ($L_{\rm EM}$)} & {1.48} & {1.48} & {1.5} & {1.47} & {1.48} & {1.46} & {1.47} & {1.46} & {1.5} \\ 
    {Disk VMF (w/ circ.) ($\%$)} & {67.98} & {40.20} & {44.74} & {34.76} & {78.25} & {46.72} & {40.77} & {38.22} & {30.13} \\ 
    {Disk VMF (w/o circ.) ($\%$)} & {55.49} & {24.41} & {23.51} & {15.44} & {66.31} & {20.17} & {19.10} & {15.34} & {11.07} \\ 
    {Avg. $S_{\rm disk}$ (w/ circ.) (J/kg/K)} & {7828.76} & {5808.71} & {5925.92} & {5359.48} & {7810.75} & {5491.1} & {5223.9} & {5134.91} & {4778.81} \\ 
    {Avg. $S_{\rm disk}$ (w/o circ.) (J/kg/K)} & {7194.66} & {5200.23} & {4986.43} & {4459.3} & {7265.32} & {4687.06} & {4601.57} & {4505.28} & {4298.12} \\ 
    {Avg. $\Delta S_{\rm circ}$ (J/kg/K)} & {632.71} & {608.01} & {938.24} & {899.9} & {542.83} & {804.02} & {622.23} & {629.52} & {480.32} \\ 
    {$M_{\rm M}$ ($M_{\rm L}$)} & {0.67} & {0.85} & {0.29} & {0.95} & {0.51} & {1.1} & {0.94} & {1.06} & {1.39} \\ 
    {Disk Theia Mass Fraction ($\%$)} & {60.32} & {67.9} & {61.42} & {71.14} & {56.68} & {71.91} & {69.31} & {69.19} & {76.57} \\ 
    {Avg. $T_{\rm disk}$ (K)} & {8536.22} & {6554.56} & {6325.06} & {4882.66} & {9504.66} & {6970.22} & {6872.69} & {6911.39} & {5886.74}
    \enddata
\tablecomments{End-state simulation outputs (50 simulated hours) corresponding to an impact parameter $b=0.75$ with setups given by Table \ref{table:GI_initial_conditions}.  See the caption of Table \ref{table:gi_b073_outputs} for an explanation of variables.}
\end{deluxetable*}
\begin{deluxetable*}{r c c c c c c c c c}
    \tabletypesize{\scriptsize}
    \tablewidth{0pt}
    \tablecaption{Secondary Impactor And Tail Results ($b = 0.73$) \label{table:si_and_tail_stats_b073}}
    \tablehead{
        \colhead{} & \colhead{5b073S} & \colhead{5b073S-high} & \colhead{500b073S} & \colhead{1000b073S} & \colhead{2000b073S} & \colhead{5b073N} & \colhead{500b073N} & \colhead{1000b073N} & \colhead{2000b073N}
    }
    \startdata
    {Disk From Protoearth ($\%$)} & {22.78} & {7.80} & {1.61} & {0.61} & {0.45} & {17.29} & {15.22} & {15.18} & {1.66}
 \\ 
    {Disk From Secondary Impactor ($\%$)} & {0.90} & {0.26} & {0.27} & {0.12} & {0.03} & {1.01} & {15.73} & {26.05} & {1.46}
 \\ 
    {Disk From Tail ($\%$)} & {70.17} & {88.24} & {89.19} & {91.98} & {94.59} & {65.47} & {61.59} & {47.70} & {87.29}
 \\ 
    {Disk From Other Debris ($\%$)} & {6.15} & {3.69} & {8.93} & {7.29} & {4.94} & {16.24} & {7.45} & {11.06} & {9.60}
 \\ 
    {Secondary Impactor Mass ($M_{\rm L}$)} & {3.38} & {3.66} & {3.80} & {3.76} & {3.07} & {3.80} & {4.66} & {5.04} & {3.88} 
 \\ 
    {Tail Mass ($M_{\rm L}$)} & {2.25} & {2.11} & {1.83} & {1.97} & {2.32} & {1.63} & {0.95} & {0.55} & {1.66}
 \\ 
    {Secondary Impactor Angular Momentum ($L_{\rm EM}$)} & {0.25} & {0.30} & {0.33} & {0.30} & {0.23} & {0.35} & {0.51} & {0.60} & {0.37}
 \\ 
    {Tail Angular Momentum ($L_{\rm EM}$)} & {0.50} & {0.47} & {0.42} & {0.45} & {0.49} & {0.36} & {0.25} & {0.16} & {0.37}
    \enddata
\tablecomments{Results of the secondary impactor and tail formed in the aftermath of the primary impact of runs with an impact parameter of $b=0.73$.  Rows 1-4 show how much of the disk is sourced from the protoearth, the secondary impactor, the tail, or elsewhere in the primary impact debris structure.  Also shown are the mass and angular momentum of the secondary impactor and the tail.  These results show that N-SPH M-ANEOS runs produce larger secondary impactors and smaller tails than Stewart M-ANEOS.  As disk particles are mainly sourced from the tail, Stewart M-ANEOS produces larger disks than N-SPH M-ANEOS within this simulation set.  These patterns are not observed in the $b=0.75$ simulation set, as secondary impactors form without a massive debris tail across runs from both EoS.}
\end{deluxetable*}
\begin{figure}[h]
\centering
\includegraphics[width=1\textwidth]{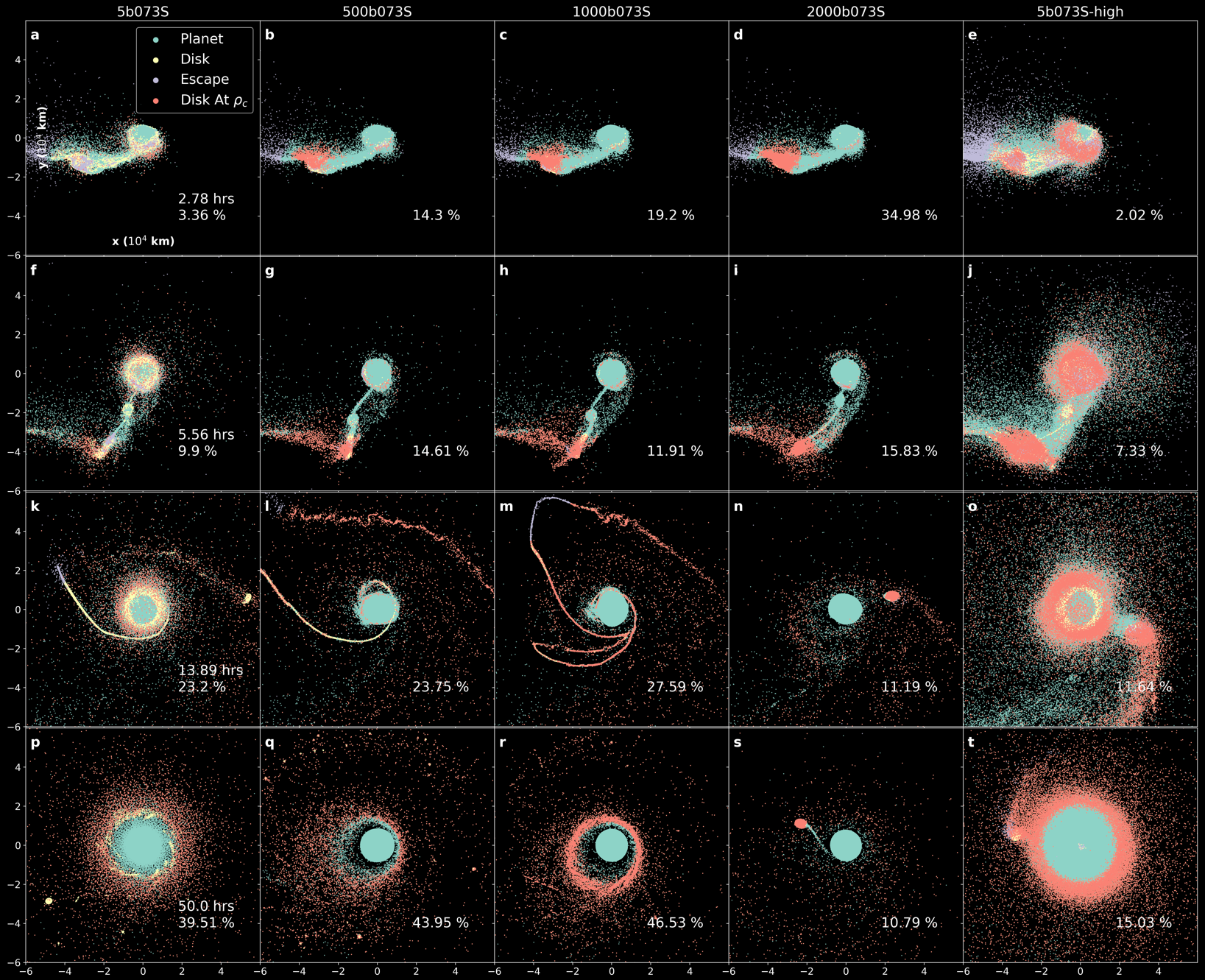}
\caption{Evolution of the giant impact and the protolunar disk for simulations using Stewart M-ANEOS with an initial impact parameter of $b=0.73$.  The axes give the x-y spatial coordinates relative to the protoearth's center of mass given in increments of $10^4$ km.  Particles are sorted by their z-axis values and colors indicate whether the particle is part of the protoearth (green), the disk (yellow), or escaping (purple).  Red particles indicate final disk particles that have density values equal to $\rho_c$ at the given time, and the percentage of final disk particles that are at the value of $\rho_c$ at each time is annotated in the lower-right corner of each subplot.  Each column corresponds to a unique run and each row is a snapshot of the simulation at \textbf{(a-e)} 2.78 hours, \textbf{(f-j)} 5.56 hours, \textbf{(k-o)} 13.89 hours, \textbf{(p-t)} 50 hours.}
\label{fig:source_scenes_b073_new}
\end{figure}
\begin{figure}[h]
\centering
\includegraphics[width=1\textwidth]{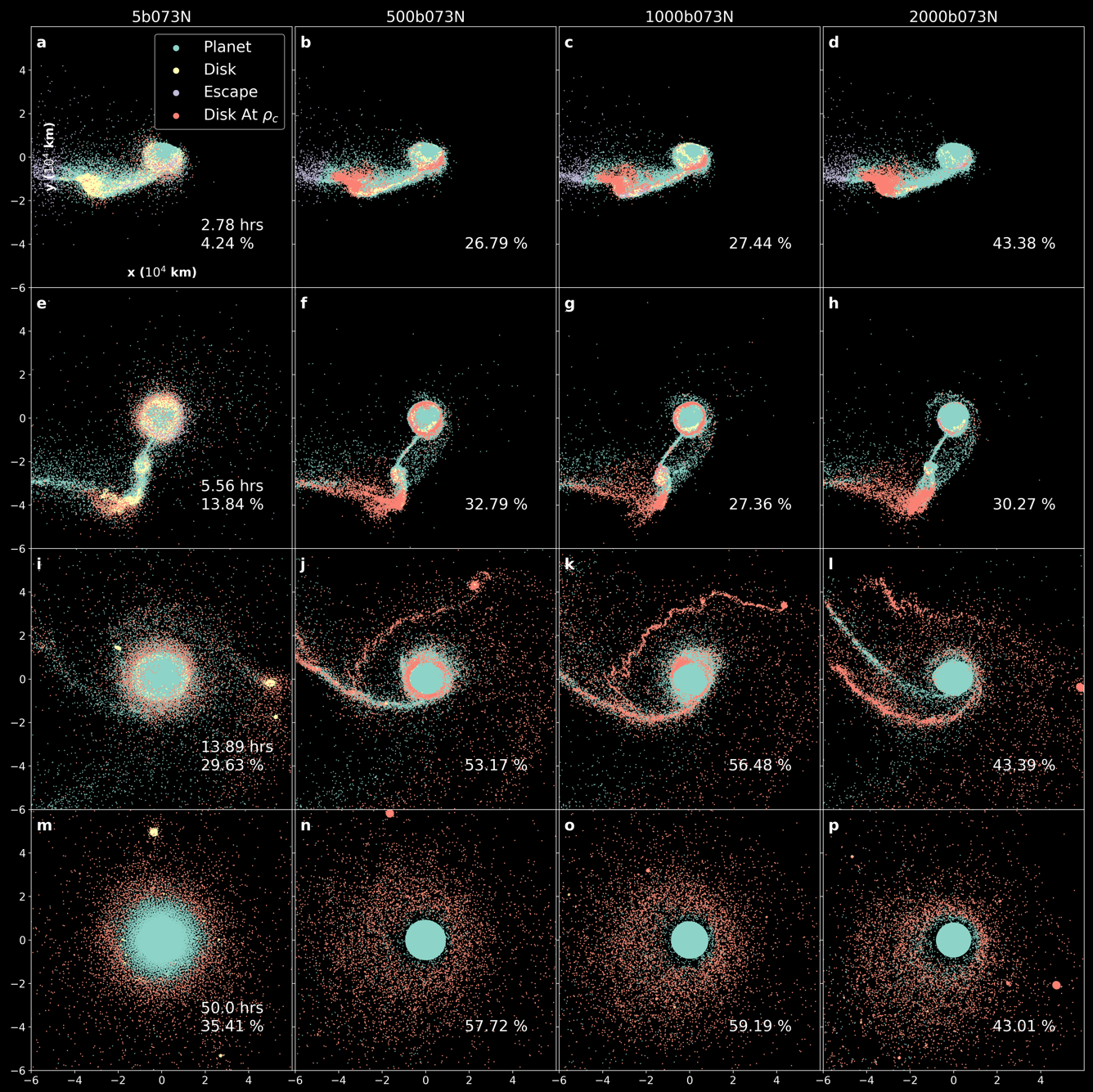}
\caption{Evolution of the giant impact and the protolunar disk for simulations using N-SPH M-ANEOS with an initial impact parameter of $b=0.73$.  The axes give the x-y spatial coordinates relative to the protoearth's center of mass given in increments of $10^4$ km.  Particles are sorted by their z-axis values and colors indicate whether the particle is part of the protoearth (green), the disk (yellow), or escaping (purple).  Red particles indicate final disk particles that have density values equal to $\rho_c$ at the given time, and the percentage of final disk particles that are at the value of $\rho_c$ at each time is annotated in the lower-right corner of each subplot.  Each column corresponds to a unique run and each row is a snapshot of the simulation at \textbf{(a-d)} 2.78 hours, \textbf{(e-h)} 5.56 hours, \textbf{(i-l)} 13.89 hours, \textbf{(m-p)} 50 hours.}
\label{fig:source_scenes_b073_old}
\end{figure}
\begin{figure}[h]
\centering
\includegraphics[width=1\textwidth]{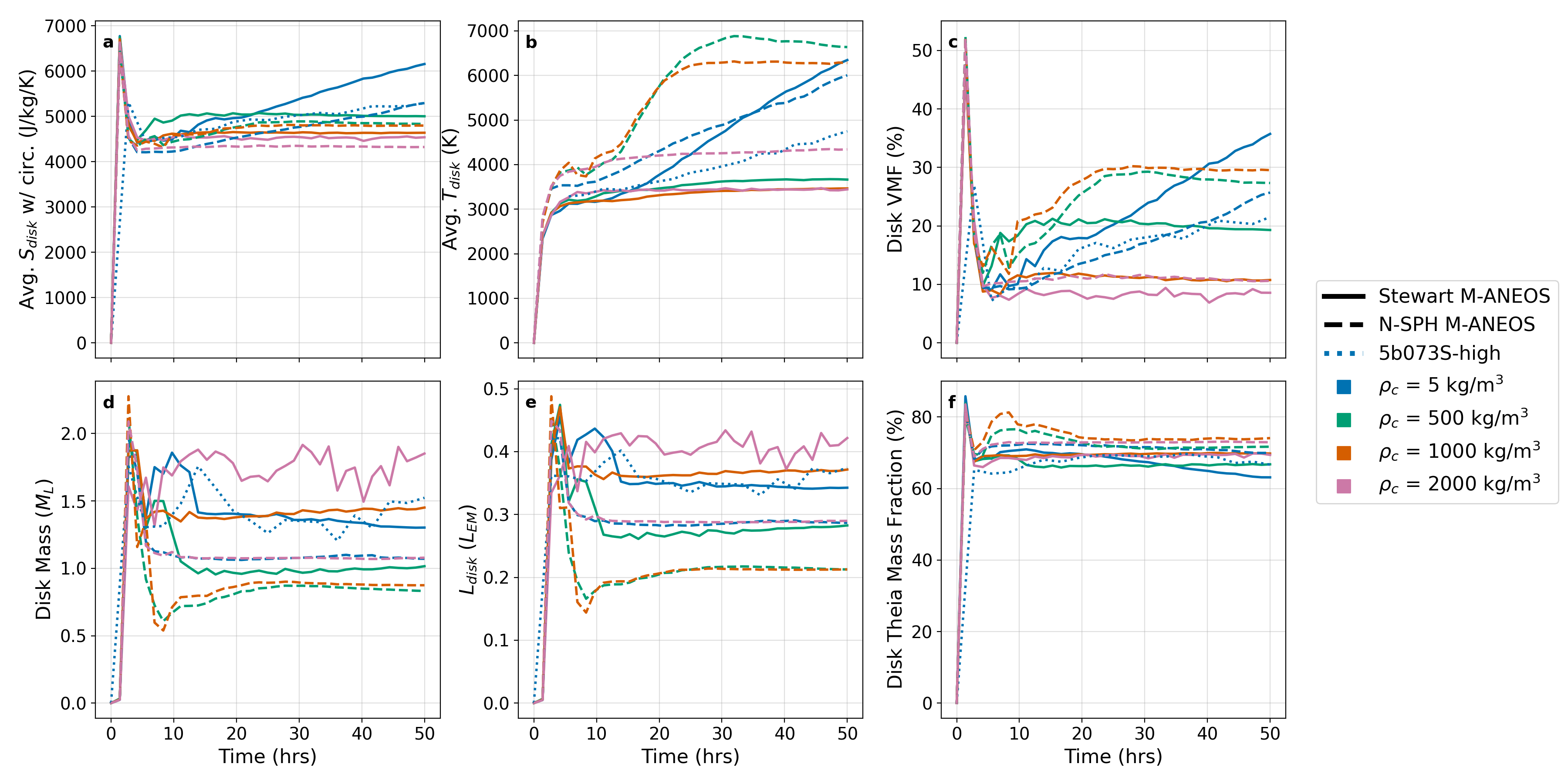}
\caption{Time evolution of the disk for simulations with impact parameter $b=0.73$.  Each cutoff density represents a distinct line color.  The solid lines represent Stewart M-ANEOS simulations, the dashed lines represent N-SPH M-ANEOS simulations, and the dotted lines corresponds to our high resolution run (5b073S-high).  \textbf{(a)} Average disk entropy with entropy gain due to orbital circularization assessed at each timestep included, \textbf{(b)} average disk temperature, \textbf{(c)} disk VMF calculated with effects of orbital circularization included, \textbf{(d)} disk mass, \textbf{(e)} disk angular momentum, \textbf{(f)} mass fraction of material in the disk originating from Theia.  A version of this figure without the effects of orbital circularization of the disk is available in Appendix \ref{appendix_C}, Figure \ref{fig:b073_disk_entropy_and_vmf_wo_circ}.}
\label{fig:b073_disk_entropy_and_vmf}
\end{figure}
\begin{figure}[h]
\centering
\includegraphics[width=1\textwidth]{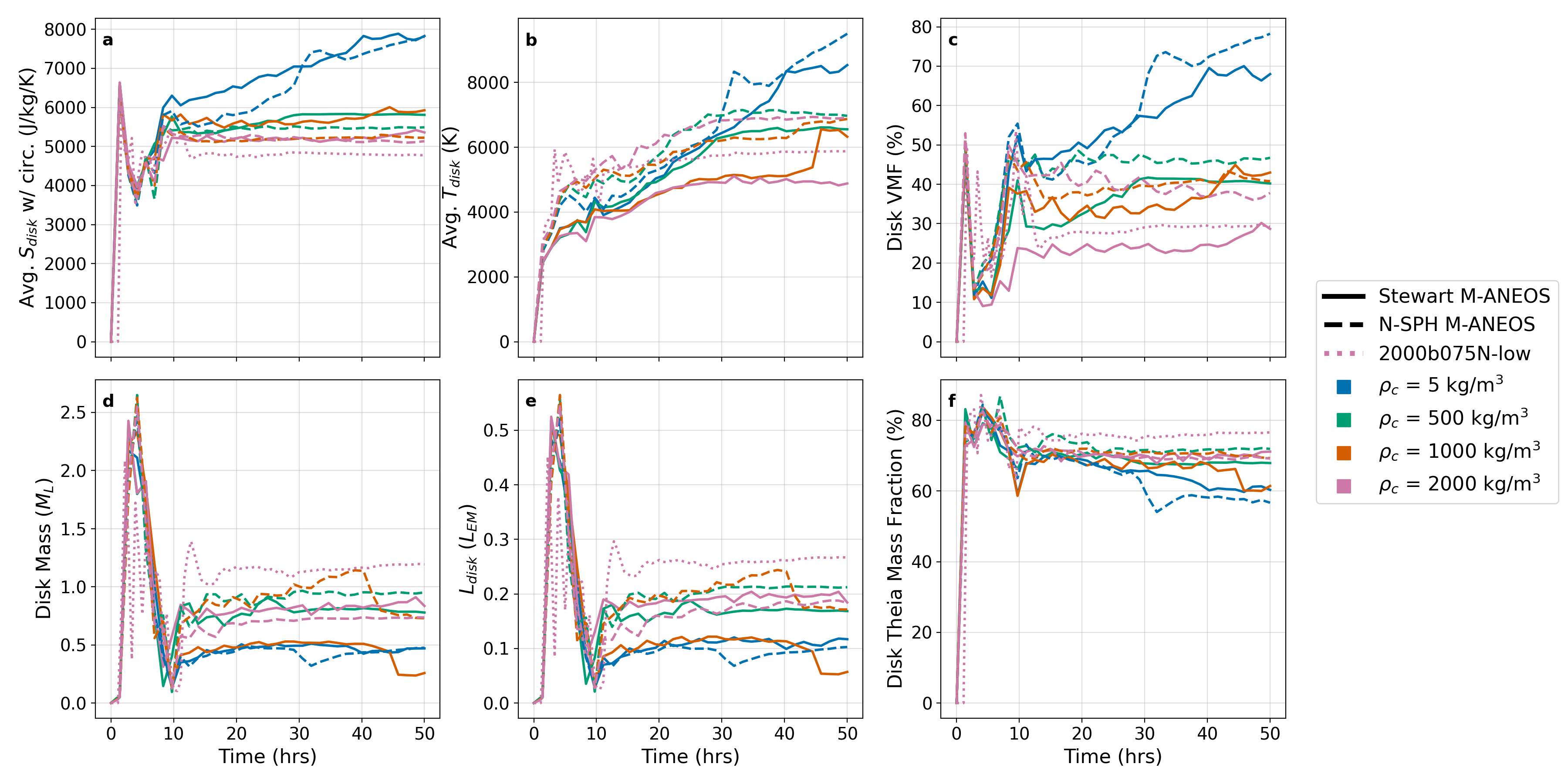}
\caption{Time evolution of the disk for simulations with impact parameter $b=0.75$.  Each cutoff density represents a distinct line color.  The solid lines represent Stewart M-ANEOS simulations, the dashed lines represent N-SPH M-ANEOS simulations, and the dotted lines corresponds to our high resolution run (5b073S-high).  \textbf{(a)} Average disk entropy with entropy gain due to orbital circularization assessed at each timestep included, \textbf{(b)} average disk temperature, \textbf{(c)} disk VMF calculated with effects of orbital circularization included, \textbf{(d)} disk mass, \textbf{(e)} disk angular momentum, \textbf{(f)} mass fraction of material in the disk originating from Theia.  A version of this figure without the effects of orbital circularization of the disk is available in Appendix \ref{appendix_C}, Figure \ref{fig:b075_disk_entropy_and_vmf_wo_circ}.}
\label{fig:b075_disk_entropy_and_vmf}
\end{figure}
\begin{figure}[h]
\centering
\includegraphics[width=1\textwidth]{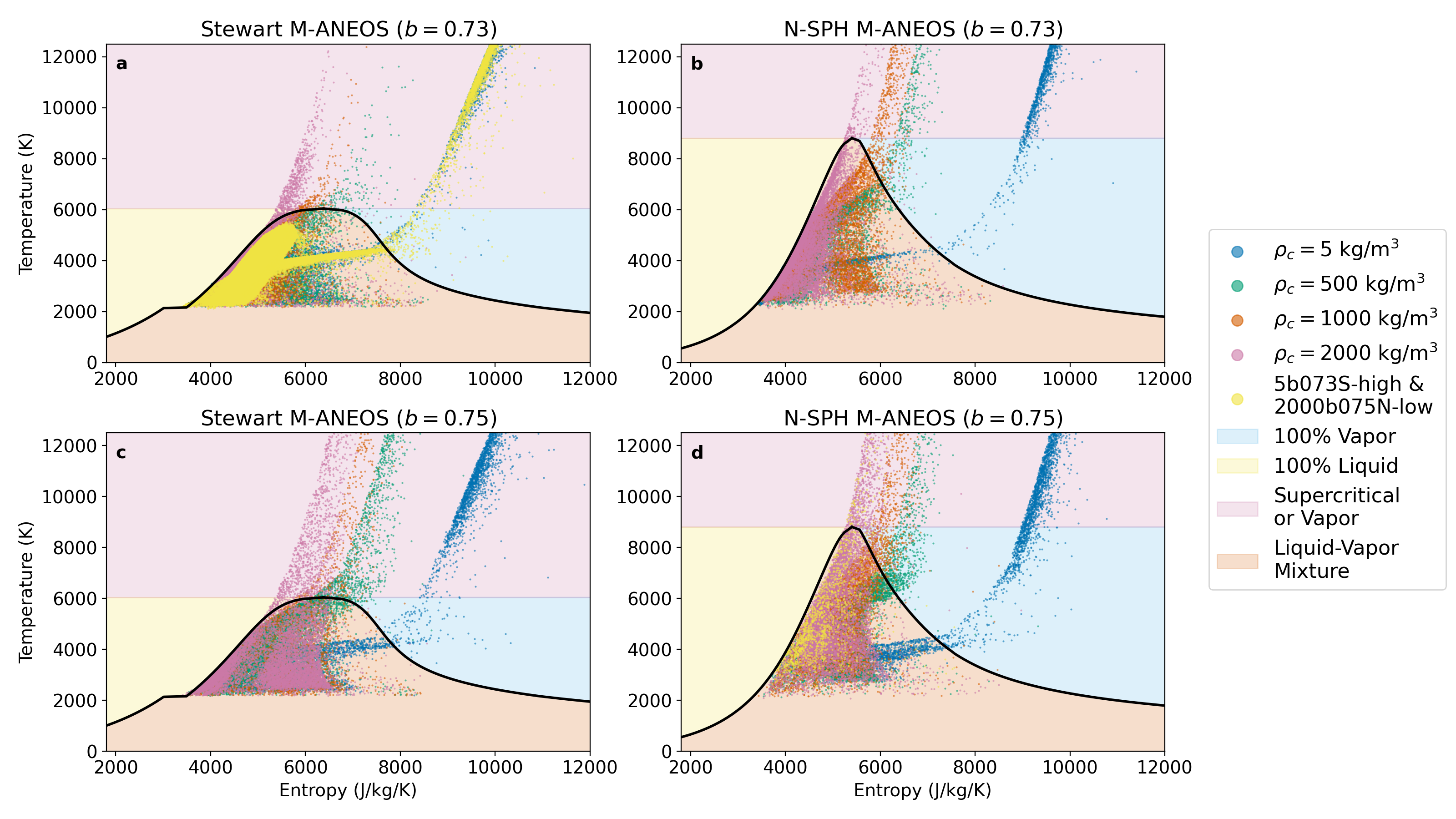}
\caption{The liquid-vapor phase curves for both Stewart (left column) and N-SPH (right column) M-ANEOS forsterite.  The top row corresponds to $b=0.73$ simulations and the bottom row corresponds to $b=0.75$ simulations.  The disk particles for each $\rho_{c}$ are plotted with their entropy gains associated with orbital circularization included.  Yellow points correspond to simulation 5b073S-high in subplot \textbf{(a)} and 2000b075N-low in subplot \textbf{(d)}.  Regions of the plots are color-coded to represent their EoS-specific phase state, with the orange region between the liquid and vapor phase curves representing a mixture of liquid and vapor.  The peak of the phase curves represents the critical point, above which the phase obtains either supercritical state or is otherwise vapor at low pressure.  All supercritical particles are treated as $100\%$ vapor in our VMF calculations.  The minimum temperatures shown here reflect the initial conditions.  A version of this figure without the effects of orbital circularization of the disk is available in Appendix \ref{appendix_C}, Figure \ref{fig:phase_curves_no_circ}.}
\label{fig:phase_curves}
\end{figure}

\subsection{Outcomes of $b=0.73$ Giant Impact Simulations} \label{outcomes-b073}

Table \ref{table:gi_b073_outputs} shows the results of runs with impact parameter $b=0.73$ and Figures \ref{fig:source_scenes_b073_new} and \ref{fig:source_scenes_b073_old} show the evolution of simulations with Stewart M-ANEOS and N-SPH M-ANEOS, respectively.  With the exception of runs 5b073N and 1000b073N, all of these runs produce a disk with mass $\geq 1$ $M_{\rm L}$ and angular momentum $\geq 0.18$ $L_{\rm EM}$.  In this particular setup, the final disk mass and angular momentum is set by the evolution of the debris cloud created by the initial impact between Earth and Theia.  Following the primary impact, the remnants of Theia and sheared-off debris from the protoearth form a debris cloud that is gravitationally-bound to the protoearth (Figures \ref{fig:source_scenes_b073_new}a-e, Figures \ref{fig:source_scenes_b073_old}a-d).  The radial interior of the debris cloud is primarily composed of the remnants of Theia’s iron core, whereas the exterior is dominated by remnants of Theia’s silicate mantle.  By 5.5 simulated hours, the debris begins to clump and a secondary impactor forms at the interior of the debris cloud.  The secondary impactor then begins its terminal impact trajectory towards the protoearth as a trailing debris ``tail'', composed of the outer portions of the debris cloud, follows behind the impactor (Figures \ref{fig:source_scenes_b073_new}f-j, Figures \ref{fig:source_scenes_b073_old}e-h).  While most of the secondary impactor accretes to the protoearth following its impact, the trailing debris tail is tidally stretched by the close encounter with the protoearth and is ejected into orbit (Figures \ref{fig:source_scenes_b073_new}k-o, Figures \ref{fig:source_scenes_b073_old}i-l), where the tail material eventually constitutes the majority of the disk mass.

While the total mass of the post-impact debris cloud is consistent within $\sim 0.3$ $M_{\rm L}$ in this simulation set, the distribution of this consistent ejecta mass between the secondary impactor and the tail varies systematically between Stewart and N-SPH M-ANEOS.  N-SPH M-ANEOS produces secondary impactors that are, on average, $\sim 20\%$ larger than those produced by Stewart M-ANEOS, whereas Stewart M-ANEOS produces more massive debris tails than N-SPH M-ANEOS (Table \ref{table:si_and_tail_stats_b073}).  Since the debris tails are generally the source of the majority of the disk mass, the EoS-dependent mass distribution between the secondary impactor and the debris tail is the cause Stewart M-ANEOS’s ability to produce more massive disks than N-SPH M-ANEOS within this simulation set (Figure \ref{fig:b073_disk_entropy_and_vmf}d).

An additional EoS-dependent phenomenon is evident in the evolution of the debris tail as it is ejected into orbit around the protoearth.  Simulation 2000b073S is notable in that a particularly large clump forms in the debris tail and survives tidal destruction, eventually achieving a stable orbit and comprising the majority of the disk mass (Figure \ref{fig:source_scenes_b073_new}s). In contrast, the debris tails produced in rest of the Stewart M-ANEOS runs in this simulation set are tidally stretched and rope out, forming coherent spirals in the disk (Figures \ref{fig:source_scenes_b073_new}k-m).  By the end of the simulation, these spirals coalesce into tight, coherent disks with asymmetric orbits around the protoearth.  This is most clearly seen in simulations 500b073S and 1000b073S (Figures \ref{fig:source_scenes_b073_new}q-r).  Runs 500b073N and 1000b073N are also unique in that the hot surface of the protoearth is ripped up and inserted into orbit, thereby substantially increasing the temperature of the disk (Figure \ref{fig:source_scenes_b073_old}j-k).  Therefore, while trends in tail and secondary impactor mass distribution are consistent across EoS, dynamical differences caused by varying $\rho_c$ can result in markedly different structures during disk evolution.  In contrast, the debris tails formed by N-SPH M-ANEOS are ejected into orbit and quickly lose coherence, evolving to form more diffuse, symmetric disks by the end of the simulation (Figures \ref{fig:source_scenes_b073_old}m-p).

\subsection{Outcomes of $b=0.75$ Giant Impact Simulations} \label{outcomes-b075}

Table \ref{table:gi_b075_outputs} and Figure \ref{fig:b075_disk_entropy_and_vmf} show the results of our $b=0.75$ simulation set.  All of these runs, regardless of EoS, fail to produce a sufficiently massive disk with high angular momentum, with simulation 500b075N coming the closest at $M_{\rm disk} = 0.95$ $M_{\rm L}$ and $L_{\rm disk} = 0.21$ $L_{\rm EM}$.  This stands in contrast to the findings of \cite{canup2004simulations} and \cite{nakajima2014investigation}, who both produce successful disks using $b=0.75$.  A notable difference between our simulations and those by \cite{canup2004simulations} and \cite{nakajima2014investigation} is that our simulations have particle resolutions that are 1-2 orders of magnitude higher at $10^6$ particles.  To test whether our diverging results are caused by the increase in particle resolution, we ran a lower-resolution simulation with settings similar to that of \cite{nakajima2014investigation}, using $\sim 10^5$ particles, N-SPH M-ANEOS, and $\rho_{c} = 2000$ $\rm kg/m^3$.  Table \ref{table:gi_b075_outputs} shows the results of this run, labeled 2000b075-low, which produces a successful disk with $M_{\rm disk} = 1.19$ $M_{\rm L}$ and $L_{\rm disk} = 0.27$ $L_{\rm EM}$.  These results are in good agreement with those shown in Table 2 of \cite{nakajima2014investigation}.  Additionally, between runs 2000b075N and 2000b075N-low, where the only difference is an order of magnitude in particle resolution, the VMF can change by $\sim 10\%$.  Therefore, our results show that an order of magnitude increase in particle resolution can change the mass of the disk by tens of percent and alter the VMF of the disk by up to at least $\sim 10\%$, although these trends may not be consistent across different EoS and impact configurations.  Another notable difference is in the SPH kernel implementation, whereby 2000b075N-low uses a Wendland C6 kernel and the SPH code employed by \cite{nakajima2014investigation} uses the cubic spline kernel recommended by \cite{monaghan1992smoothed}.  Therefore, the implementation of $\rho_c$ does not significantly alter results across these two commonly-used kernels.

This simulation set plays out similarly across both EoS and values of $\rho_{c}$.  As this impact setup is more grazing than the $b=0.73$ runs, less momentum is transferred from Theia to the protoearth during the initial impact and much of Theia remains intact immediately after the initial impact.  However, enough momentum is lost from Theia to keep its remnants from escaping the system, instead placing these remnants on course for a highly-grazing secondary impact with the protoearth.  At this point, a considerable amount of secondary impact debris is gravitationally accelerated as they slingshot around the protoearth and ejected from the system above escape velocity.  These runs have notably higher values of escaping mass than their counterparts in the $b=0.73$ simulation set, with escaping mass values ranging from $0.27$-$0.9$ $M_{\rm L}$ (Table \ref{table:gi_b075_outputs}).  Additionally, the tidal destruction of Theia’s remnants during the secondary impact and increased entropy in the disk due to the circularization of disk particles on more eccentric orbits produce entropy values $\sim 1000$ J/kg/K and temperature values $1000$ - $3000$ K greater than simulations at $b = 0.73$, in part helping these simulations obtain higher VMF values (Figures \ref{fig:b075_disk_entropy_and_vmf}a-c).

We do not conclude that impacts shallower than $b=0.75$ result in more successful disks.  As seen in our entire body of impact simulations, the final state of the disk is highly variable and is a result of complex interactions and dynamics that extend beyond just the initial impact.  Moreover, our results show that changing the particle resolution can significantly alter the final state of the disk.  Therefore, it is difficult to place upper and lower bounds on $b$ to constrain successful impact setups.  A more extensive search of the relationship between $b$, particle resolution, and the resulting disk is needed to define broad trends about how these variables influence giant impact simulations and the disk.  Additional analysis of the $b=0.75$ simulation set can be found in Appendix \ref{appendix_B}.

\subsection{Effects of $\rho_{c}$ on the Disk} \label{effects-rho-c}

Our results show that the final disk mass, angular momentum, and thermodynamic state are all dependent on the choice of $\rho_{c}$.  Disks with low particle resolutions are ubiquitous among SPH simulations of the giant impact as the protoearth dominates in mass.
The formation of the disk is a complex process that is set in motion by the initial impact between the protoearth and Theia.  Following the initial impact, low resolution structures immediately begin to form in the debris and $\rho_{c}$ begins to influence the simulation.  It is difficult to discern broad dynamical patterns caused by changing $\rho_{c}$.  Our lowest value of $\rho_{c}$, $\rho_{c} = 5$ $\rm kg/m^3$, seems to broadly produce ``puffier'' protoearths and disks with no discernible gap between the two.  In contrast, one run (2000b073S) from our highest value of $\rho_{c}$, $\rho_{c} = 2000$ $\rm kg/m^3$, is characterized by the survival of a large clump that survives tidal destruction and dominates the disk.  To first order, at least some of the dynamical differences between runs of varying $\rho_{c}$ can be attributed to decompression and the relationship between $P$ and $\rho$ in the evolving disk.  This is covered in more detail in Section \ref{eos-differences-lunar-chem}.  By 50 hours, upwards of $\sim 40\%$ of disk particles reach $\rho_c$.  Across all particles in the simulation at 50 hours, about $\sim 1\%$ of particles reach $\rho_c$ at $\rho_c = 5$ $\rm kg/m^3$ and upwards of $\sim 10\%$ do at $\rho_c = 2000$ $\rm kg/m^3$.

Perhaps the most striking result from varying $\rho_{c}$ came from our lowest value, $\rho_{c} = 5$ $\rm kg/m^3$.  Regardless of choice of EoS or $b$, these disks fail to achieve a thermodynamic steady state as the disks' average entropy, temperature, and VMF enter states of runaway growth (Figures \ref{fig:b073_disk_entropy_and_vmf}a-c and Figures \ref{fig:b075_disk_entropy_and_vmf}a-c).  At such a low value of $\rho_{c}$, disk particles that experience brief encounters with one another can produce quick density jumps slightly greater than $5$ $\rm kg/m^3$, effectively causing artificial shocks in the disk over the duration of the simulation and causing a thermodynamic runaway state.  These phenomena are explored in more detail in Section \ref{treatment-rho-c}.  In addition to the thermodynamic issues at $\rho_{c} = 5$ $\rm kg/m^3$, we also see that $\rho_{c}$ introduces a systematic effect in the resulting temperature and VMF of the disk.  Across both $b=0.73$ and $b=0.75$ simulations, increasing values of $\rho_{c}$ result in cooler disks that depress the temperature and VMF of the disk without significantly altering the average disk entropy (Figures \ref{fig:b073_disk_entropy_and_vmf}a-c and Figures \ref{fig:b075_disk_entropy_and_vmf}a-c).

To test whether increasing the particle resolution of the disk could mitigate the artificial disk shocks for low values of $\rho_{c}$, we ran an additional simulation, 5b073S-high, at a total particle resolution of $\sim 10^7$, $\rho_{c} = 5$ $\rm kg/m^3$, $b=0.73$, and using Stewart M-ANEOS.  By increasing the total resolution of the simulation by an order of magnitude, we are also able to boost the particle resolution of the disk by an order of magnitude (Table \ref{table:gi_b073_outputs}).  Relative to the corresponding lower-resolution simulation (5b073S), we find that the increase in disk resolution slows, but does not stop, the runaway growth in entropy, temperature, and VMF (Figures \ref{fig:b073_disk_entropy_and_vmf}a-c).

\subsection{Effects of EoS on the Disk} \label{effects-eos-discussion}
As suggested in Section \ref{outcomes-b073}, there are broad thermodynamic trends in our results that show an EoS dependency in addition to its effect on the disk mass and angular momentum.  The increased heat capacities given by Stewart M-ANEOS would intuitively lead to cooler disks than their N-SPH M-ANEOS counterparts, as more internal energy would be required to raise the temperature of the disk.  While disk temperature can also be affected by the stochastic evolution of the simulation, our results indeed broadly show that N-SPH M-ANEOS produces higher disk temperatures on average than Stewart M-ANEOS by approximately $\sim 10 \%$.  Conversely, disk entropy values between Stewart and N-SPH M-ANEOS insignificantly differ within $6 \%$.

In addition to the lower disk temperatures, our results show that the VMF values produced by Stewart M-ANEOS are generally lower than N-SPH M-ANEOS by $\sim 10-20 \%$ depending on choice of $b$ (Figure \ref{fig:b073_disk_entropy_and_vmf}c and Figure \ref{fig:b075_disk_entropy_and_vmf}c).  The exception to this trend is in run 5b073S, which experiences greater thermodynamic instability in the interior of the disk.  Figure \ref{fig:phase_curves} shows offsets in the location of the liquid-vapor phase curves between both EoS, which is caused by their inherent differences in treatment of heat capacity and their underlying experimental Hugoniot.  The critical point of the Stewart M-ANEOS phase curves is at a lower temperature relative to N-SPH M-ANEOS by $\sim 3000$ K (Figure \ref{fig:phase_curves}) and a greater entropy by $\sim 1000$ J/kg/K.  Stewart M-ANEOS also features a broader mixed-phase region along the entropy axis than N-SPH M-ANEOS, leading to less vapor production at temperatures below the critical point upon application of the lever rule.  While it is intuitive to expect the lower critical point of Stewart M-ANEOS to result in more supercritical particles and a higher VMF, both EoS produce roughly similar numbers of supercritical particles on average across all values of $\rho_c$.  Instead, the majority of particles ($\sim 70$-$98\%$) exist in the mixed-phase region, where the tighter phase curves of N-SPH M-ANEOS result in consistently higher values of VMF.

%% file: discussion.tex
\begin{figure}[h]
\centering
\includegraphics[width=1\textwidth]{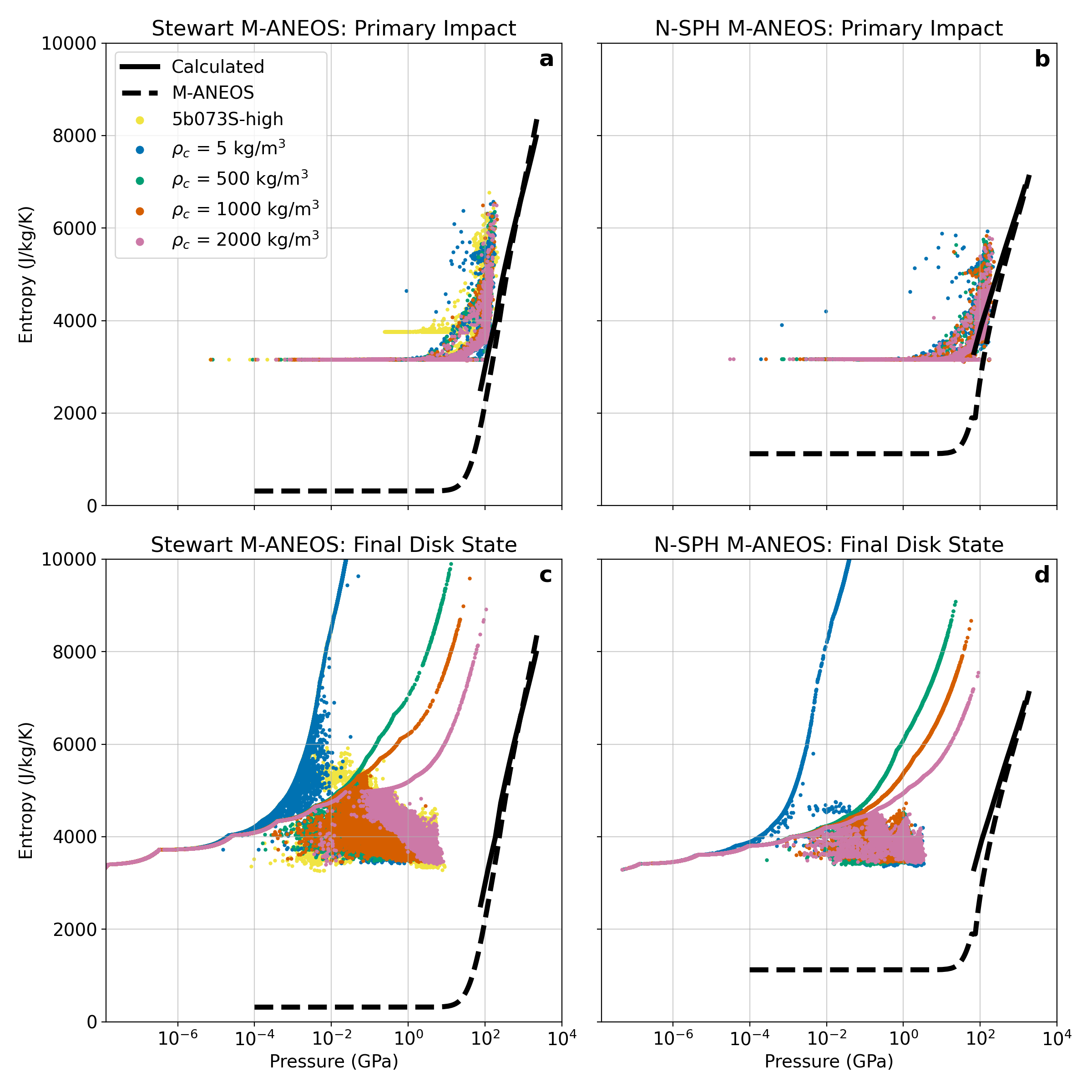}
\caption{Disk particles for simulations with impact parameter $b=0.73$ relative to the EoS-specific Hugoniot curve (dashed curve) and the calculated Hugoniot curve (solid curve) given by the Rankine-Hugoniot relations.  \textbf{(a)} Disk particles from Stewart M-ANEOS runs at the moment of the initial impact between Earth and Theia, \textbf{(b)} Disk particles from N-SPH M-ANEOS runs at the moment of the initial impact between Earth and Theia, \textbf{(c)} Disk particles from Stewart M-ANEOS runs at 50 simulated hours, \textbf{(d)} Disk particles from N-SPH M-ANEOS runs at 50 simulated hours.  The disk particles by 50 hours are not well approximated by the Hugoniot and cannot be used to constrain a value of $\rho_{c}$ for the disk.  The wave-like features in panels \textbf{c-d} are the result of particles that have reached their respective value of $\rho_c$.}
\label{fig:b073_hugoniot_with_max_pressure_vals}
\end{figure}

\subsection{Treatment of $\rho_{c}$ and Artificial Disk Shocks} \label{treatment-rho-c}

Our results show that the chosen value of $\rho_{c}$ influences the final disk mass, angular momentum, and thermodynamic state (Figure \ref{fig:b073_disk_entropy_and_vmf} and Figure  \ref{fig:b075_disk_entropy_and_vmf}).  The most extreme effects are seen in runs using our lowest value of $\rho_{c}$, as the disk entropy, temperature, and VMF experienced runaway growth and the disks experienced dynamic instability, both of which can be attributed to the low particle resolution of the disk.  Low resolution structures in SPH, like the disk, are composed of comparatively few, widely spaced particles that are isolated within their kernel volume with few to no neighboring particles.  Particle isolation within a kernel can occur when the smoothing length is capped to prevent it from becoming unphysically large, as we do here with our implementation of $\rho_c$.  As seen in Equation \ref{sph-density-dis}, the density of each particle is resolved by summing over all the neighboring particles, with non-zero values terms only being added if the particle falls within the effective radius of the kernel.  In low resolution structures, the density may fall to the limit of the density contribution of just a few neighboring particles, if not just the isolated particle only.  Such isolated particles, with few to no neighbors, then begin to lose fluid-like behavior and instead behave more like distant N-bodies.  In all of our simulations, we find that most disk particles acquire values of $\rho_{c}$ owing to this phenomenon.  Assuming a cylindrical disk of radius of $3.5$ $R_\oplus$ and height $0.5$ $R_\oplus$, we expect the bulk density of the disk to be between approximately $10$ to $20$ $\rm kg/m^3$, assuming that the liquid and vapor are well mixed.

If not supported by a high enough value of $\rho_{c}$, the density calculation in low resolution structures introduces two problems into SPH simulations which can explain both the disk instability and the runaway thermodynamics of our runs with $\rho_{c} = 5$ $\rm kg/m^3$.  First, as described in Section \ref{intro-fdps-sph}, calculated values of $h$ are inversely proportional to $\rho^{1/3}$.  Therefore, values of $h$ in the disk increase as particle densities fall, often to the limit defined by $\rho_c$, causing disk particles to become susceptible to kernel overlaps and hydrodynamic interactions with increasingly distant particles.  These additional interactions increase the exchange of angular momentum between disk particles, thereby scattering particles on unrealistic timescales.  Secondly, as disk particles become isolated and their densities and temperatures approach $0$, they become increasingly sensitive to brief encounters with other particles.  In our simulations with $\rho_{c} = 5$ $\rm kg/m^3$, we find that these brief encounters often produce small density jumps in a substantial number of disk particles that act as artificial shocks which irreversibly increases the particle and bulk disk entropy as well as the temperature and VMF of the disk.  We also observe that some increases in entropy are not directly correlated with density jumps and therefore might be associated with the so-called ``$E_0$'' error, which is a type of low-order error introduced into the SPH momentum equation caused by insufficient smoothing in the kernel \citep{read2012sphs, hopkins2015new}.  It is also possible that the density jumps are occurring, but our file output intervals (typically $\sim 100$ simulated seconds) are too coarse to capture them.  We find that disk particles that experience these artificial shocks usually do so as they pass through the interior part of the disk, which is typically more particle-dense than the outer regions and therefore presents an increased chance for encounters with neighboring particles.  Given our viscosity implementation in SPH, our results demonstrate that the effect of $\rho_c$ is stronger than the moderating effects of viscosity in SPH.  The artificial shocking is contained to isolated particles in the disk, but theses particles contribute to increasing the average entropy of the disk and the resulting VMF.  At our values of $\rho_c \geq 500$ $\rm kg/m^3$, the SPH-calculated particle density is well below $\rho_c$ and therefore the small density jumps are not felt by the disk particles.  We expect that these issues for low values of $\rho_{c}$ would disappear if SPH simulations could achieve particle resolutions that currently exceed the reasonable capabilities of modern computational hardware, as evidenced by the slowing of artificial shocking of the disk in run 5b073S-high using $10^7$ particles and its order of magnitude increase in particle resolution.  Should higher-resolution runs be achievable, we expect every order of magnitude increase in disk particle resolution to roughly halve the distance between each particle and halve the smoothing length necessary to have a given number of neighboring particles on average, thereby increasing neighboring particle counts naturally and promoting thermodynamic stability in the simulation.

Our runs with values of $\rho_{c} \geq 500$ $\rm kg/m^3$ do not exhibit the same numerical issues as our $\rho_{c} = 5$ $\rm kg/m^3$ runs, placing an upper bound on a value of $\rho_{c}$ that achieves numerical stability in giant impact disks.  However, given the propensity of most disk particle densities to fall to the defined value of $\rho_{c}$ and influence the evolution of the disk, this technique for resolving the low resolution density problem remains contrived, especially if $\rho_{c}$ is not physically motivated or if the simulation’s particle resolution is not high enough to offset the issue.  Even though disk entropies are fairly consistent across all runs with $\rho_c > 5$ $\rm kg/m^3$, higher $\rho_c$ values tend to be associated with slightly lower disk entropy values, which in turn suppress the VMF of the disk as more particles remain closer to the liquid phase curve.  However, differences in disk evolution and disk features as a function of $\rho_c$, like the emergence of clumps, sometimes complicate a direct comparison.  High $\rho_c$ values are also likely not realistic for the protolunar disk.  Ultimately, the reliance of the disk's thermodynamic state on a hardcoded parameter like $\rho_{c}$ is an issue for reconciling dynamical simulations with the geochemistry of the Moon, as the resulting differences in disk temperature and VMF change the predicted volatile composition and MVE isotope fractionation of the Moon.  We further assess the location and extent of numerical shocks in the disk in Appendix \ref{appendix_C}.

Initially, one may think that the Hugoniot curve can provide a physically-motivated value of ejecta entropy after the shock, from which the choice of $\rho_c$ may be justified.  The Hugoniot curve gives all of the possible solutions to the Rankine-Hugoniot equations, which predict the conditions behind the shock front given the initial conditions.  If the disk is well approximated by the Hugoniot, then the minimum shock density provided by the Hugoniot may provide a reasonable value of $\rho_{c}$.  Figures \ref{fig:b073_hugoniot_with_max_pressure_vals} and \ref{fig:b075_hugoniot_with_max_pressure_vals} show the pressures and entropies of particles that eventually compose the end-state disk at the time of the primary impact and at 50 simulated hours for both $b=0.73$ and $b=0.75$ simulation sets, respectively.  While the pressure-entropy conditions of the disk particles are reasonably approximated by the Hugoniot at the time of the primary impact regardless of the choice of $\rho_c$, the disk is not well described by the Hugoniot by 50 hours, indicating that secondary impacts and further evolution after the secondary impact continue to increase the entropy of the disk.  Like \cite{nakajima2015melting} and \cite{carter2020energy}, we observe that most disk particles experience peak shock conditions after the primary giant impact due to secondary impacts and the subsequent disk evolution which sets the final thermodynamic state of the disk.  Therefore, the Hugoniot curve is not well-suited for constraining a physically-motivated $\rho_c$.  We recommend that future giant impact SPH studies with implementations of $\rho_c$ ensure the thermodynamic stability of the disk as a function of time and particle resolution.

Our $b=0.73$ simulation set also shows that the survival of large clumps in orbit is affected by $\rho_{c}$.  Notably large clumps survive and achieve stable orbits in both simulations using $\rho_{c} = 2000$ $\rm kg/m^3$ (Figure \ref{fig:source_scenes_b073_new}t and Figure \ref{fig:source_scenes_b073_old}p) but not in our runs with $\rho_{c} < 2000$ $\rm kg/m^3$, which either produce no clumping or small clumps.  In both cases, these clumps form on the back edge of the secondary impact tail and survive tidal destruction as they are ejected into the disk.  Clumping in SPH as an EoS or resolution dependent phenomenon has been previously investigated as increasing computational power allows for higher resolution simulations \citep[e.g.,][]{canup2013lunar, hosono2017unconvergence, shimoni2022influence}.  Recently, \cite{kegerreis2022immediate} ran simulations using $b=0.71$ and found that the formation of large satellites in the secondary impact structure is a function of numerical resolution, with simulations below $10^{6.5}$ particles forming only a secondary impactor and simulations above $10^{6.5}$ particles consistently forming both a secondary impactor and a trailing, large clump.  Our runs suggest that the formation and survival of large satellites in the secondary impact debris is also a potential result of $\rho_{c}$ implementation.

Previously, \cite{nakajima2014investigation} and \cite{nakajima2015melting} have produced simulations with $\rho_c$ values near 1500-1600 $\rm kg/m^3$.  \cite{genda2011merging} also employs $\rho_c$ and uses it in setting a maximum smoothing length.  While their smoothing length is adaptive such that there are $\sim 64$ neighboring particles within $2h$, they adopted to save computational time by setting a maximum smoothing length defined by setting $\rho_c = 5 \times 10^{-3}$ $\rm kg/m^3$.  As \cite{genda2011merging} did not attempt to form a protolunar disk nor study the thermodynamics of the system, it is unknown whether their simulations experienced the artificial shocks that we document here.  It is possible that $\rho_c$ and a maximum smoothing length are implemented but undocumented in other SPH giant impact studies as well, and that underlying issues with artificial shocks are unaddressed as they would only be apparent with analysis of the thermodynamic state of the disk as a function of time.

\cite{thacker2000smoothed} briefly notes in their comparison of SPH implementations that smoothing over too few particles can result in unphysical shocking, which is the phenomenon we observe here.  They also note that a very small timestep can help to mitigate this issue, as there is a better chance that a new particle entering a sparsely-populated kernel may be initially ``felt'' towards the edge of the smoothing length and hence more gradually introduced into the density calculation.  However, the computational efficiency is likely to become unfeasible, especially at higher particle resolutions.  They also recommend an algorithmic smoothing length to balance computational speed with a desired number of neighboring particles, although in the context of a moon-forming giant impact it still may suffer from the caveats of smoothing lengths that become too large.

Other methods may provide solutions to the numerical issues associated with $\rho_c$ and large smoothing lengths.  \cite{read2012sphs} and \cite{canup2013lunar} used a kernel method that forced the smoothing length to overlap a defined number of neighboring particles to reduce low-order error in SPH, whereby the kernel-dependent value is at least $\sim 32$ particles for a cubic spline kernel and upwards of 400 for a Wendland C6 kernel \citep{dehnen2012improving, hopkins2015new}.   Such a method with uncapped smoothing lengths may improve the numerical stability of low resolution regions in SPH and reduce error associated with noise in the conservation equations \citep{read2012sphs, dehnen2012improving}, while at the same time resulting in disk particle scattering issues associated with large values of $h$, typically disk particle densities that are smaller than the bulk disk density (see Appendix \ref{appendix_A}), and reducing computational efficiency \citep{hopkins2015new}.  Moreover, fast-moving particles with near-instantaneous close encounters with other disk particles may still impart artificial shocks.

Particle splitting methods have also been implemented to resolve low resolution SPH features.  Particle splitting divides a particle’s mass into many smaller-mass particles that can locally boost the resolution in SPH \citep{kitsionas2002smoothed, canup2018origin}.  Particle splitting may be a less-contrived method for dealing with low resolution issues that plays to the strengths of SPH.  However, the application of particle splitting to giant impact simulations is new, therefore requiring some additional scrutiny in its results.  Additionally, particle splitting may also still fail to boost the resolution of the disk sufficiently to alleviate low resolution issues.  As seen in our high resolution simulation (5b073S-high), even an order of magnitude increase to the disk particle resolution to $10^7$ particles still results in numerical instability, although to a lesser degree than our $10^6$ particle runs, without a sufficiently large value of $\rho_{c}$.  Lastly, the use of the density-independent formulation of SPH (DISPH) \citep{saitoh2013density, hosono2013density, hosono2016giant} may mitigate some density-dependent issues.  DISPH is unique in that its formulation of the conservation laws depend on pressure gradients rather than density gradients.  Future investigations of DISPH on low resolution structures in SPH may stabilize some of the disk’s dependency on $\rho_{c}$, although the density dependency of the kernel may continue to pose issues for the numerical stability of the disk.  Moreover, the validity of DISPH simulations should be interpreted with skepticism until the discordance of EoS solutions between SSPH and DISPH can be reconciled \citep{canup2021origin}.

\subsection{EoS Differences in Disk Outcomes and Lunar Geochemistry} \label{eos-differences-lunar-chem}

The effects of EoS in SPH simulations have been previously discussed in the literature, mainly in the comparison between the ANEOS and the Tillotson EoS \citep{tillotson1962metallic}.  Tillotson is an older, less robust EoS compared to ANEOS and was developed to provide an extensive EoS solution space for metals.  However, Tillotson does not include phase boundaries or a self-consistent treatment of temperature and entropy, and includes a large interpolated region in $P$-$U$ phase space that can lead to erroneous particle dynamics \citep[][their Figure 2]{stewart2020shock}.  \cite{benz1989origin} provided comparisons between ANEOS and Tillotson in early, low resolution SPH simulations of giant impacts and noticed that Tillotson EoS resulted in a compressed protoearth that was $8\%$ smaller than its ANEOS counterpart.  Given that momentum is conserved across all of the similar initial setups of the unique ANEOS and Tillotson giant impact simulations, this predictably resulted in dynamical differences between EoS.  \cite{emsenhuber2018sph} ran a similar study for a smaller Martian giant impact and found that, while Tillotson and ANEOS converged in their mass distributions, Tillotson overestimated temperatures due to a lack of treatment for the latent heat of vaporization.

Our simulations compare the newer “Stewart M-ANEOS” to the older “N-SPH M-ANEOS”, the former including an improved treatment of heat capacity, an expanded experimental Hugoniot, and the inclusion of a melt curve.  Given that there are no structural differences between these two EoS aside from these differences, a comparison between these two EoS is less complicated than between ANEOS and Tillotson.  We observe some difference in the initial conditions of the target and impactor, where the bodies produced by N-SPH M-ANEOS are slightly more compressed than those produced by Stewart M-ANEOS (Table \ref{table:GI_initial_conditions}).  Our results show that there are differences in the final thermodynamic state of the disks produced by both EoS.  The cooler disk temperatures and higher VMFs produced by Stewart M-ANEOS relative to N-SPH M-ANEOS are a first-order effect of the improved treatment of heat capacity and the expanded pressure range of the underlying experimental data.  These improvements lead to notable offsets in the location of the liquid-vapor phase curves and the critical point, above which all material becomes supercritical and is treated like a vapor for the purposes of our VMF calculations.

Developing robust EoS to accurately constrain the amount of vapor produced by the giant impact is critical to bridge experiments and models with the measured lunar geochemistry.  Geochemical observations of the Moon indicate that the BSM is strongly depleted in volatile elements relative to both the Earth and chondrites \citep{ringwood1977basaltic, wolf1980moon, jones2000geochemical, day2014evaporative}, indicating that the lunar source material underwent an episode of vaporization and devolatilization during the Moon’s formation.  The timing of this episode is inferred from \ch{Rb/Sr} and \ch{U/Pb} radioactive decay, where the Moon’s low \ch{^{87}Sr/^{86}Sr} and high \ch{^{238}U/^{204}Pb} relative to the BSE indicate that moderately volatile \ch{Rb} and \ch{Pb} were lost prior to the final accretion of the Moon \citep{day2014evaporative}.

In addition to volatile element depletion, the Moon’s enrichment of heavy MVE isotopes also suggest an episode of vaporization.  In an environment like the high-temperature disk generated by the giant impact, isotopes kinetically fractionate due to differing reaction rate constants between the heavy and light isotopes of a given element during chemical reactions or phase changes.  Rayleigh fractionation models and experimental observations of kinetic isotope fractionation predict that heavy isotopes remain in the condensate during evaporation and remain in the vapor during condensation \citep{bourdon2020isotope}, making the Moon’s heavy MVE enrichment a likely additional signal of vaporization.  Previous studies have proposed evaporative mechanisms for kinetic MVE fractionation, including vapor drainage from a $99\%$ saturated disk to the protoearth \citep{nie2019vapor, charnoz2021tidal} and evaporation from the lunar magma ocean \citep{day2014evaporative}, although fractionation under the latter regime may be throttled by a vapor buildup and subsequent near-equilibrium conditions at the surface of the magma ocean \citep{tang2020evaporation}.  Regardless, an unknown but presumably significant amount of vapor production is required to impart both the observed volatile depletion and MVE heavy isotope enrichment of the Moon.

The connection between VMF, volatile loss, and MVE isotope fractionation is not straightforward.  Each element varies in volatility in part due to its condensation temperature \citep{lodders2003solar}, changing the element’s partial pressure in the vapor phase and, critically for the MVEs, changing the evaporative isotope flux \citep{richter2004timescales}.  Modelled vapor pressures can be combined with experimental equilibrium pressures to predicted MVE isotope fractionation in the Moon \citep{richter2004timescales, dauphas2015planetary}.  Our results suggest that Stewart M-ANEOS, with its lower VMF values relative to N-SPH M-ANEOS, would promote less devolatilization as less vapor would be available to escape from the protolunar materials.  Similarly, assuming that vapor could be lost at a rate that maintains net evaporative flux from the liquid phase, Stewart M-ANEOS would also predict less MVE isotope fractionation than N-SPH M-ANEOS.  Future work on reconciling models of VMF, volatile loss, and MVE fractionation with lunar geochemistry is required to place constraints on the thermodynamic state of the protolunar disk and validate EoS solutions.

In addition to the thermodynamics of the disk, our results demonstrate that there are notable differences in dynamics of the evolution of the disk between both EoS.  This was most clearly seen in the $b=0.73$ simulation set, in which the secondary impact structure systematically differed and resulted in larger disk masses for Stewart M-ANEOS.  To first order, some dynamical differences between EoS emerging from a shock-inducing impact may be explained by the relationship between particle velocity, $u_{\rm p}$, and the $P$-$V$ gradient during isentropic decompression following the giant impact \citep{stewart2020shock},
\begin{equation}
    u_{p} = u_{p, H} \pm \int_{P_{H}}^{P} \frac{dP}{\sqrt{(\partial P / \partial V)_{S}}} = u_{p, H} \pm \int_{P_{H}}^{P} \frac{V}{c_{s}} dP, \label{1D-decompression-velocity}
\end{equation}
where $c_s$ is the sound speed, $P_{\rm H}$ is the shock pressure, $P$ is an ambient pressure, and $V = 1/\rho$ is the specific volume.  However, it should be noted that Equation \eqref{1D-decompression-velocity} assumes a one dimensional model and this may not capture the complex nature of the impact event. Variations in the $P$-$V$ relationship can be attributed to many facets of the given EoS, including the influence of its underlying experimental Hugoniot or the location of the phase curve, which alters the compressibility of the material and therefore the dynamics and physical evolution of the simulation.

While both our $b=0.73$ and $b=0.75$ simulation sets produced a wide range of disk masses, Theia's mass fraction contribution was similar across all simulations, consistently ranging between $60$-$80\%$ of the mass of the disk.  At such a large fraction, it is difficult to reconcile our canonical giant impact results with the remarkable similarity in notable isotope abundances between the Earth and Moon.  Therefore, it is desirable to find trends that can decrease Theia's contribution to the disk while preserving the agreeable dynamical results of the canonical impact.  While interesting EoS-dependent trends in disk mass were visible in our results, our simulations do not show a clear mechanism to decrease Theia's disk mass contribution.

%% file: conclusions.tex
We ran 18 SPH simulations of the canonical giant impact to compare the effects of the newly-developed Stewart M-ANEOS against the older N-SPH M-ANEOS.  The improved experimental Hugoniot and treatment of heat capacity found in Stewart M-ANEOS results in systematically cooler disks with lower vapor mass fraction (VMF) values than N-SPH M-ANEOS.  In our simulation set using an impact parameter of $b=0.73$, Stewart M-ANEOS also forms less-massive secondary impactors, thereby resulting in more massive disks than counterpart N-SPH M-ANEOS runs.  Previously, the behavior of clumping and the formation of secondary impactors was identified as a resolution-dependent feature of SPH simulations of giant impacts \citep[e.g.,][]{canup2013lunar, hosono2017unconvergence, meier2021eos, shimoni2022influence, kegerreis2022immediate}.  Our results add to these findings by showing that the distribution of mass during the evolution of the disk is an EoS-dependent feature.

Additionally, we explored the influence of the hard-coded minimum ``cutoff'' density, $\rho_{c}$, on the disk.  Protolunar disks produced by the canonical giant impact are low resolution features, thereby causing a breakdown in SPH's resolution-dependent density calculations in the disk.  Without a sufficient amount of particles in proximity to one another, disk densities can become very small, with most disk particles immediately attaining density values equal to $\rho_{c}$.  Modest differences in disk entropy, temperature, VMF, mass, and angular momentum occur as a function of changing $\rho_{c}$, with higher values of $\rho_{c}$ suppressing vapor production regardless of EoS choice.  The most dramatic differences are seen in runs with our lowest value of $\rho_{c}$, $\rho_{c} = 5$ $\rm kg/m^3$.  These disks experience continuous artificial shocking throughout the entirety of the simulated time, which we attribute to small density jumps \textbf{or ``$E_0$'' error} in particles composing the low resolution disk.  Our additional high resolution simulation ($N \sim 10^7$) was able to increase the disk resolution by an order of magnitude and slowed the effects of these artificial shocks, but ultimately failed to stop it.  Our results also show that changes in particle resolution can alter both the dynamical and thermodynamic outcomes of the disk, which is most clearly seen in the low resolution run using $b=0.75$.

Previous work has shown that devolatilization, MVE isotope fractionation, and the dynamical evolution of the Moon are all functions of the disk VMF, thereby motivating the development of EoS for complex geological materials that can accurately capture the liquid-vapor phase assemblage over an extreme range of temperatures and pressures.  In addition to the VMF differences between the two EoS investigated here, we also find that variations in $\rho_{c}$ and particle resolution produce disks with VMF that can vary by upwards of $30\%$.  Such a wide range of VMF solutions would most likely result in compositionally unique disks owing to the varying propensities for devolatilization and MVE isotope fractionation, as well as have implications for the ability of lunar seeds to remain on stable orbits \citep{nakajima2022large}.  Future work will incorporate these findings to quantitatively constrain the composition of the vapor and residual protolunar melt, the degree of heavy MVE enrichment, and amount of vapor that is likely to be lost through hydrodynamic escape following the giant impact.

%% file: acknowledgement.tex
This work was supported in part by the National Aeronautics and Space Administration (NASA) grant numbers 80NSSC19K0514 and 80NSSC21K1184. Partial funding was provided by the Center for Matter at Atomic Pressures (CMAP), the National Science Foundation (NSF) Physics Frontier Center under Award PHY-2020249, as well as by EAR-2237730. Any opinions, findings, conclusions or recommendations expressed in this material are those of the authors and do not necessarily reflect those of the National Science Foundation. This work was also supported in part by the Alfred P. Sloan Foundation under grant number G202114194 and the Computational Infrastructure for Geodynamics (CIG), which is supported by the National Science Foundation award NSF-1550901, and by JSPS KAKENHI Grant Number 19K14826.

%% file: appendix_A.tex
\begin{figure}[h]
\centering
\includegraphics[width=1\textwidth]{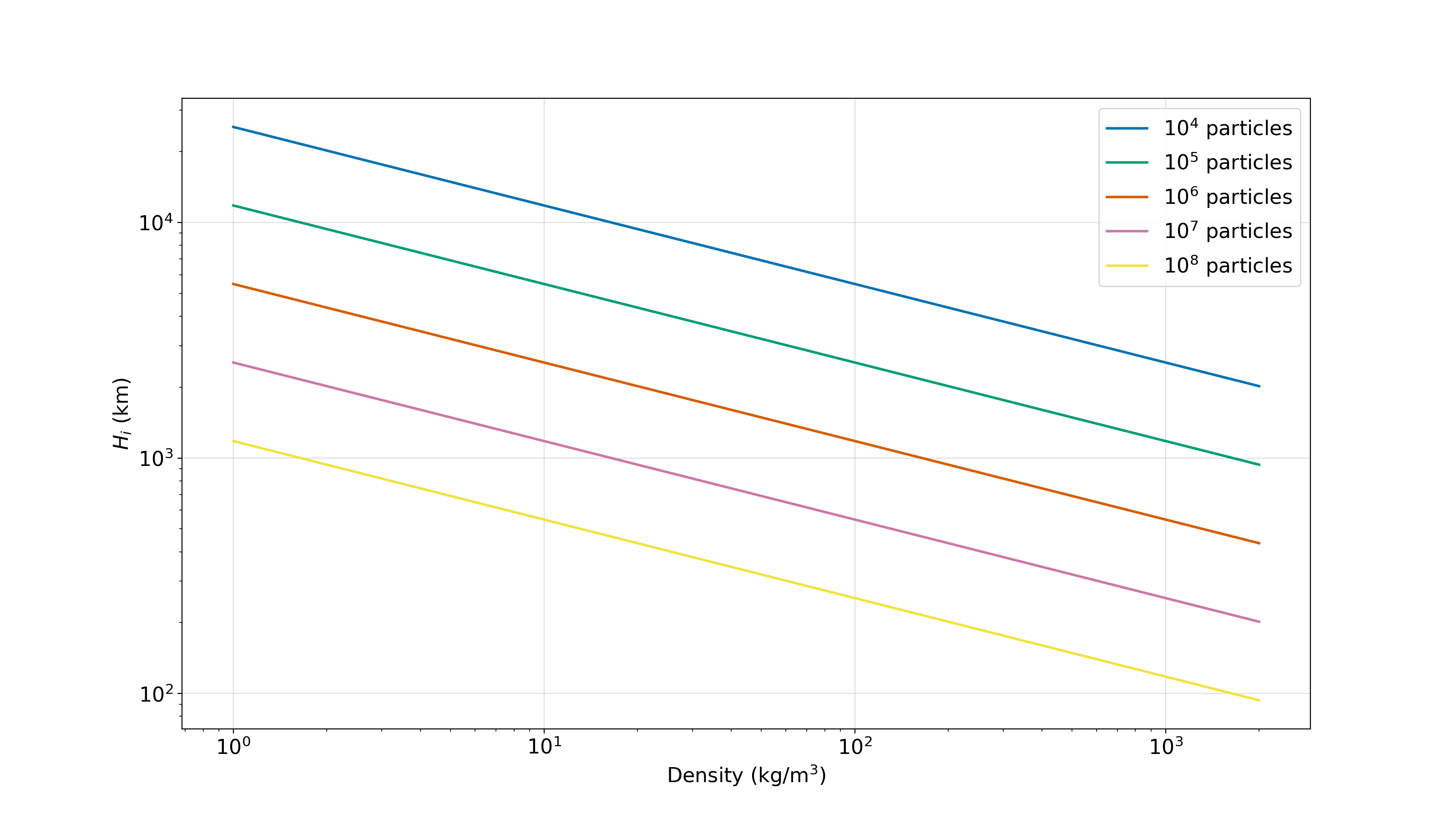}
\caption{The supported smoothing length $H_i$ as a function of particle density, where each line represents a different simulation particle resolution.  The functional dependency of $h$ on particle mass $m$ scales as a function of particle count and therefore particle mass is given as the total mass of the canonical giant impact, $M_{\rm tot} = 1.019 M_\oplus$, divided by the total number of particles in the simulation.}
\label{fig:smth_length}
\end{figure}
\begin{figure}[h]
\centering
\includegraphics[width=1\textwidth]{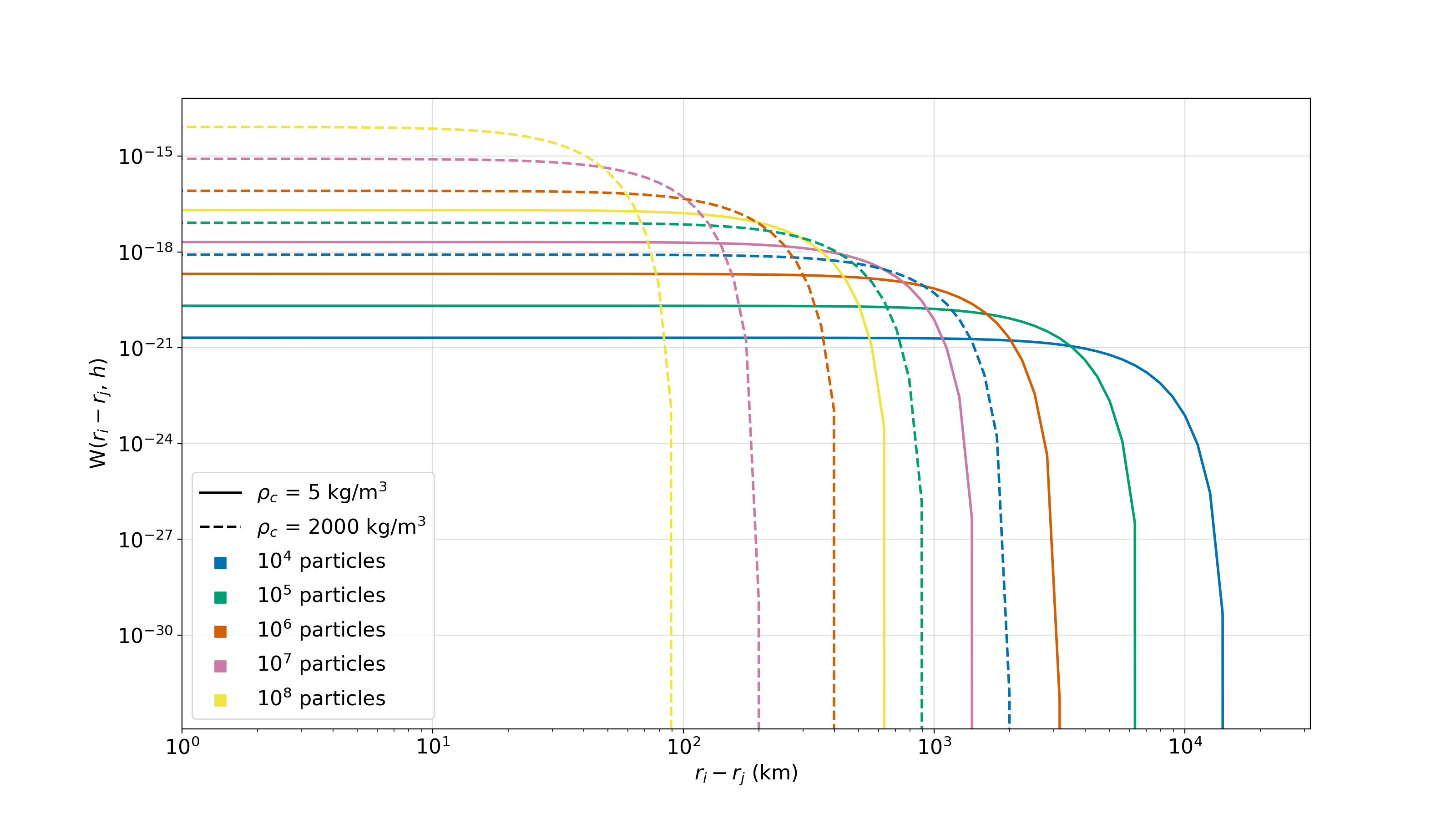}
\caption{The Wendland C6 SPH kernel as a function of particle distance.  The kernel is given for our minimum and maximum values of $\rho_{c}$, values: $\rho_{c} = \rm 5$ $\rm kg/m^3$ (solid lines) and $\rho_{c} = \rm 2000$ $\rm kg/m^3$ (dashed lines).  Line colors represent different particle resolutions.}
\label{fig:kernel}
\end{figure}
Density for a given particle $i$ in SPH is calculated by a summation across all surrounding particles $j$,
\begin{equation}
    \rho_i = \sum_{j}^N m_j W (\mathbf{r}_i - \mathbf{r}_j, h_{i}) \label{particle-density-eq},
\end{equation}
where $m$ is the particle mass, $W$ is the kernel, $\mathbf{r}_i - \mathbf{r}_j$ is the distance between two particles, and $h_{i}$ is the particle-dependent smoothing length.  The SPH kernel $W$ gives weight to particles $j$ with close proximity to particle $i$ by producing scalar weighting values that fall off monotonically to 0 as $|\mathbf{r}_i - \mathbf{r}_j| \rightarrow \infty$.  We employ a Wendland C6 kernel in FDPS SPH, which is given in 3D as,
\begin{equation}
    W(s) = \frac{1365 \sigma}{64 \pi}
    \begin{cases} 
      (1 - s)^8 (1 + 8s + 25s^2 + 32s^3) & 0 \leq s \leq  1 \\
      0 & s > 1,
   \end{cases} \label{wendland-c6}
\end{equation}
where $\sigma = 1 / H_{i}^3$, and $s = |\mathbf{r}_i - \mathbf{r}_j| / H_{i}$, where $H_{i} = 2.5 h_{i}$ includes a support radius.  Here, the 3D smoothing length is,
\begin{equation}
    h_{i} = \xi \left( \frac{m_i}{\rho_i} \right )^{1/3}, \label{appendix-a-smth}
\end{equation}
where $\xi = 1.2$ is the smoothing coefficient and $\rho_i$ is the particle density.  We note that values of $h_{i} = 1.866 \left (m_{i} / \rho_{i} \right )^{1/3}$ and $H_{i} = 2.449 h_{i}$ have been previously proposed for the Wendland C6 kernel \citep{dehnen2012improving}.  Equation \eqref{appendix-a-smth} also contains a particle resolution dependency, as standard SPH formulations require all particles to have equivalent masses and therefore particle mass decreases with increasing particle resolution.  Figures \ref{fig:smth_length} and \ref{fig:kernel} show the effects of SPH particle density and particle resolution on the smoothing length and the kernel, respectively.

Low resolution SPH structures like the protolunar disk are composed of comparatively few, widely-spaced particles.  Equations \eqref{particle-density-eq} and \eqref{wendland-c6} show that widely spaced particles in low-resolution environments can obtain very small densities associated with the mass and volume contribution of the given, single particle as $W \rightarrow 0$.  Moreover, without a high value of $\rho_{c}$, brief encounters between two disk particles where $W \neq 0$ can produce momentary shocks in the disk, driving up the entropy and vapor mass fraction (VMF) of the disk well beyond the impact phases of the simulation.  Such behavior can be suppressed by setting a high value of $\rho_{c}$ in the SPH simulation.  A high value of $\rho_{c}$ has the effect of both lowering the volumetric effect of the kernel and thereby lowering the chances of any two disk particles exerting an influence on one another, and is a density value higher than that which would be predicted by equation \eqref{particle-density-eq} given a small number of particles in the summation.  However, this means that the properties of the disk are entirely controlled by the choice of $\rho_{c}$.

Some previous SPH studies have enforced smoothing lengths to overlap a certain number of neighboring particles for stability.  We crudely estimate the typical particle density required for the particle's stable smoothing length, $h_{i, s}$, to overlap a certain number of neighboring particles, $N_{\rm neighbor}$.  We assume that the disk is cylindrical with a radius of $3.5 R_\oplus$ with the volume of the Earth subtracted, such that the disk volume is $V_{\rm disk} = 11.25 \pi R_\oplus^2 d$, where $d$ is the height of the disk which we assume to be $d \sim R_\oplus$.  The number of neighboring particles in a spherical smoothing volume, $V_{i, s}$, is given by,
\begin{equation}
    N_{\rm neighbor} = V_{i, s} \rho_{\rm particle} = \frac{4}{3} \pi h_{i, s}^3 \frac{N_{\rm disk}}{11.25 \pi R_\oplus^3} ,
\end{equation}
where $\rho_{\rm particle}$ is the disk particle density (i.e., $\rm particles/m^3$) and $N_{\rm disk}$ is the number of particles in the disk.  We solve for $h_{i,s}$,
\begin{equation}
    h_{i,s} = \left (\frac{33.75 R_\oplus^3 N_{\rm neighbor}}{4 N_{\rm disk}} \right )^{1/3} .
\end{equation}
%
The corresponding density of the particle defined by $h_{i, s}$ is,
\begin{equation}
    \rho_{i, s} \sim \frac{M_{\rm total}}{N_{\rm total} h_{i, s}^3} = \frac{4 M_{\rm total} N_{\rm disk}}{33.75 N_{\rm total} N_{\rm neighbor} R_\oplus^3} ,\label{appendix_a_rho_i_s_eq}
\end{equation}

%
We find that the canonical impact disk is usually $\sim 1\%$ of the total mass of the simulation ($N_{\rm disk} = 0.01 N_{\rm total}$), which when substituted into equation \eqref{appendix_a_rho_i_s_eq} makes $\rho_{i,s}$ independent of the particle resolution of the simulation.  Therefore, $\rho_{i, s}$ is given as,
\begin{equation}
    \rho_{i, s} \sim \frac{0.04 M_{\rm total}}{33.75 N_{\rm neighbor} R_\oplus^3} . \label{appendix_a_rho_i_s_eq_simplified}
\end{equation}
Values of $N_{\rm neighbor} \sim 30$ particles for a cubic spline kernel and $N_{\rm neighbor} \sim 400$ particles for a Wendland C6 kernel have previously been suggested \citep{dehnen2012improving, hopkins2015new}, and therefore $\rho_{i, s} = 0.93$ $\rm kg/m^3$ and $\rho_{i, s} = 0.07$ $\rm kg/m^3$, respectively.  $\rho_{i, s}$ is smaller than the bulk disk density ($\sim 10$-$20$ $\rm kg/m^3$), and therefore our rough approximation suggests that the smoothing length with an enforced neighboring particle count is too large to reproduce the expected density of the disk regardless of particle resolution as $N_{\rm total}$ and $N_{\rm disk}$ cancel out in equation \eqref{appendix_a_rho_i_s_eq_simplified}.

%% file: appendix_B.tex
We report on the results on 9 SPH simulations with impact parameter $b=0.75$ across varying values of $\rho_{c}$, particle resolution, and EoS choice.  The setup of each simulation is given in Table \ref{table:GI_initial_conditions} and the results are given in Table \ref{table:gi_b075_outputs}.  As discussed in the main text, none of the 8 simulations of resolution $10^6$ particles achieved a disk capable of forming the Moon, which is in contrast to the findings of previous work \citep{nakajima2014investigation}.  To investigate the cause of this difference, we applied the settings from \cite{nakajima2014investigation} in run 2000b075N-low, which is a simulation of resolution $10^5$ particles using N-SPH M-ANEOS and $\rho_{c} = 2000$ $\rm kg/m^3$.  This run closely matches the findings of \citep{nakajima2014investigation} and shows that the resulting disk is sensitive to the particle resolution of the simulation.

\begin{figure}[h]
\centering
\includegraphics[width=1\textwidth]{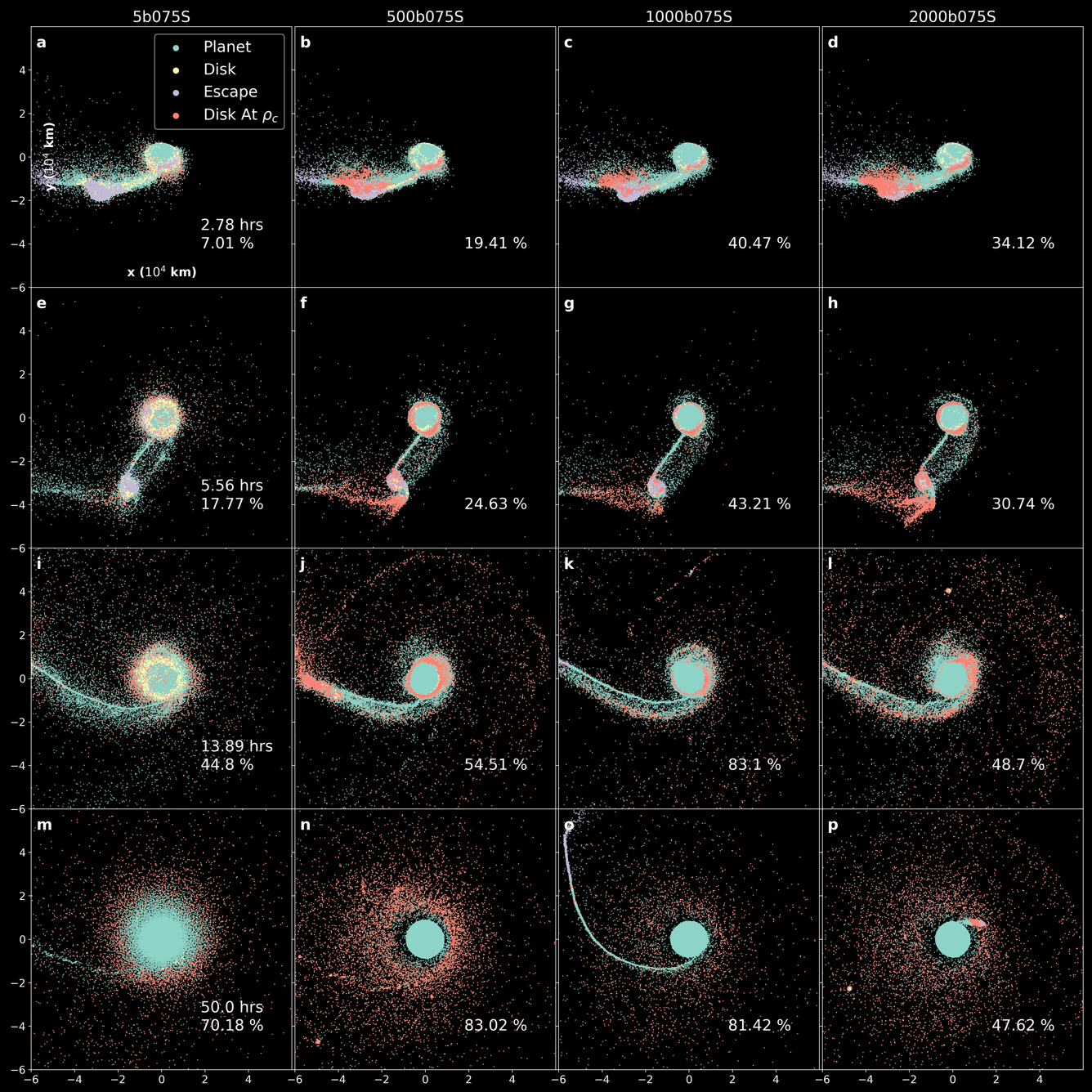}
\caption{Evolution of the giant impact and the protolunar disk for simulations using Stewart M-ANEOS with an initial impact parameter of $b=0.73$.  The axes give the x-y spatial coordinates relative to the protoearth's center of mass given in increments of $10^4$ km.  Particles are sorted by their z-axis values and colors indicate whether the particle is part of the protoearth (green), the disk (yellow), or escaping (purple).  Red particles indicate final disk particles that have density values equal to $\rho_c$ at the given time, and the percentage of final disk particles that are at the value of $\rho_c$ at each time is annotated in the lower-right corner of each subplot.  Each column corresponds to a unique run and each row is a snapshot of the simulation at \textbf{(a-d)} 2.78 hours, \textbf{(e-h)} 5.56 hours, \textbf{(i-l)} 13.89 hours, \textbf{(m-p)} 50 hours.}
\label{fig:source_scenes_b075_new}
\end{figure}

\begin{figure}[h]
\centering
\includegraphics[width=1\textwidth]{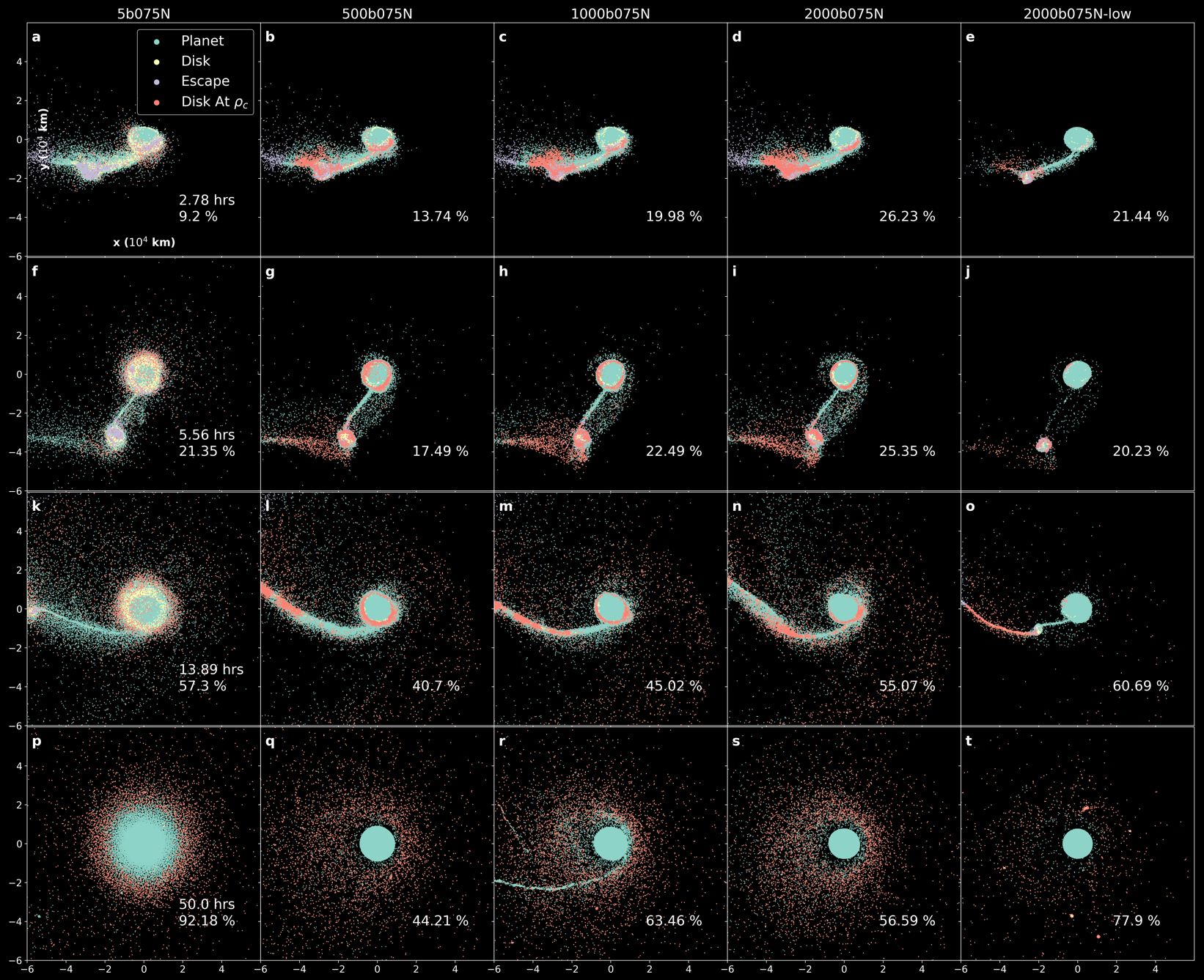}
\caption{Evolution of the giant impact and the protolunar disk for simulations using N-SPH M-ANEOS with an initial impact parameter of $b=0.73$.  The axes give the x-y spatial coordinates relative to the protoearth's center of mass given in increments of $10^4$ km.  Particles are sorted by their z-axis values and colors indicate whether the particle is part of the protoearth (green), the disk (yellow), or escaping (purple).  Red particles indicate final disk particles that have density values equal to $\rho_c$ at the given time, and the percentage of final disk particles that are at the value of $\rho_c$ at each time is annotated in the lower-right corner of each subplot.  Each column corresponds to a unique run and each row is a snapshot of the simulation at \textbf{(a-e)} 2.78 hours, \textbf{(f-j)} 5.56 hours, \textbf{(k-o)} 13.89 hours, \textbf{(p-t)} 50 hours.}
\label{fig:source_scenes_b075_old}
\end{figure}

\begin{figure}[h]
\centering
\includegraphics[width=1\textwidth]{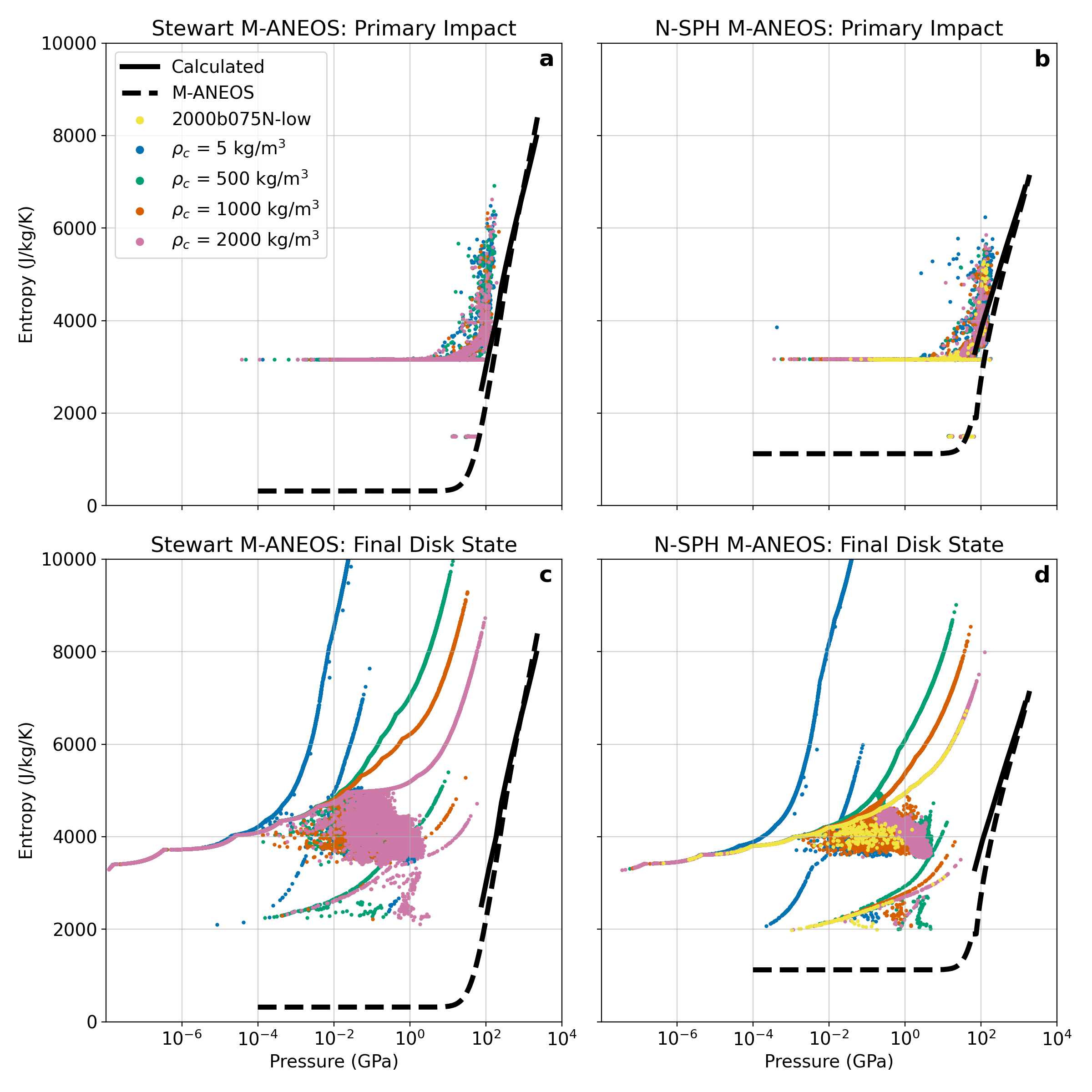}
\caption{Disk particles for simulations with impact parameter $b=0.75$ relative to the EoS-specific Hugoniot curve (dashed curve) and the calculated Hugoniot curve (solid curve) given by the Rankine-Hugoniot relations.  \textbf{(a)} Disk particles from Stewart M-ANEOS runs at the moment of the initial impact between Earth and Theia, \textbf{(b)} Disk particles from N-SPH M-ANEOS runs at the moment of the initial impact between Earth and Theia, \textbf{(c)} Disk particles from Stewart M-ANEOS runs at 50 simulated hours, \textbf{(d)} Disk particles from N-SPH M-ANEOS runs at 50 simulated hours.  The disk particles by 50 hours are not well approximated by the Hugoniot and cannot be used to constrain a value of $\rho_{c}$ for the disk.  The wave-like features in panels \textbf{c-d} are the result of particles that have reached their respective value of $\rho_c$.}
\label{fig:b075_hugoniot_with_max_pressure_vals}
\end{figure}

Figures \ref{fig:source_scenes_b075_new} and \ref{fig:source_scenes_b075_old} show the evolution of the giant impact simulations with $b=0.75$ and variable value of $\rho_{c}$ across both Stewart and N-SPH M-ANEOS, respectively.  The properties of the resulting disks are displayed in Figure \ref{fig:b075_disk_entropy_and_vmf}.  Additionally, the disk particles are plotted against the M-ANEOS Hugoniot and the calculated Hugoniot both right after the primary impact and at the simulation end at 50 hours in Figure \ref{fig:b075_hugoniot_with_max_pressure_vals}.

%% file: appendix_C.tex
\begin{figure}[h]
\centering
\includegraphics[width=1\textwidth]{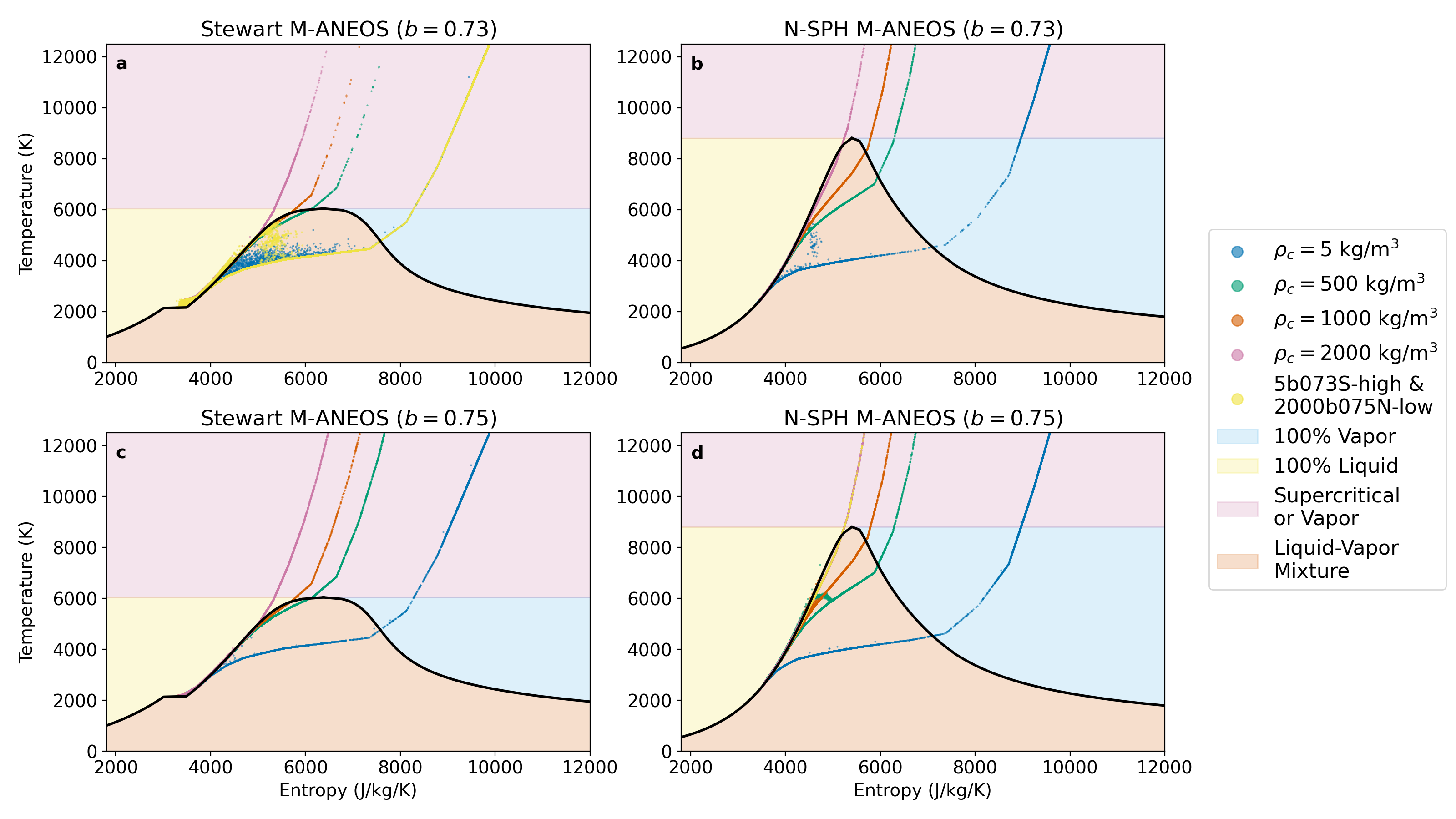}
\caption{The liquid-vapor phase curves for both Stewart (left column) and N-SPH (right column) M-ANEOS forsterite, not including the entropy gain associated with orbital circularization, in contrast with the companion Figure \ref{fig:phase_curves}.  The top row corresponds to $b=0.73$ simulations and the bottom row corresponds to $b=0.75$ simulations.  Yellow points correspond to simulation 5b073S-high in subplot \textbf{(a)} and 2000b075N-low in subplot \textbf{(d)}.  Regions of the plots are color-coded to represent their EoS-specific phase state, with the orange region between the liquid and vapor phase curves representing a mixture of liquid and vapor.  The peak of the phase curves represents the critical point, above which the phase obtains a supercritical state  or is otherwise vapor at low pressure.  All supercritical particles are treated as $100\%$ vapor in our VMF calculations.  The minimum temperatures shown here reflect the initial conditions.}
\label{fig:phase_curves_no_circ}
\end{figure}
\begin{figure}[h]
\centering
\includegraphics[width=1\textwidth]{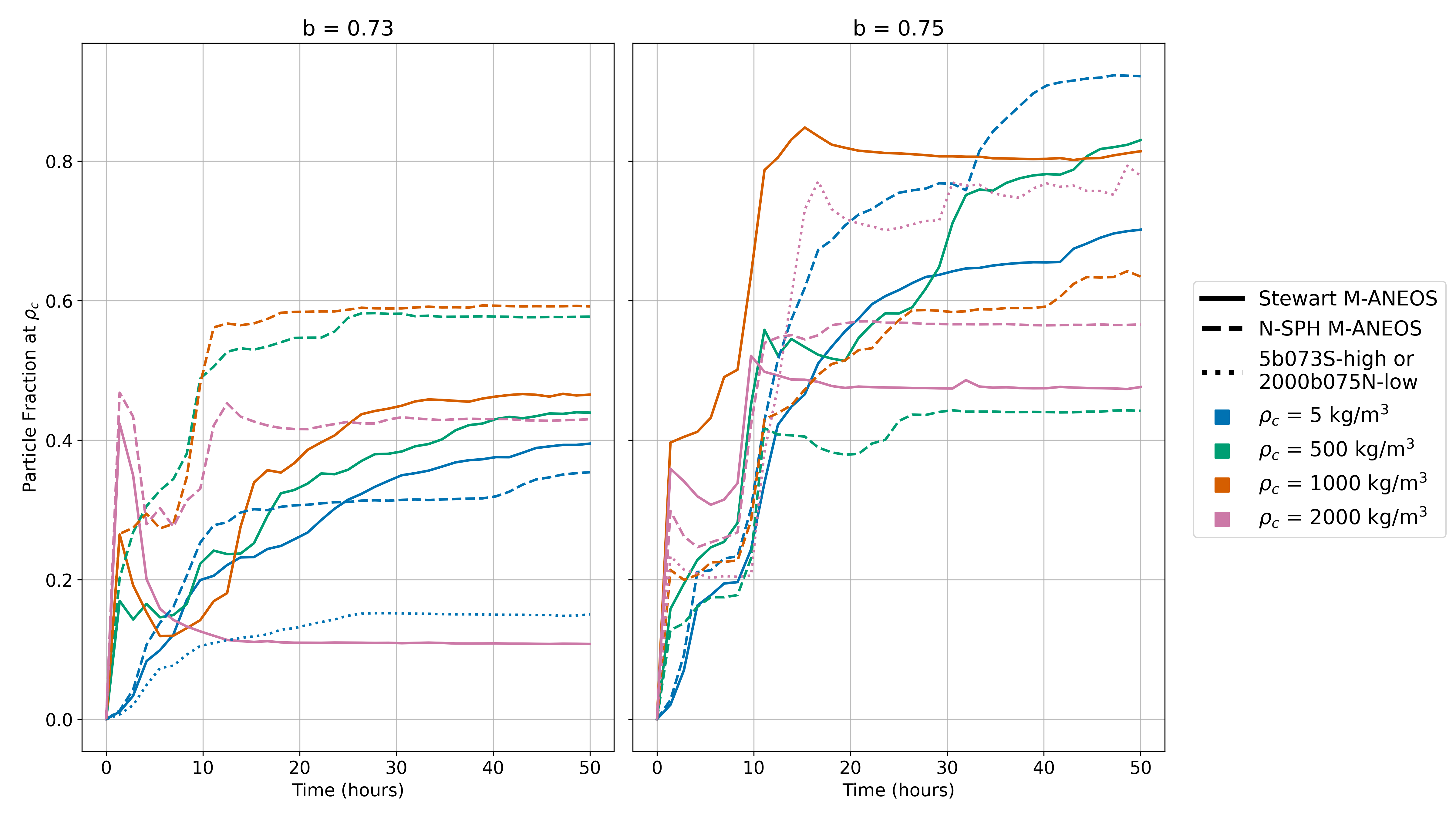}
\caption{Fraction of final disk particles with densities equal to $\rho_c$ as a function of simulated time.  Line colors correspond to the value of $\rho_c$ set in the simulation, solid lines are simulations using Stewart M-ANEOS, and dashed lines are simulations using N-SPH M-ANEOS.  The thinly-dashed lines correspond to simulation 5b073S-high in subplot \textbf{a} or 2000b075N-low in subplot \textbf{b}.  \textbf{(a)} Runs with $b=0.73$, \textbf{(b)} runs with $b=0.75$.}
\label{fig:particle_fraction_at_rho_cutoff}
\end{figure}
\begin{figure}[h]
\centering
\includegraphics[width=1\textwidth]{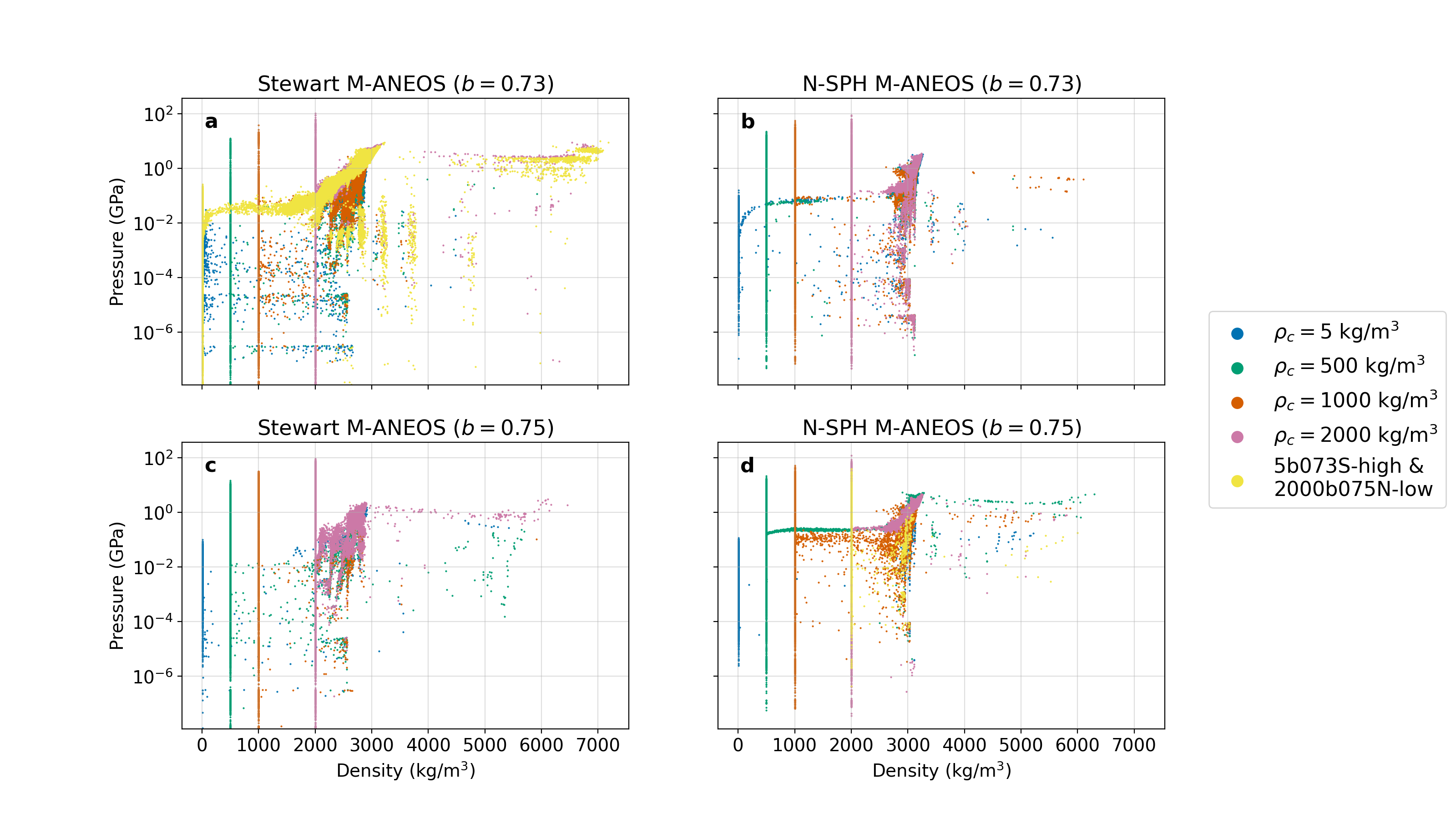}
\caption{Density vs. pressure for disk particles at 50 simulated hours.  Lower values of $\rho_c$ tend to be capped at slightly lower pressures than those at high $\rho_c$.  \textbf{(a-b)} Simulations with $b=0.73$, \textbf{(c-d)} simulations with $b=0.75$.  The left column corresponds to simulations using Stewart M-ANEOS and the right column corresponds to simulations using N-SPH M-ANEOS.}
\label{fig:paper1_disk_density_vs_pressure}
\end{figure}
\begin{figure}[h]
\centering
\includegraphics[width=1\textwidth]{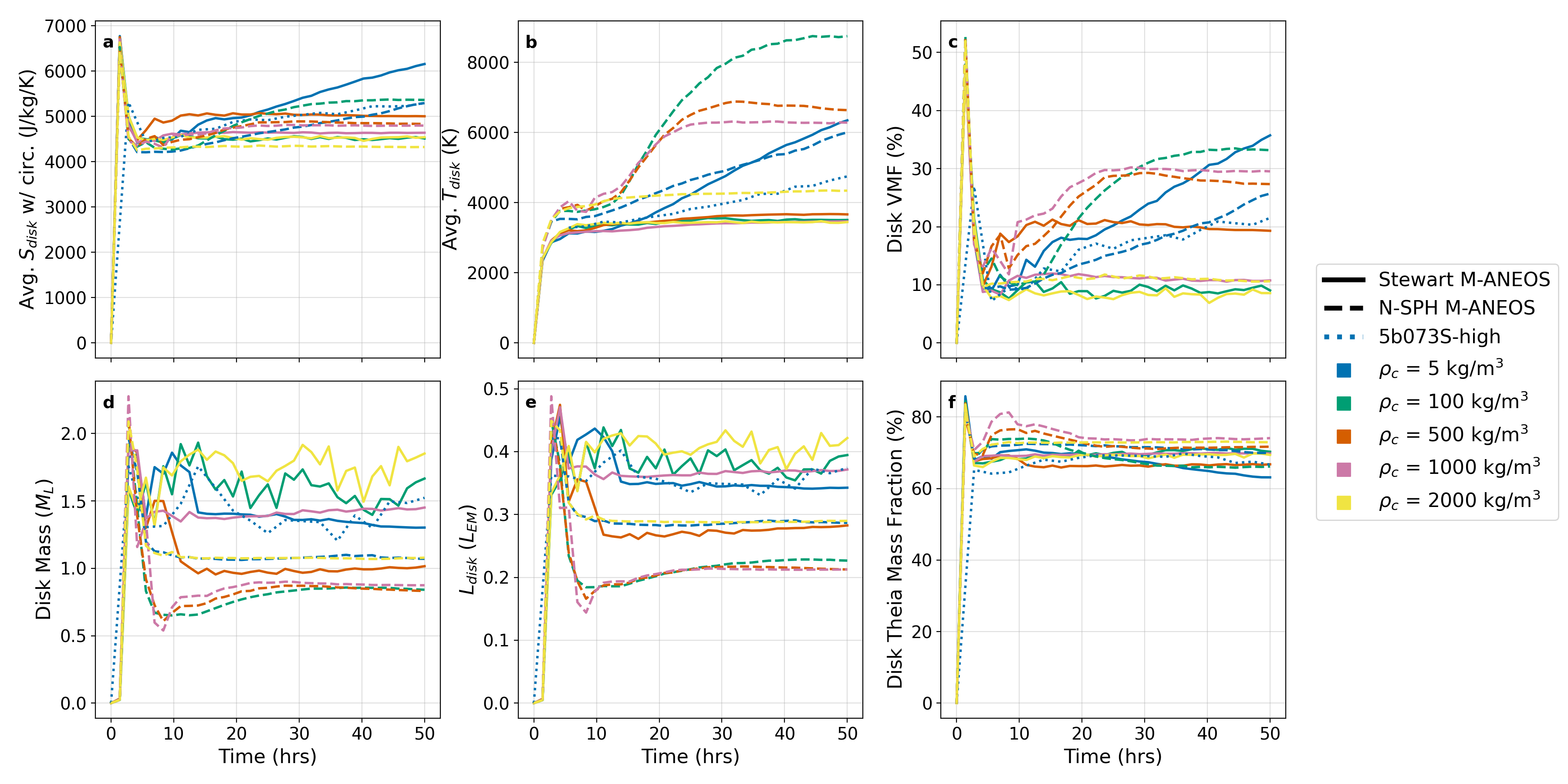}
\caption{Time evolution of the disk for simulations with impact parameter $b=0.73$.  This figure differs from Figure \ref{fig:b073_disk_entropy_and_vmf} in that it includes runs with $\rho_c = 100$ $\rm kg/m^3$.  Each cutoff density represents a distinct line color.  The solid lines represent Stewart M-ANEOS simulations, the dashed lines represent N-SPH M-ANEOS simulations, and the dotted lines corresponds to our high resolution run (5b073S-high).  \textbf{(a)} Average disk entropy with entropy gain due to orbital circularization assessed at each timestep included, \textbf{(b)} average disk temperature, \textbf{(c)} disk VMF calculated with effects of orbital circularization included, \textbf{(d)} disk mass, \textbf{(e)} disk angular momentum, \textbf{(f)} mass fraction of material in the disk originating from Theia.}
\label{fig:b073_disk_entropy_and_vmf_w_100}
\end{figure}
\begin{figure}[h]
\centering
\includegraphics[width=1\textwidth]{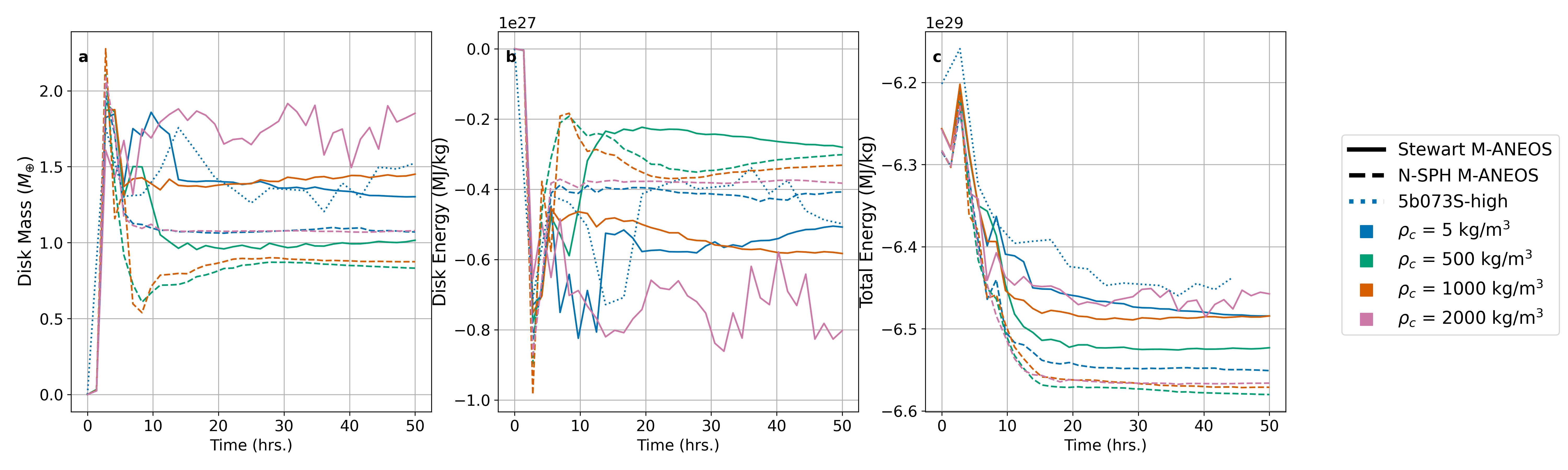}
\caption{Energy conservation for simulations with $b=0.73$. FDPS SPH conserves energy within $\sim 1\%$, showing that the numerical issues associated with small values of $\rho_c$ do not violate energy conservation beyond typical numerical noise in SPH.  \textbf{(a)}. Mass in the disk, \textbf{(b)}. Total specific energy in the disk.  Some variance is associated with the changing mass of the disk as a function of time, \textbf{(c)}. Total specific energy of the entire simulation, including the protoearth, disk, and escaping particles.}
\label{fig:energy_conservation_b073}
\end{figure}
\begin{figure}[h]
\centering
\includegraphics[width=1\textwidth]{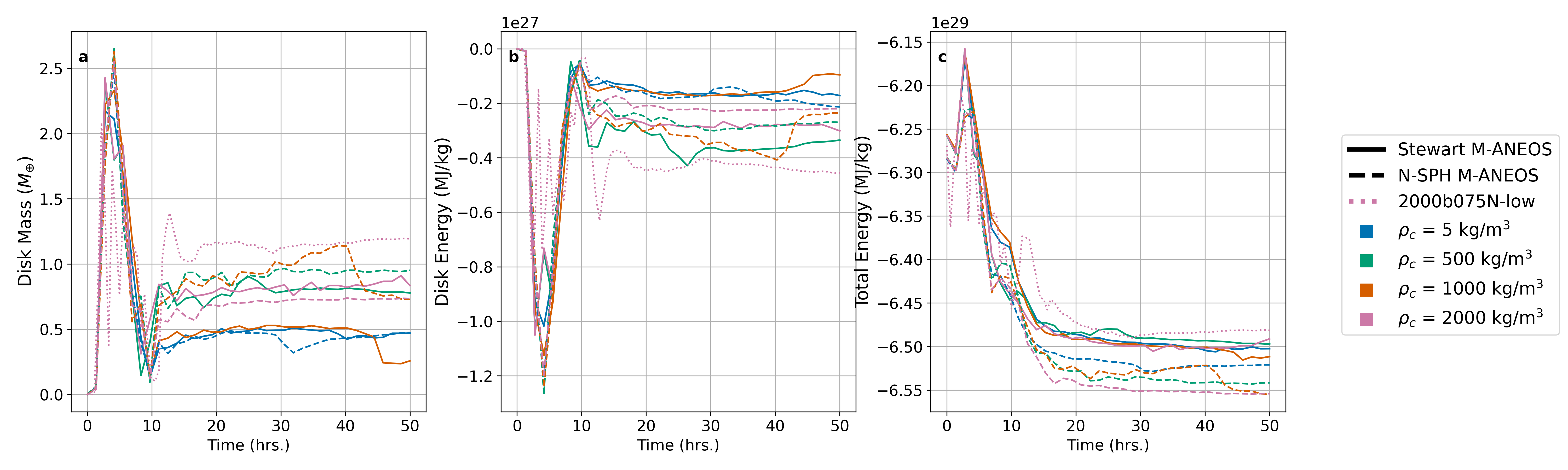}
\caption{Energy conservation for simulations with $b=0.75$.  FDPS SPH conserves energy within $\sim 1\%$, showing that the numerical issues associated with small values of $\rho_c$ do not violate energy conservation beyond typical numerical noise in SPH.  \textbf{(a)}. Mass in the disk, \textbf{(b)}. Total specific energy in the disk.  Some variance is associated with the changing mass of the disk as a function of time, \textbf{(c)}. Total specific energy of the entire simulation, including the protoearth, disk, and escaping particles.}
\label{fig:energy_conservation_b075}
\end{figure}
\begin{figure}[h]
\centering
\includegraphics[width=1\textwidth]{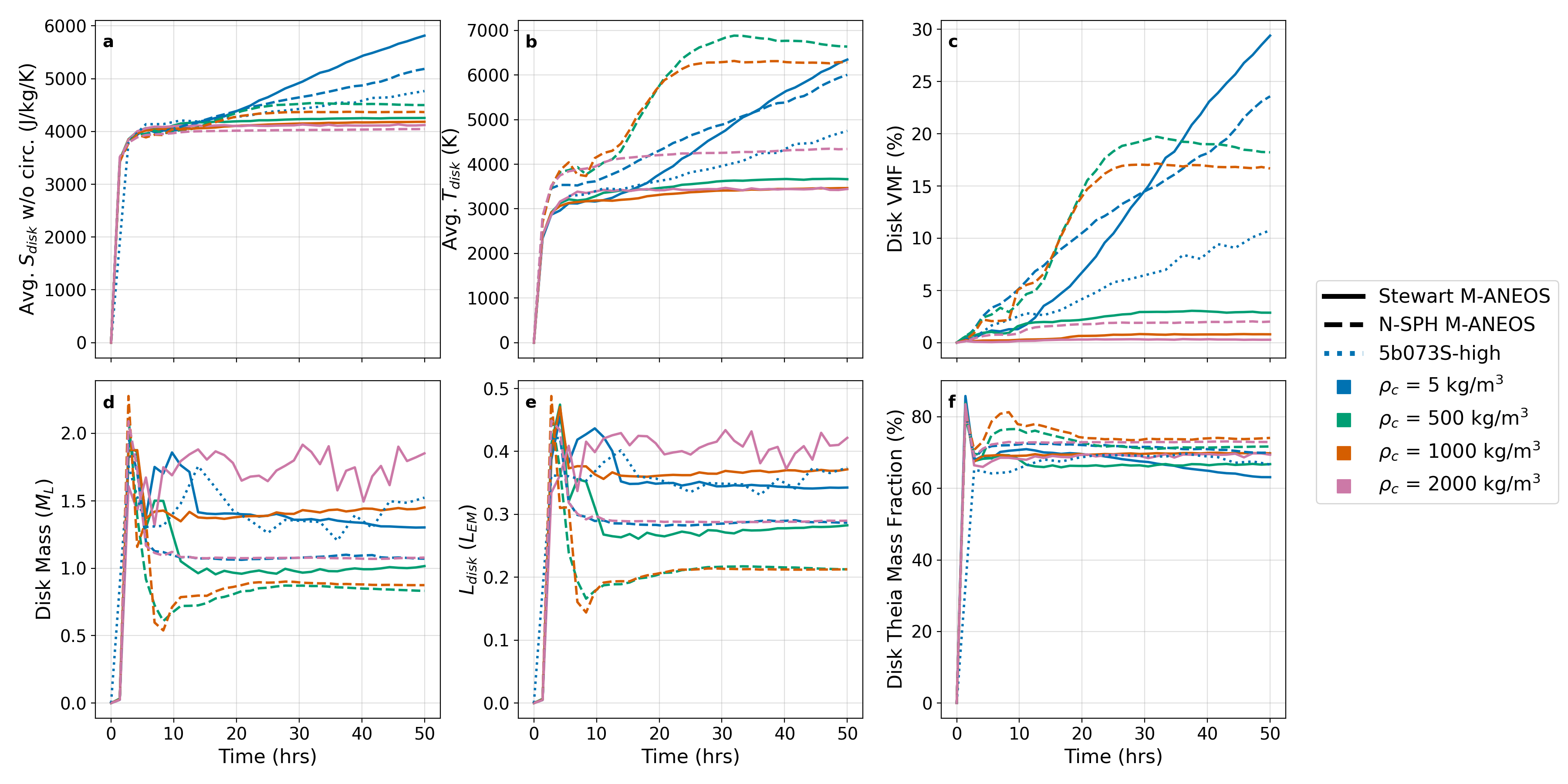}
\caption{Companion figure to Figure \ref{fig:b073_disk_entropy_and_vmf} for simulations with $b=0.73$, but the entropy gain and related increase in VMF associated with orbital circularization of the disk are not included in subplots (a) and (c).}
\label{fig:b073_disk_entropy_and_vmf_wo_circ}
\end{figure}
\begin{figure}[h]
\centering
\includegraphics[width=1\textwidth]{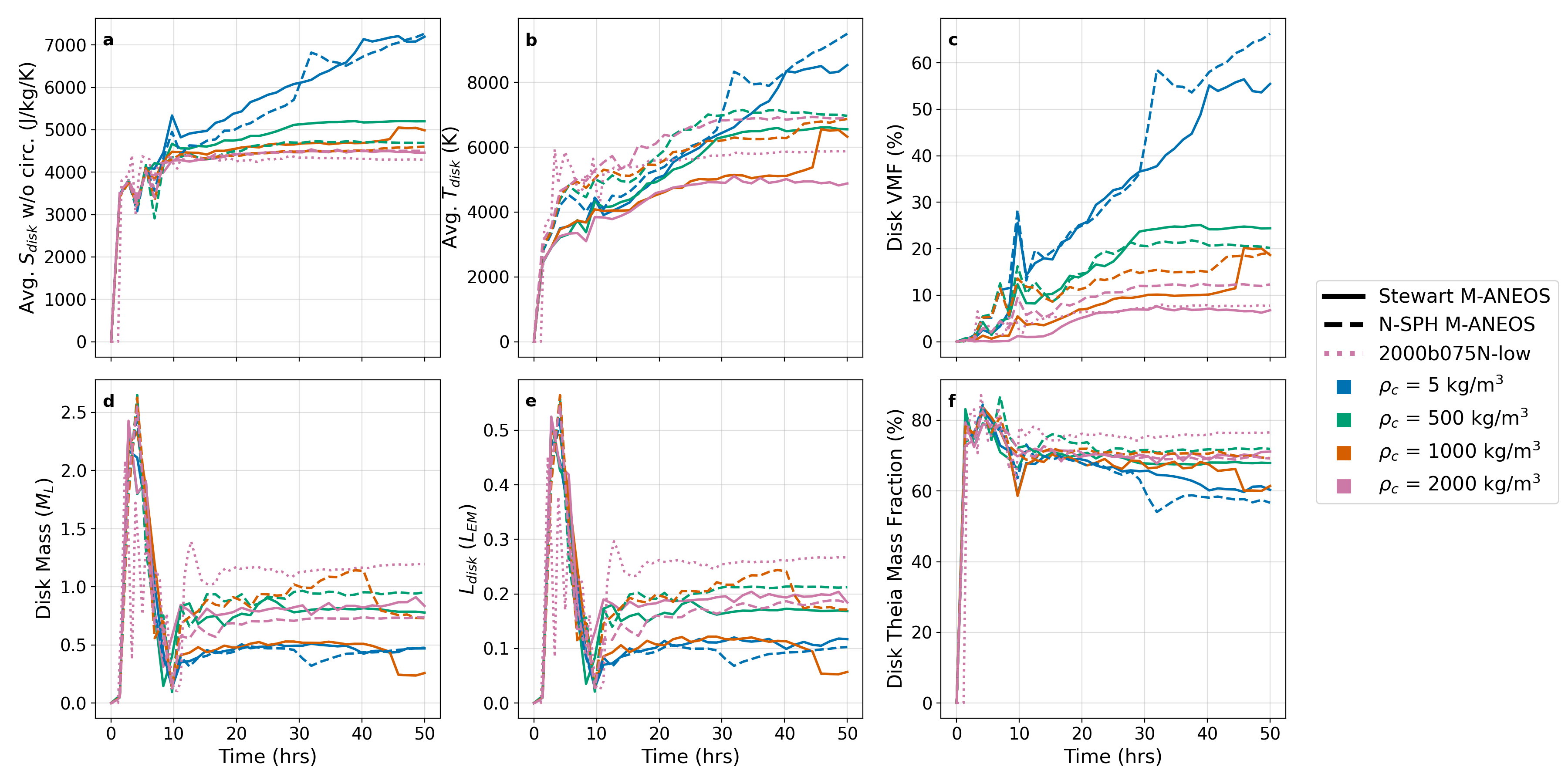}
\caption{Companion figure to Figure \ref{fig:b075_disk_entropy_and_vmf} for simulations with $b=0.75$, but the entropy gain and related increase in VMF associated with orbital circularization of the disk are not included in subplots (a) and (c).}
\label{fig:b075_disk_entropy_and_vmf_wo_circ}
\end{figure}

%
Figure \ref{fig:phase_curves_no_circ} is a companion to Figure \ref{fig:phase_curves} in the main text, showing disk particles on the phase diagram but without the entropy gain associated with orbital circularization. Figure \ref{fig:particle_fraction_at_rho_cutoff} shows the fraction of final disk particles with densities equal to $\rho_c$ as a function of simulated time.  We show the pressure as a function of density for disk particles in each simulation in Figure \ref{fig:paper1_disk_density_vs_pressure}.  To further test numerical stability beneath $\rho_c = 500$ $\rm kg/m^3$, we include the results of $\rho_c = 100$ $\rm kg/m^3$ at a $b=0.73$ impact angle in Figure \ref{fig:b073_disk_entropy_and_vmf_w_100}.  These runs with $\rho_c = 100$ $\rm kg/m^3$ show numerical stability through the duration of the simulation as this value remains above the instantaneous density jump due to briefly-encountering particles in the disk.  Figures \ref{fig:energy_conservation_b073} and \ref{fig:energy_conservation_b075} show that specific energy is conserved in FDPS SPH within an order of magnitude and that the numerical instability of low-$\rho_c$ simulations does not cause significant energy conservation problems within the entirety of the simulated system.  Figures \ref{fig:b073_disk_entropy_and_vmf_wo_circ} and \ref{fig:b075_disk_entropy_and_vmf_wo_circ} are companions to Figures \ref{fig:b073_disk_entropy_and_vmf} and \ref{fig:b075_disk_entropy_and_vmf}, but without the entropy gain and additional vapor production associated with orbital circularization of the disk.

\begin{figure}[h]
\centering
\includegraphics[width=1\textwidth]{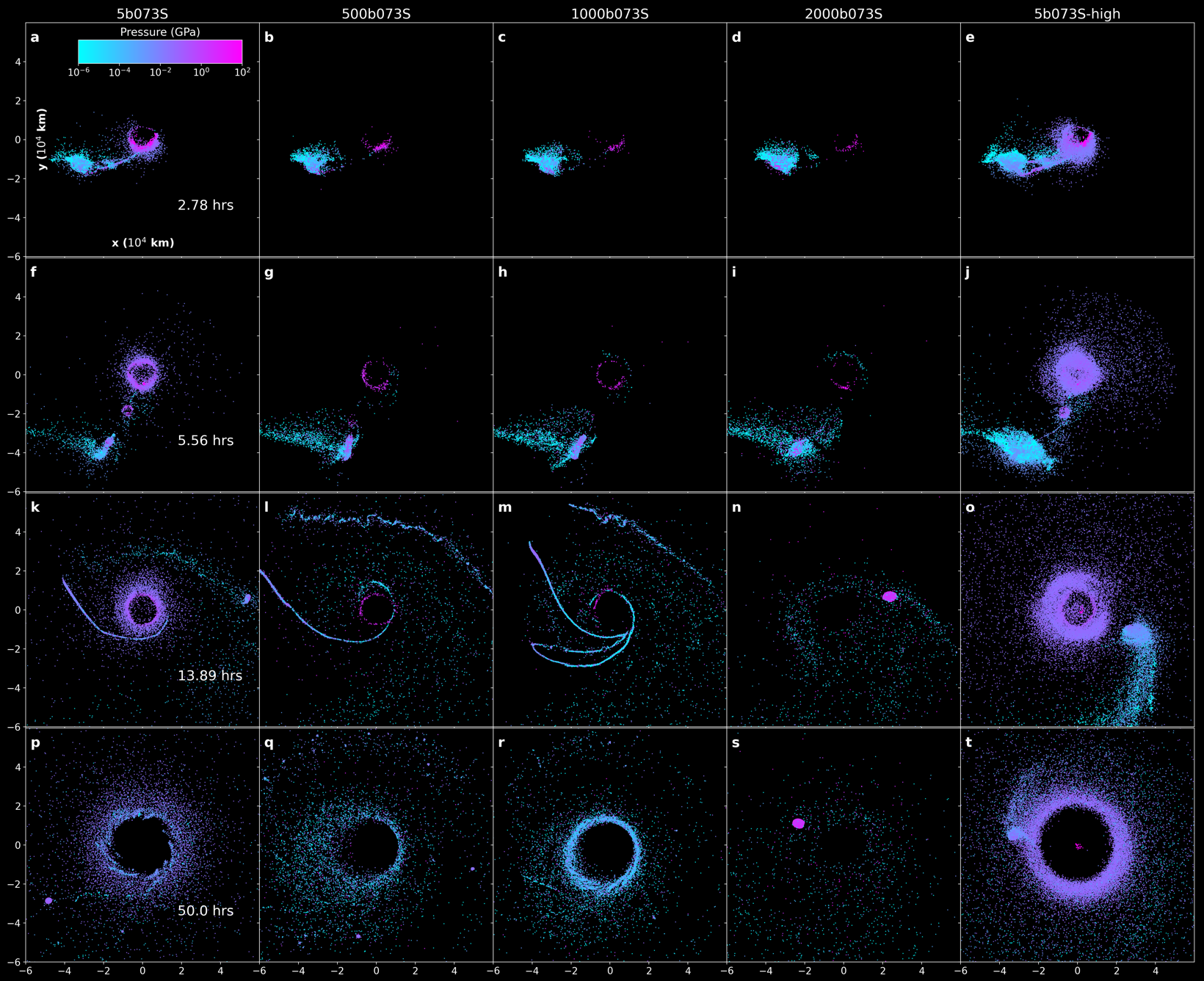}
\caption{Disk particles colored as a function of pressure for simulations using Stewart M-ANEOS with $b=0.73$.  The axes give the x-y spatial coordinates relative to the protoearth's center of mass given in increments of $10^4$ km.  Particles are sorted by their z-axis values.  Each column corresponds to a unique run and each row is a snapshot of the simulation at \textbf{(a-e)} 2.78 hours, \textbf{(f-j)} 5.56 hours, \textbf{(k-o)} 13.89 hours, \textbf{(p-t)} 50 hours.}
\label{fig:pressure_source_scenes_b073_new}
\end{figure}

\begin{figure}[h]
\centering
\includegraphics[width=1\textwidth]{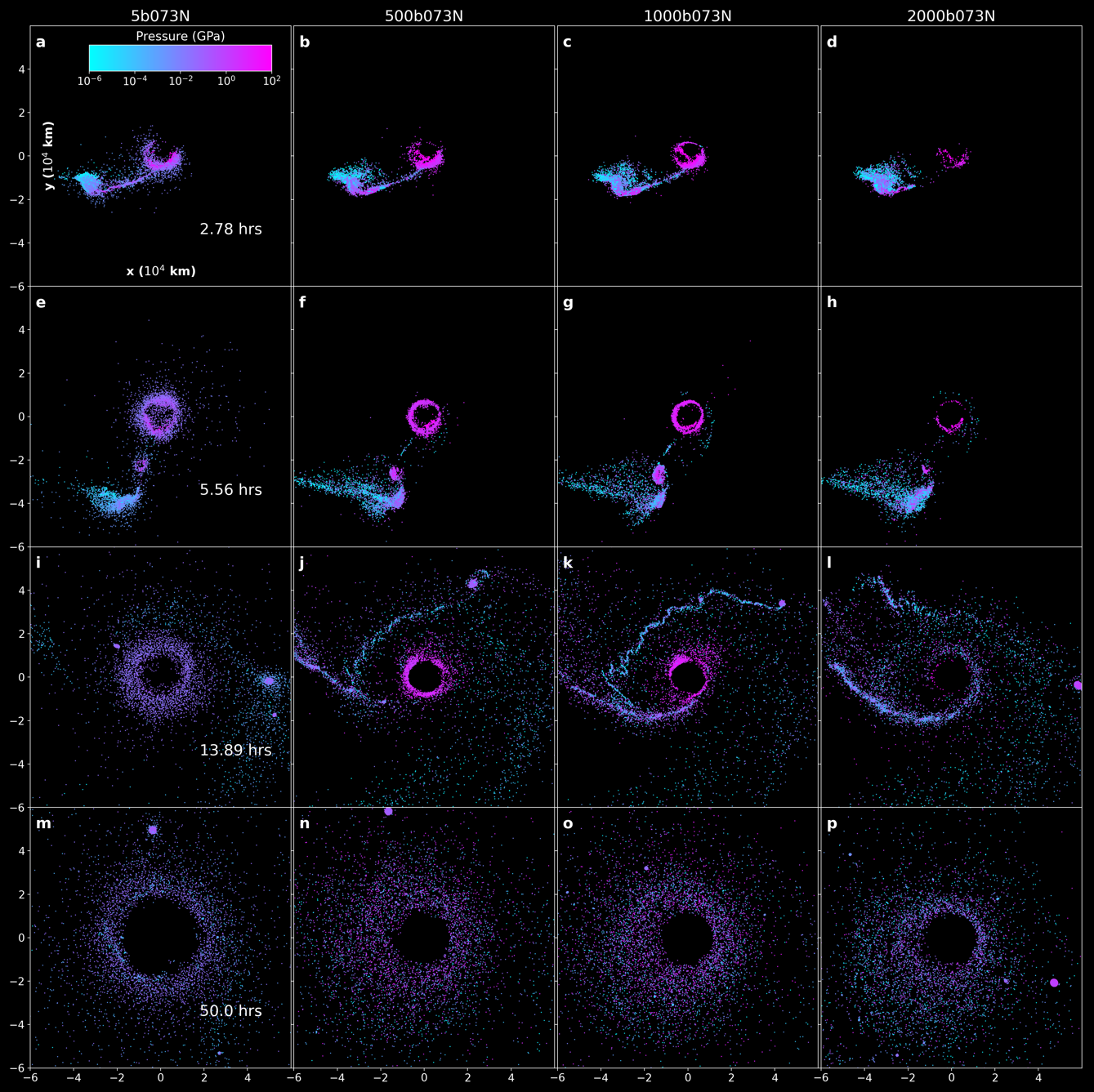}
\caption{Disk particles colored as a function of pressure for simulations using N-SPH M-ANEOS with $b=0.73$.  The axes give the x-y spatial coordinates relative to the protoearth's center of mass given in increments of $10^4$ km.  Particles are sorted by their z-axis values.  Each column corresponds to a unique run and each row is a snapshot of the simulation at \textbf{(a-d)} 2.78 hours, \textbf{e-h)} 5.56 hours, \textbf{(i-l)} 13.89 hours, \textbf{(m-p)} 50 hours.}
\label{fig:pressure_source_scenes_b073_old}
\end{figure}

\begin{figure}[h]
\centering
\includegraphics[width=1\textwidth]{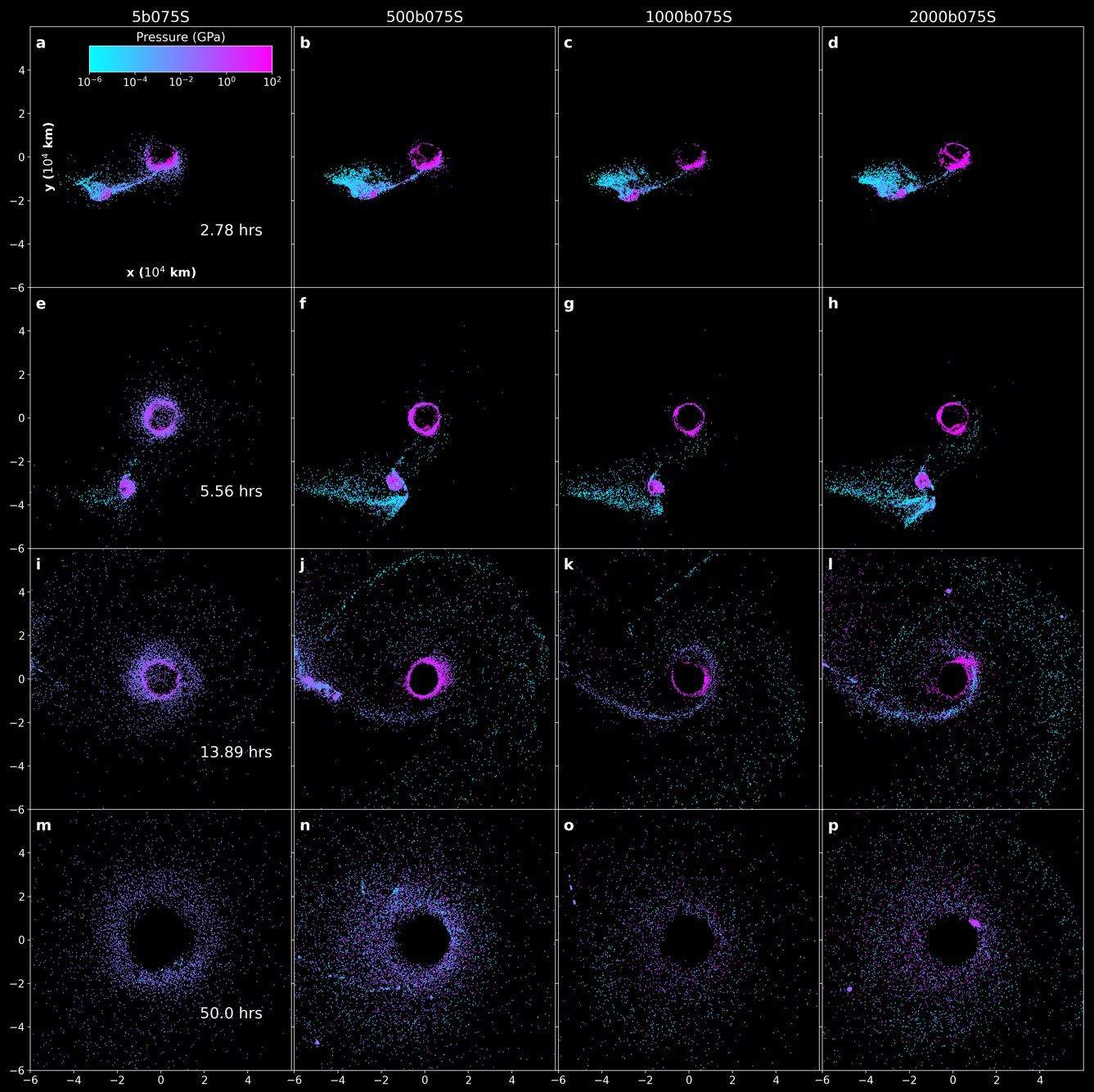}
\caption{Disk particles colored as a function of pressure for simulations using Stewart M-ANEOS with $b=0.75$.  The axes give the x-y spatial coordinates relative to the protoearth's center of mass given in increments of $10^4$ km.  Particles are sorted by their z-axis values.  Each column corresponds to a unique run and each row is a snapshot of the simulation at \textbf{(a-d)} 2.78 hours, \textbf{e-h)} 5.56 hours, \textbf{(i-l)} 13.89 hours, \textbf{(m-p)} 50 hours.}
\label{fig:pressure_source_scenes_b075_new}
\end{figure}

\begin{figure}[h]
\centering
\includegraphics[width=1\textwidth]{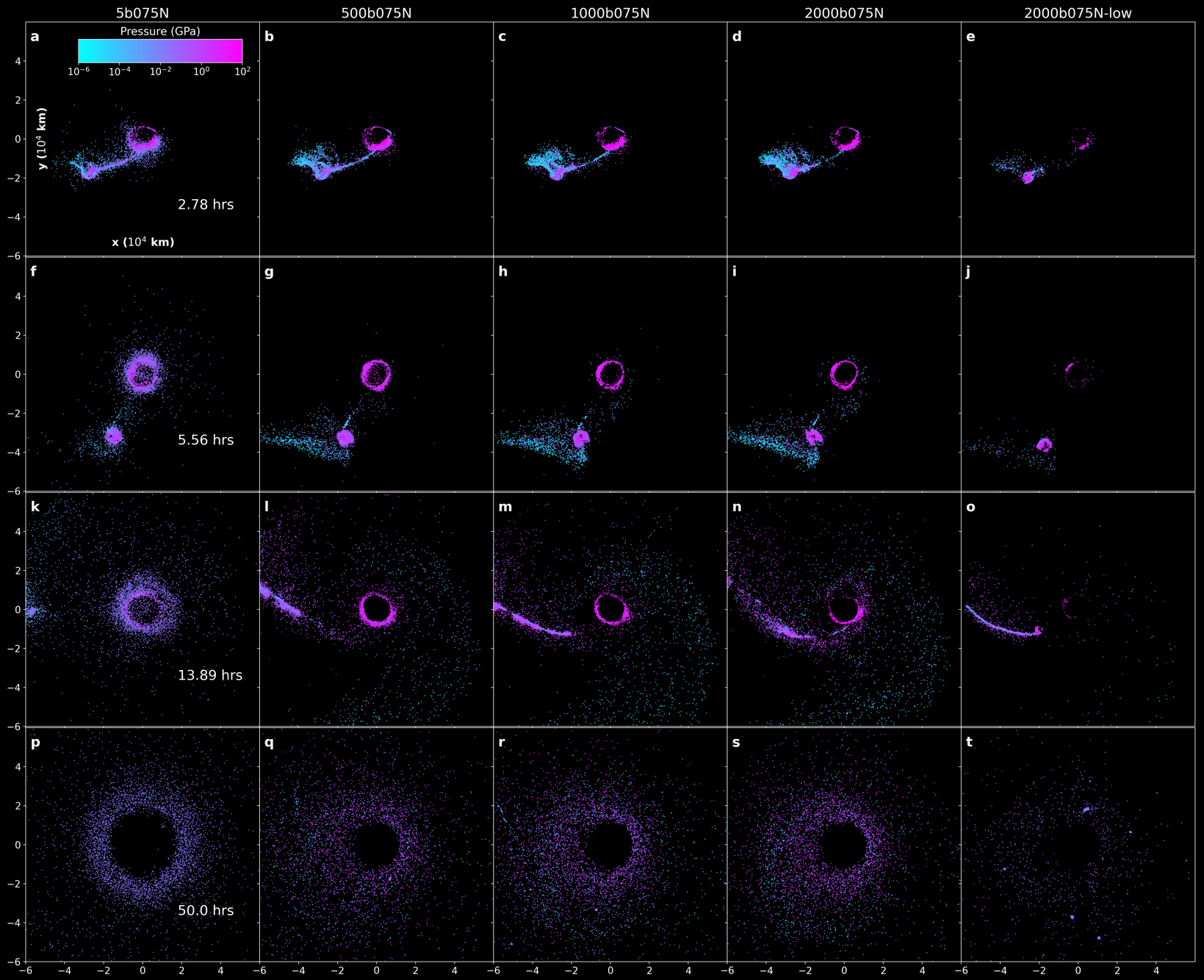}
\caption{Disk particles colored as a function of pressure for simulations using N-SPH M-ANEOS with $b=0.75$.  The axes give the x-y spatial coordinates relative to the protoearth's center of mass given in increments of $10^4$ km.  Particles are sorted by their z-axis values.  Each column corresponds to a unique run and each row is a snapshot of the simulation at \textbf{(a-e)} 2.78 hours, \textbf{(f-j)} 5.56 hours, \textbf{(k-o)} 13.89 hours, \textbf{(p-t)} 50 hours.}
\label{fig:pressure_source_scenes_b075_old}
\end{figure}

\begin{figure}[h]
\centering
\includegraphics[width=1\textwidth]{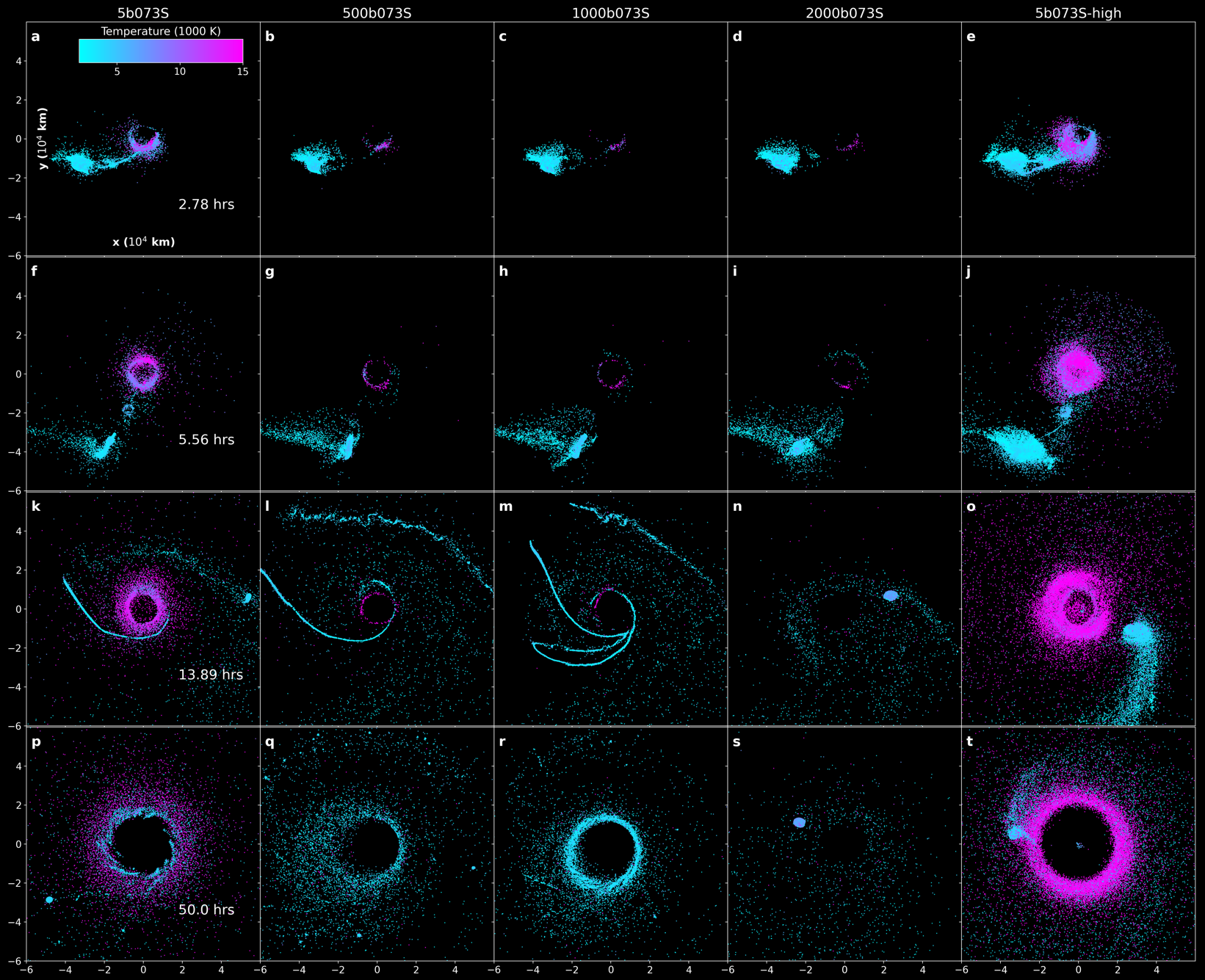}
\caption{Disk particles colored as a function of temperature for simulations using Stewart M-ANEOS with $b=0.73$.  The axes give the x-y spatial coordinates relative to the protoearth's center of mass given in increments of $10^4$ km.  Particles are sorted by their z-axis values.  Each column corresponds to a unique run and each row is a snapshot of the simulation at \textbf{(a-e)} 2.78 hours, \textbf{(f-j)} 5.56 hours, \textbf{(k-o)} 13.89 hours, \textbf{(p-t)} 50 hours.}
\label{fig:temperature_source_scenes_b073_new}
\end{figure}

For all simulations, we visualize disk pressures in Figures \ref{fig:pressure_source_scenes_b073_new}, \ref{fig:pressure_source_scenes_b073_old}, \ref{fig:pressure_source_scenes_b075_new}, and \ref{fig:pressure_source_scenes_b075_old}.  We find that all disks, including those with $\rho_c = 5$ $\rm kg/m^3$, experience pressure forces.  Runs 5b073S and 5b073S-high are notable in their increased disk pressures relative to the other runs.  Both are associated with increased temperatures in regions of the disk sourced from the original impact site (Figure \ref{fig:temperature_source_scenes_b073_new}).  While a characteristic feature of run 5b073S-high is the survival of a large clump in orbit, run 5b073S is notable for its stratified disk.  The hotter, higher-pressure parts of the disk sourced from around the impact site spread outwards to compose the outer regions of the disk, whereas the roping structure produced by the tidal stretching and breakup of a secondary clump (i.e., Figures \ref{fig:pressure_source_scenes_b073_new}f and \ref{fig:pressure_source_scenes_b073_new}k) form the tight, coherent, and compact interior region of the disk, which takes on a spiral-like shape by 50 hours (e.g., Figure \ref{fig:pressure_source_scenes_b073_new}k).  The different outcomes of disks across all runs are both combined functions of $\rho_c$ and EoS, and the stratification of the disk produced by 5b073S is a direct result of the chosen EoS and its influence of the post-impact evolution of the disk described in the main text results (Section \ref{results}).  Across all runs, we also find that disk pressures are broadly low enough to support both vapor and supercritical phases (see \cite{caracas2023no}, their Figure 3).

\begin{figure}[h]
\centering
\includegraphics[width=1\textwidth]{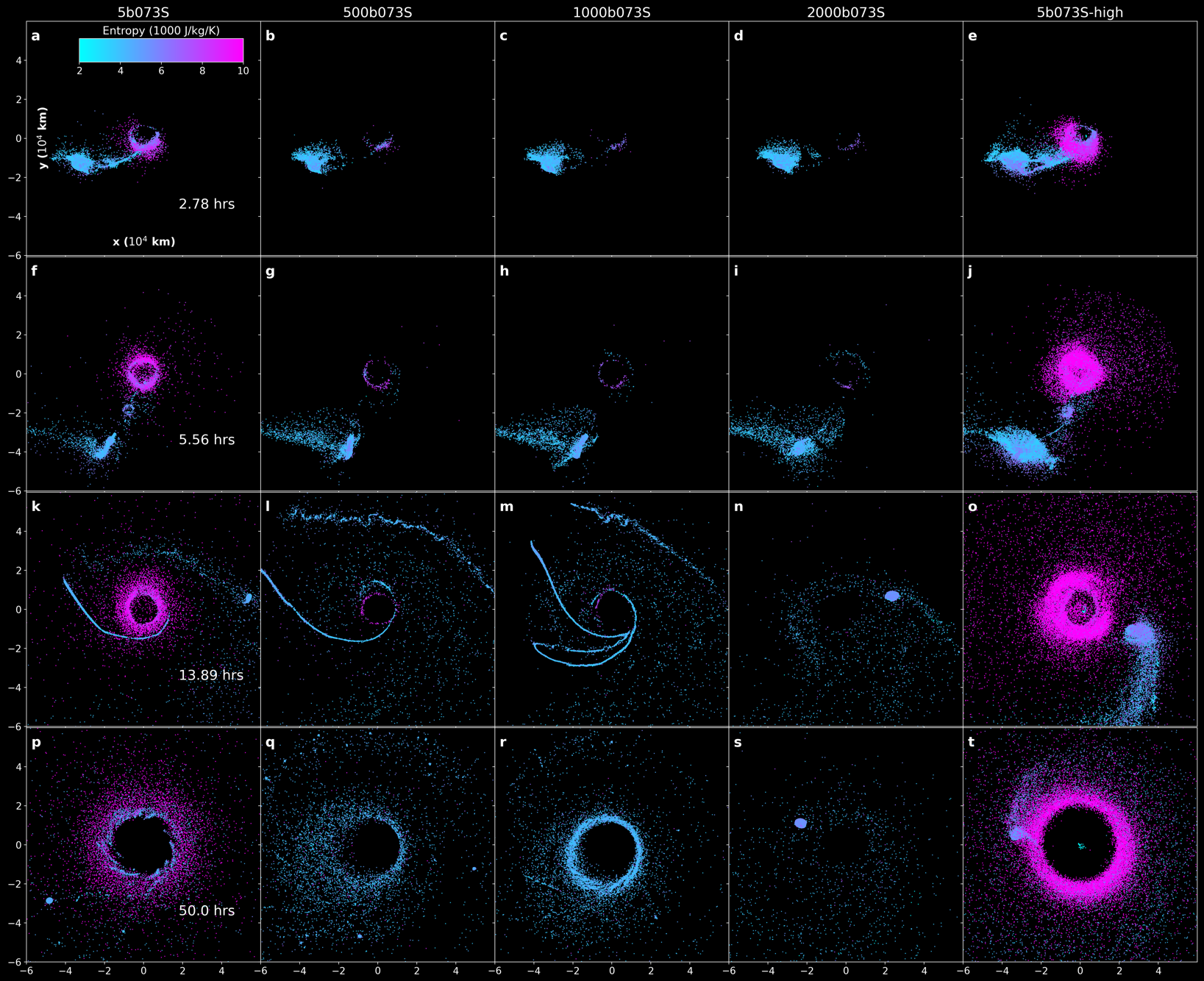}
\caption{Disk particles colored as a function of entropy for simulations using Stewart M-ANEOS with $b=0.73$.  The axes give the x-y spatial coordinates relative to the protoearth's center of mass given in increments of $10^4$ km.  Particles are sorted by their z-axis values.  Each column corresponds to a unique run and each row is a snapshot of the simulation at \textbf{(a-e)} 2.78 hours, \textbf{(f-j)} 5.56 hours, \textbf{(k-o)} 13.89 hours, \textbf{(p-t)} 50 hours.}
\label{fig:entropy_source_scenes_b073_new}
\end{figure}

\begin{figure}[h]
\centering
\includegraphics[width=1\textwidth]{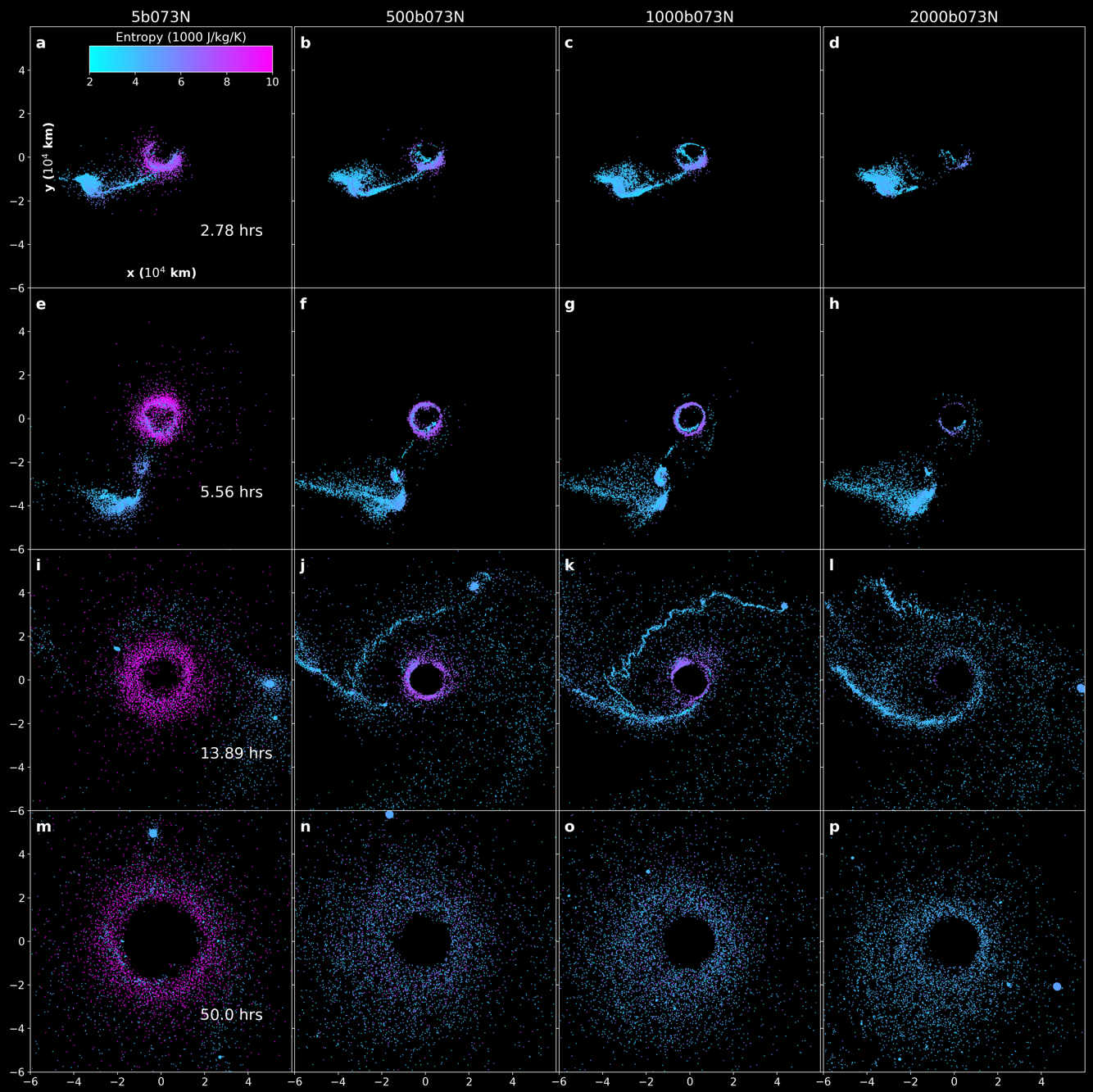}
\caption{Disk particles colored as a function of entropy for simulations using N-SPH M-ANEOS with $b=0.73$.  The axes give the x-y spatial coordinates relative to the protoearth's center of mass given in increments of $10^4$ km.  Particles are sorted by their z-axis values.  Each column corresponds to a unique run and each row is a snapshot of the simulation at \textbf{(a-d)} 2.78 hours, \textbf{e-h)} 5.56 hours, \textbf{(i-l)} 13.89 hours, \textbf{(m-p)} 50 hours.}
\label{fig:entropy_source_scenes_b073_old}
\end{figure}

\begin{figure}[h]
\centering
\includegraphics[width=1\textwidth]{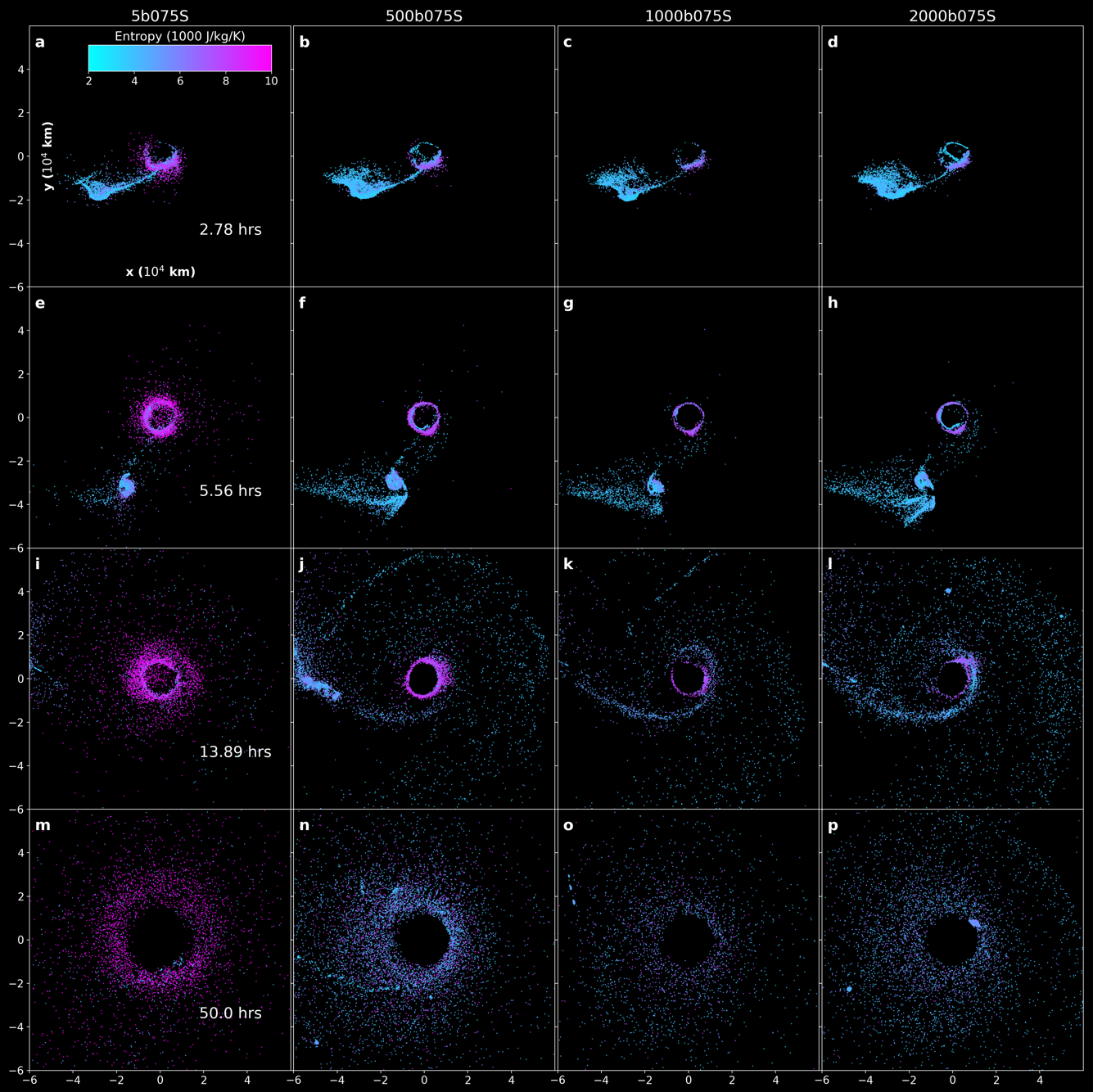}
\caption{Disk particles colored as a function of entropy for simulations using Stewart M-ANEOS with $b=0.75$.  The axes give the x-y spatial coordinates relative to the protoearth's center of mass given in increments of $10^4$ km.  Particles are sorted by their z-axis values.  Each column corresponds to a unique run and each row is a snapshot of the simulation at \textbf{(a-d)} 2.78 hours, \textbf{e-h)} 5.56 hours, \textbf{(i-l)} 13.89 hours, \textbf{(m-p)} 50 hours.}
\label{fig:entropy_source_scenes_b075_new}
\end{figure}

\begin{figure}[h]
\centering
\includegraphics[width=1\textwidth]{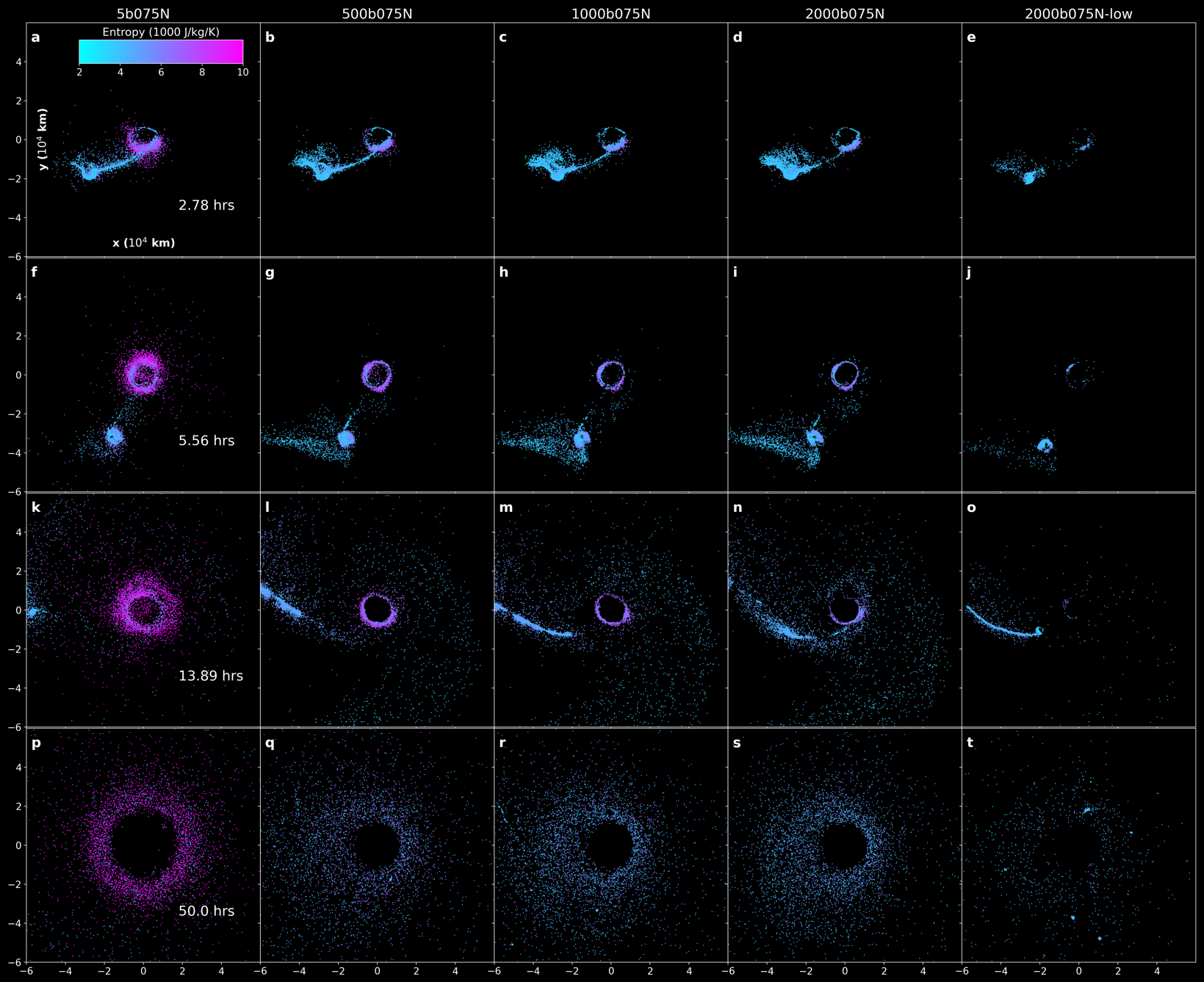}
\caption{Disk particles colored as a function of entropy for simulations using N-SPH M-ANEOS with $b=0.75$.  The axes give the x-y spatial coordinates relative to the protoearth's center of mass given in increments of $10^4$ km.  Particles are sorted by their z-axis values.  Each column corresponds to a unique run and each row is a snapshot of the simulation at \textbf{(a-e)} 2.78 hours, \textbf{(f-j)} 5.56 hours, \textbf{(k-o)} 13.89 hours, \textbf{(p-t)} 50 hours.}
\label{fig:entropy_source_scenes_b075_old}
\end{figure}

\begin{figure}[h]
\centering
\includegraphics[width=1\textwidth]{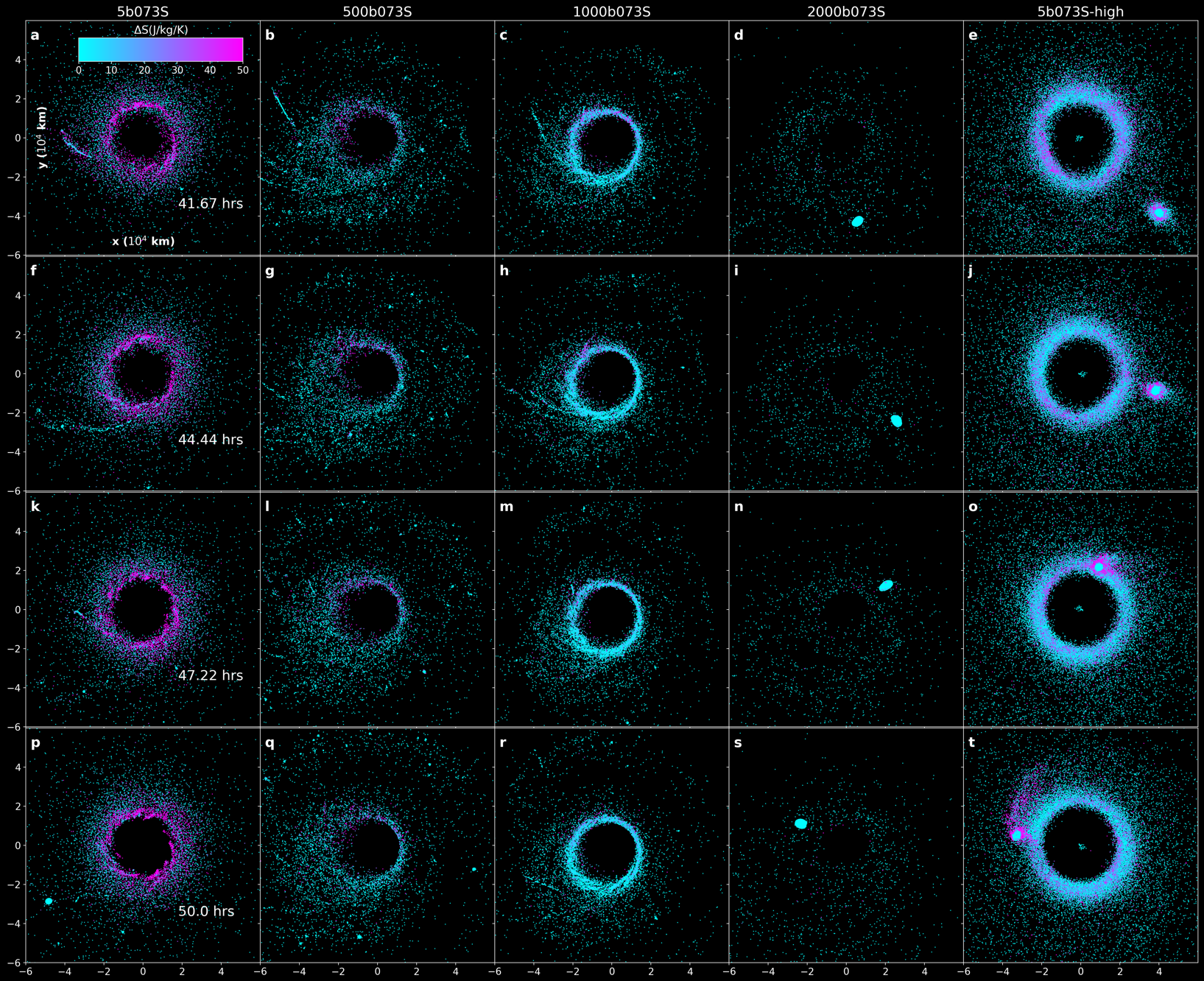}
\caption{Disk particles from runs with $b=0.73$ and using Stewart M-ANEOS colored as a function of specific entropy change ($\Delta S$) between the displayed time in each row and $\sim 3$ hours prior.  The axes give the x-y spatial coordinates relative to the protoearth's center of mass given in increments of $10^4$ km.  Particles are sorted by their z-axis values.  Each column corresponds to a unique run and each row is a snapshot of the simulation at \textbf{(a-e)} 41.67 hours, \textbf{(f-j)} 44.45 hours, \textbf{(k-o)} 47.22 hours, \textbf{(p-t)} 50 hours.}
\label{fig:delta_S_source_scenes_b073_new}
\end{figure}

\begin{figure}[h]
\centering
\includegraphics[width=1\textwidth]{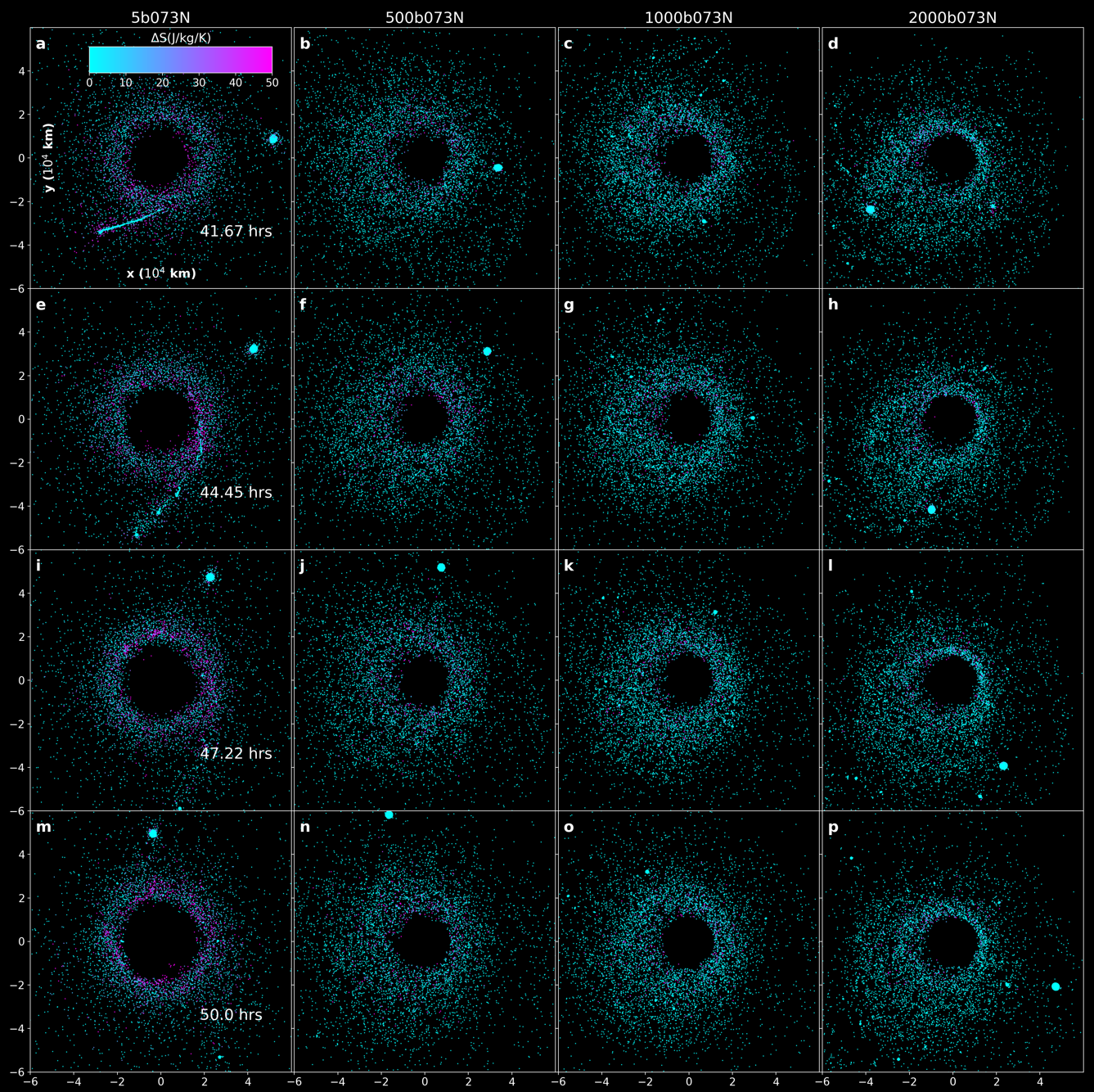}
\caption{Disk particles from runs with $b=0.73$ and using N-SPH M-ANEOS colored as a function of specific entropy change ($\Delta S$) between the displayed time in each row and $\sim 3$ hours prior.  The axes give the x-y spatial coordinates relative to the protoearth's center of mass given in increments of $10^4$ km.  Particles are sorted by their z-axis values.  Each column corresponds to a unique run and each row is a snapshot of the simulation at \textbf{(a-d)} 41.67 hours, \textbf{(e-h)} 44.45 hours, \textbf{(i-l)} 47.22 hours, \textbf{(m-p)} 50 hours.}
\label{fig:delta_S_source_scenes_b073_old}
\end{figure}

\begin{figure}[h]
\centering
\includegraphics[width=1\textwidth]{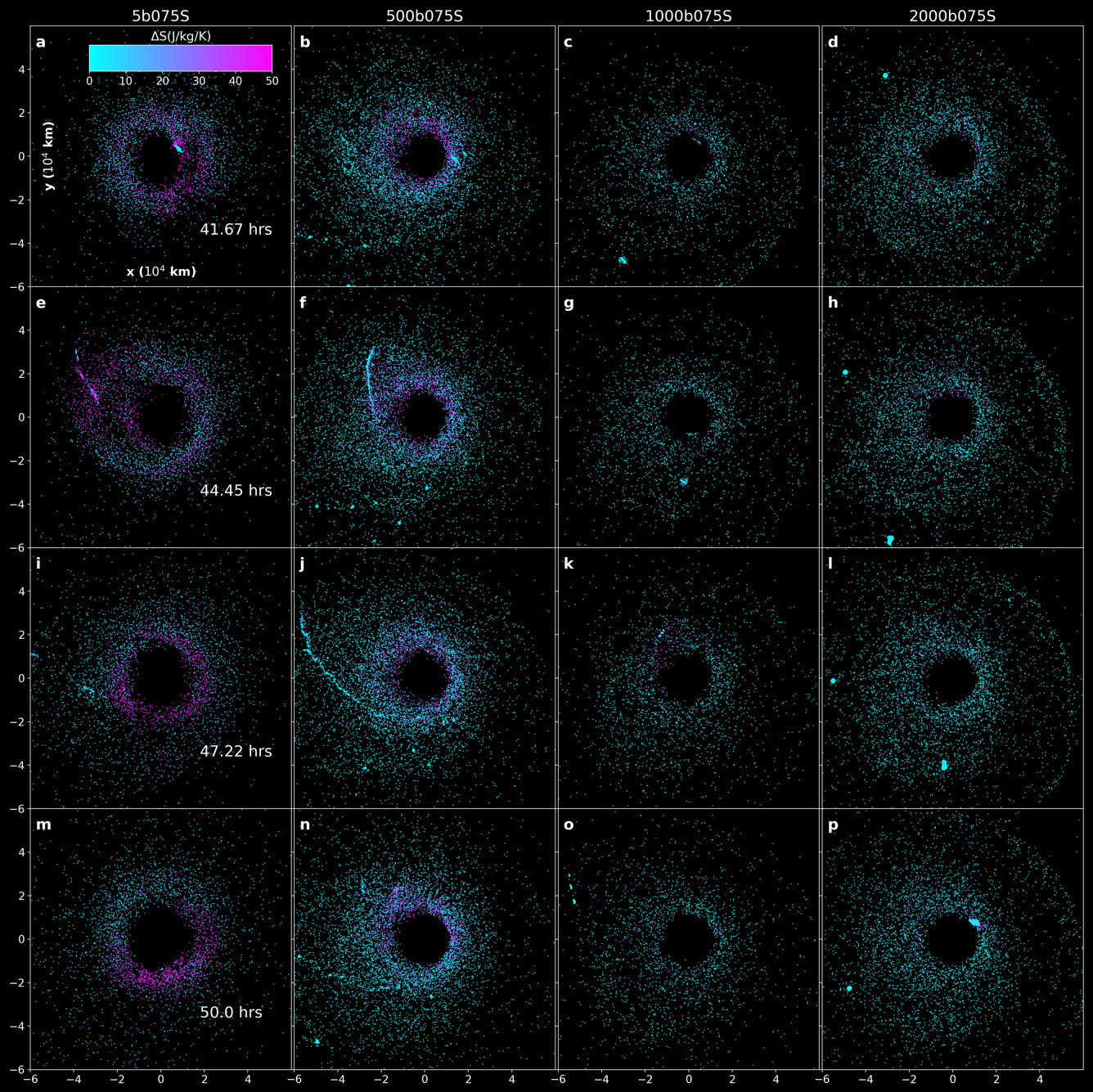}
\caption{Disk particles from runs with $b=0.75$ and using Stewart SPH M-ANEOS colored as a function of specific entropy change ($\Delta S$) between the displayed time in each row and $\sim 3$ hours prior.  The axes give the x-y spatial coordinates relative to the protoearth's center of mass given in increments of $10^4$ km.  Particles are sorted by their z-axis values.  Each column corresponds to a unique run and each row is a snapshot of the simulation at \textbf{(a-d)} 41.67 hours, \textbf{(e-h)} 44.45 hours, \textbf{(i-l)} 47.22 hours, \textbf{(m-p)} 50 hours.}
\label{fig:delta_S_source_scenes_b075_new}
\end{figure}

\begin{figure}[h]
\centering
\includegraphics[width=1\textwidth]{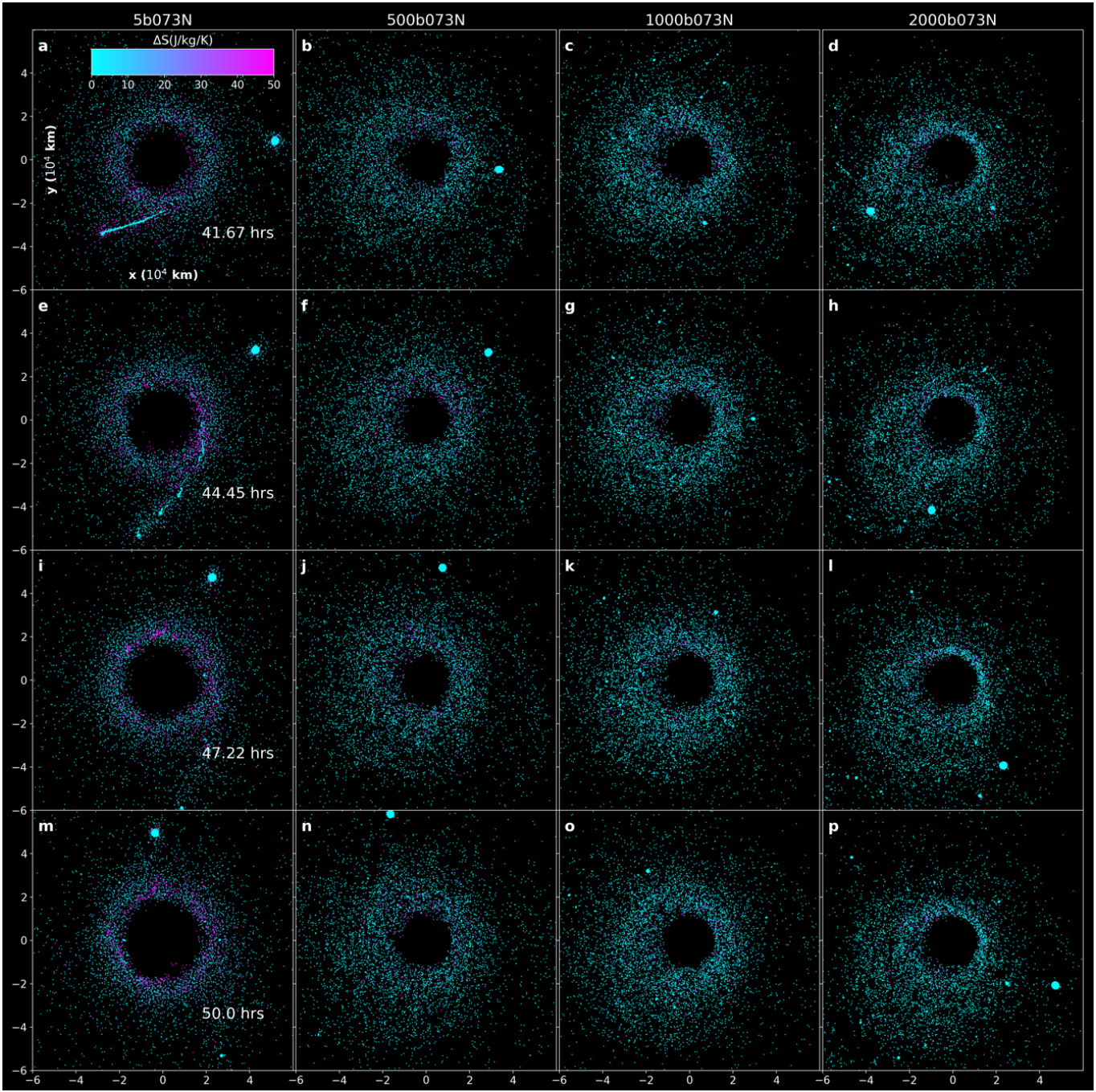}
\caption{Disk particles from runs with $b=0.75$ and using N-SPH M-ANEOS colored as a function of specific entropy change ($\Delta S$) between the displayed time in each row and $\sim 3$ hours prior.  The axes give the x-y spatial coordinates relative to the protoearth's center of mass given in increments of $10^4$ km.  Particles are sorted by their z-axis values.  Each column corresponds to a unique run and each row is a snapshot of the simulation at \textbf{(a-e)} 41.67 hours, \textbf{(f-j)} 44.45 hours, \textbf{(k-o)} 47.22 hours, \textbf{(p-t)} 50 hours.}
\label{fig:delta_S_source_scenes_b075_old}
\end{figure}

\begin{figure}[h]
\centering
\includegraphics[width=1\textwidth]{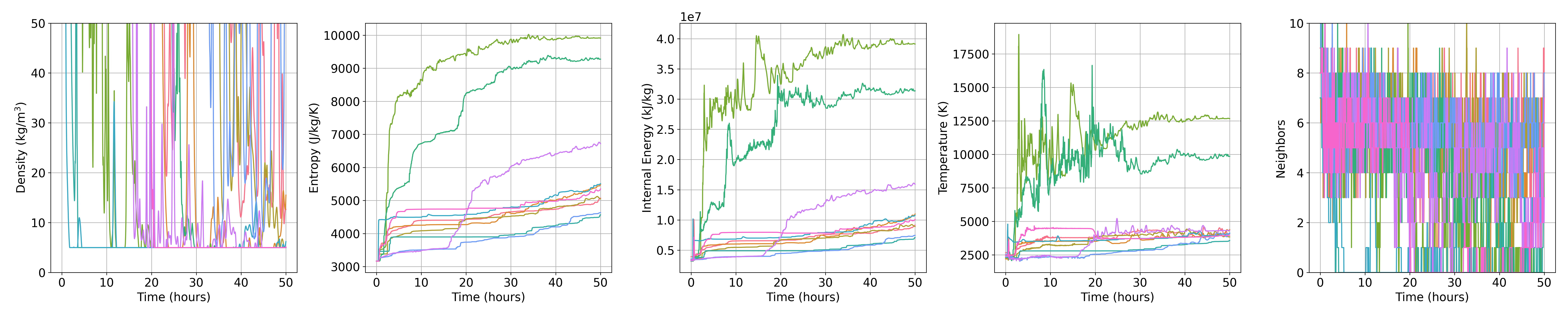}
\caption{A sample of disk particles from the 5b073S run showing numerical shocks and instability as a function of simulation time.  Left to right: Particle density, particle entropy, particle specific internal energy, particle temperature, number of neighboring particles within the smoothing length $H_i$.  The density y-axis is artificially limited to emphasize density perturbations around $\rho_c$ late in the simulation.}
\label{fig:5_b073_new_sawtooth_entropy_all_disk}
\end{figure}

\begin{figure}[h]
\centering
\includegraphics[width=1\textwidth]{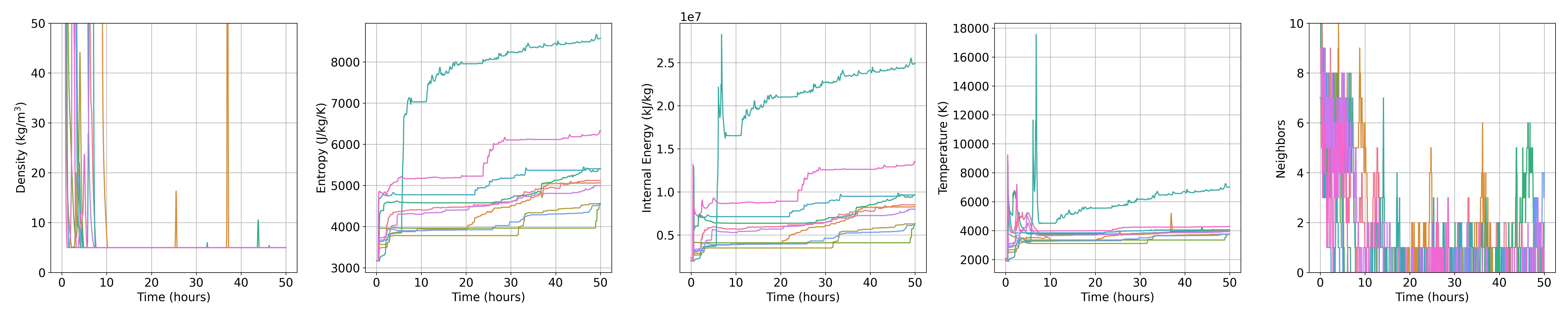}
\caption{A sample of disk particles from the 5b073N run showing numerical shocks and instability as a function of simulation time.  Left to right: Particle density, particle entropy, particle specific internal energy, particle temperature, number of neighboring particles within the smoothing length $H_i$.  The density y-axis is artificially limited to emphasize density perturbations around $\rho_c$ late in the simulation.}
\label{fig:5_b073_old_sawtooth_entropy_all_disk}
\end{figure}

To assess the behavior of disk particles that experience artificial shocks and runaway thermodynamic growth at $\rho_c$ (Section \ref{treatment-rho-c}), we plot disk entropies for all runs in Figures \ref{fig:entropy_source_scenes_b073_new}, \ref{fig:entropy_source_scenes_b073_old}, \ref{fig:entropy_source_scenes_b075_new}, and \ref{fig:entropy_source_scenes_b075_old}.  Much like our observations above with regard to temperature and pressure, we find that high-entropy regions of the disks from runs 5b073S-high and 5b073S are associated with ejecta originating from near the impact site. 
 However, higher-entropy disks are ubiquitous across all runs at $\rho_c = 5$ $\rm kg/m^3$.  To highlight where numerical instability is occurring most in the the disk at our low-$\rho_c$ runs, we also plot the increase in entropy over the last 10 hours of our 50 hour simulation, $\Delta S$, in 3 hour increments (Figures \ref{fig:delta_S_source_scenes_b073_new}, \ref{fig:delta_S_source_scenes_b073_old}, \ref{fig:delta_S_source_scenes_b075_new}, \ref{fig:delta_S_source_scenes_b075_old}).  We also randomly sample disk particles in each run within runs 5b073S and 5b073N that experience considerable entropy growth over time (Figures \ref{fig:5_b073_new_sawtooth_entropy_all_disk} and \ref{fig:5_b073_old_sawtooth_entropy_all_disk}).  In tracking these sample particles in animations of the disk, we find they become isentropic when they are isolated in the outer-regions of the disk and experience gains in entropy as they move into the more particle-dense interior region of the disk or encounter tidally-disrupted moonlets stretching and spreading through the disk (i.e., as observed in Figure \ref{fig:delta_S_source_scenes_b073_old}, column 1).  This is particularly noticeable in run 5b073S, whose coherent disk interior results in sample particles continuously passing through the coherent spiralled interior of the disk, resulting in density and thermodynamic instability (Figure \ref{fig:5_b073_new_sawtooth_entropy_all_disk}).  In contrast, the sample particles in run 5b073N are more stable in density, but entropy growth is driven by corresponding, sporadic changes in particle energy and not necessarily density jumps (Figure \ref{fig:5_b073_old_sawtooth_entropy_all_disk}).  These energy changes may be similar to the ``$E_{\rm 0}$`'' error described by \cite{read2012sphs} and \cite{hopkins2015new}, which results from the particle's kernel not being sufficiently smoothed to fully satisfy the discretized SPH conservation equations, although the total effect of this error is not apparent enough to result in any significant violation of conservation of energy across the entire simulation (Figures \ref{fig:energy_conservation_b073}c and \ref{fig:energy_conservation_b075}c).  It is also possible that the density jumps are occurring, but our file output intervals (typically $\sim 100$ simulated seconds) are too coarse to capture them.

